\documentclass[epjST]{svjour}
\usepackage{graphicx}
\usepackage{mathtools}
\usepackage{setspace}

\begin{document}

\title{BASE - The Baryon Antibaryon Symmetry Experiment}
\author{
C. Smorra\inst{1,2}\fnmsep\thanks{\email{Christian.Smorra@cern.ch}} \and 
K. Blaum \inst{3} \and
L. Bojtar \inst{2} \and
M. Borchert \inst{4} \and
K.A. Franke \inst{3} \and 
T. Higuchi \inst{1,5} \and
N. Leefer \inst{6} \and
H. Nagahama \inst{1,5} \and
Y. Matsuda \inst{5} \and
A. Mooser \inst{1} \and
M. Niemann \inst{4} \and
C. Ospelkaus \inst{4,8} \and
W. Quint \inst{9,10} \and
G. Schneider \inst{7} \and
S. Sellner \inst{1} \and
T. Tanaka \inst{5} \and
S. Van Gorp \inst{11} \and
J. Walz \inst{5,6} \and
Y. Yamazaki \inst{11} \and
S. Ulmer \inst{1}
}
\institute{
Ulmer Initiative Research Unit, RIKEN, 2-1 Hirosawa, Wako, Saitama 351-0198, Japan \and 
CERN, CH-1211 Geneva 23, Switzerland \and 
Max-Planck-Institut f\"ur Kernphysik, Saupfercheckweg 1, D-69117 Heidelberg, Germany \and
Institute of Quantum Optics, Leibniz Universit\"at Hannover, Welfengarten 1, D-30167 Hannover, Germany \and
Graduate School of Arts and Sciences, University of Tokyo, 3-8-1 Komaba, Meguro-ku, Tokyo 153-8902, Japan \and
Helmholtz-Institut Mainz, D-55099 Mainz, Germany \and
Institut f\"ur Physik, Johannes Gutenberg-Universit\"at Mainz, D-55099 Mainz, Germany \and
Physikalisch-Technische Bundesanstalt, Bundesallee 100, D-38116 Braunschweig, Germany \and
GSI-Helmholtzzentrum f\"ur Schwerionenforschung, D-64291 Darmstadt, Germany \and
Ruprecht-Karls-Universit\"at Heidelberg, D-69047 Heidelberg, Germany \and
Atomic Physics Laboratory, RIKEN, 2-1 Hirosawa, Wako, Saitama 351-0198, Japan
}
\abstract{
The Baryon Antibaryon Symmetry Experiment (BASE) aims at performing a stringent test of the combined charge parity and time reversal (CPT) symmetry by comparing the magnetic moments of the proton and the antiproton with high precision. Using single particles in a Penning trap, the proton/antiproton $g$-factors, i.e. the magnetic moment in units of the nuclear magneton, are determined by measuring the respective ratio of the spin-precession frequency to the cyclotron frequency. The spin precession frequency is measured by non-destructive detection of spin quantum transitions using the continuous Stern-Gerlach effect, and the cyclotron frequency is determined from the particle's motional eigenfrequencies in the Penning trap using the invariance theorem. By application of the double Penning-trap method we expect that in our measurements a fractional precision of $\delta g/g$ 10$^{-9}$ can be achieved. The successful application of this method to the antiproton will represent a factor 1000 improvement in the fractional precision of its magnetic moment. The BASE collaboration has constructed and commissioned a new experiment at the Antiproton Decelerator (AD) of CERN. This article describes and summarizes the physical and technical aspects of this new experiment.
} 
\maketitle
%

\begin{spacing}{1.2}
\section{Introduction}
\label{Sect:1}
The invariance of the physical interactions under the combined charge parity and time reversal (CPT) transformation is one of the basic cornerstones of the Lorentz-invariant local quantum field theories of the Standard Model (SM). It states that the physical interactions under the combined transformation of charge conjugation (C), parity transformation (P) and time reversal (T) are identical. As consequence, particles and their conjugate antiparticles have identical masses, lifetimes, charges and magnetic moments, but the latter two with opposite sign. Therefore, precise comparisons of fundamental properties of antiparticles and their matter conjugates constitute stringent tests of CPT invariance. 

\begin{figure}[htb]
        \centerline{\includegraphics[width=0.8 \textwidth,keepaspectratio]{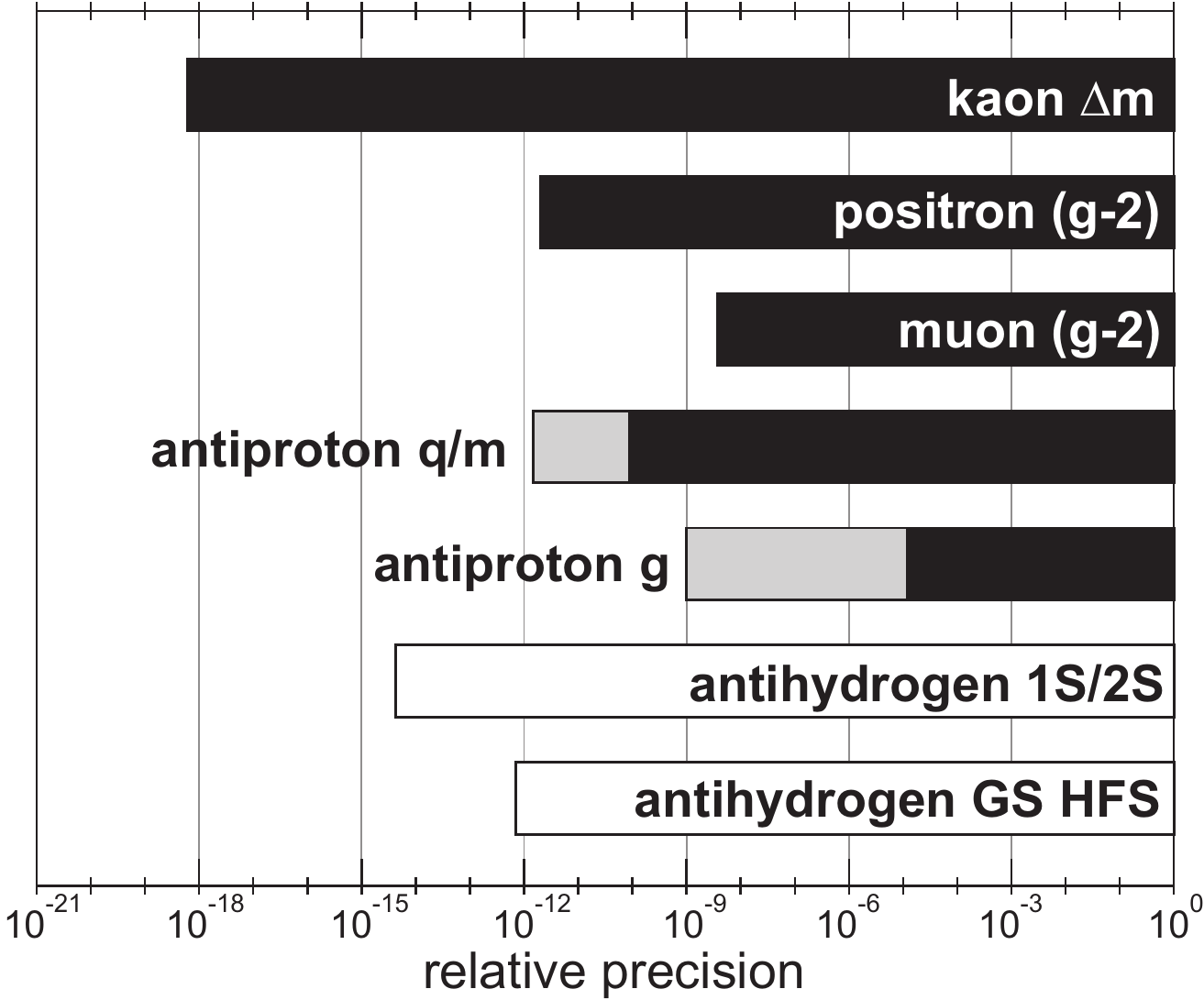}}
            \caption[Precision]{Overview of different high-precision tests of CPT invariance. The black bars represent the relative precision of performed measurements. BASE plans to improve the antiproton charge-to-mass ratio and the antiproton $g$-factor up to the precision shown by the respective gray bars. Other experiments are planning to perform CPT tests by spectroscopy of antihydrogen. The relative uncertainties shown as white bars are the best ones achieved in hydrogen for the respective transitions. For details see the text. }
						\label{fig:FigMerit}
    \end{figure}

Despite its importance in the SM, direct high-precision tests of CPT symmetry are scarce (see Fig.~\ref{fig:FigMerit}). A widely recognized CPT-test was carried out by comparing decay channels of the neutral mesons $K_0/\bar{K_0}$ to charged and neutral pions. Thereby, their relative mass difference was constrained to be less than 10$^{-18}$ \cite{KAONCPT,PDGMesons}. In other efforts, experiments with single particles in Penning traps have reported as well on tests of CPT invariance with great precision. By using the elegant continuous Stern-Gerlach effect for the non-destructive detection of spin eigenstates of single particles in Penning traps, electron and positron $(g-2)$ values were compared with better than 4 ppb precision, which allowed to compare their $g$-factors with $\delta g/g \approx 2\times10^{-12}$ uncertainty \cite{Dehmelt,DehmeltCPT}.
A recent improvement in the measurement of the electron $g-2$ value opens the exciting perspective to improve this test by at least a factor of 10 \cite{Hanneke2008}. Another precise test was performed by comparing the $g-2$ values of $\mu^+$ and $\mu^-$ in a storage ring with a fractional precision of 3.7$\times10^{-9}$ confirming CPT invariance \cite{Bennett2006,Bennett2008}. However, the muon $g$-factors deviate by 3.6 standard deviations from the Standard Model prediction, which has been interpreted to be caused by coupling to dark gauge bosons \cite{DarkPhoton,DarkPhoton2}. Currently, efforts are in progress to repeat these measurements with higher precision to resolve or confirm this deviation \cite{FNAC,Mibe}. 

The most precise comparisons of matter/antimatter pairs in the baryon sector are measurements of the proton and antiproton charge-to-mass ratios \cite{JerryAntiproton,UlmerNature2015}. These were first performed at CERN's low-energy antiproton ring LEAR by the TRAP collaboration by measuring the cyclotron-frequencies of antiprotons and protons $\nu_{c,\bar{p}}/\nu_{c,p}$ \cite{Jerry1995}. Later, a negatively-charged hydrogen ions (H$^-$) was used as a proxy of the proton, which allowed to increase the relative precision of the cyclotron frequency ratio $\nu_{c,\bar{p}}/\nu_{c,p}$ to 9.0$\times$10$^{-11}$ uncertainty \cite{JerryAntiproton}. The result, initially within a one $\sigma$ uncertainty consistent with CPT invariance, was later corrected by a polarization shift of the $H^-$ ion yielding a 1.8$\,\sigma$ deviation from the Standard Model prediction \cite{Pritchard,JerryReview2006}. Recently, we carried out a new comparison of the proton and antiproton charge-to-mass ratios in the BASE apparatus using also antiprotons and H$^-$ ions. By using adiabatic shuttling for the particle exchange \cite{SmorraIJMS2015} and sideband-coupling techniques for the cyclotron frequency measurements \cite{Cornell1990}, we were able to measure 6500 frequency ratios within 35 days and obtained $(q/m)_p/(q/m)_{\overline{p}}-1$=1(69)$\times 10^{-12}$, which is in excellent agreement with CPT invariance \cite{UlmerNature2015}. As our measurements were carried out at lower cyclotron frequencies compared to ref.~\cite{JerryAntiproton}, it provides a four times higher energy resolution to CPT violating effects \cite{UlmerNature2015,Dehmelt1999,KosteleckySME}.

In addition to these efforts, several collaborations at CERN's antiproton decelerator (AD) \cite{Maury1999HypInt} target precision spectroscopy of the electromagnetic properties of antihydrogen \cite{ALPHA,ASACUSA,ATRAP}. As its matter-counterpart hydrogen is one of the best understood composite systems in modern fundamental physics, comparisons of its properties to antihydrogen constitute a new branch of highly-sensitive CPT tests.
The 1S-2S transition frequency in hydrogen was measured with a relative uncertainty of 4.2$\times10^{-15}$ \cite{Parthey2011} using a cold beam of hydrogen atoms. First measurements of this transition in antihydrogen are planned to be carried out in magnetic gradient traps. By using hydrogen, relative precisions on the order of $10^{-12}$ have been achieved in such systems \cite{Cesar1996}. Another appealing possibility to perform high-precision tests of CPT invariance is the comparison of the ground-state hyperfine-splitting (GS-HFS) of hydrogen and antihydrogen. By using a MASER, the hydrogen GS-HFS transition has been measured with a fractional precision of 0.7$\,$ppt \cite{HydrogenMaser}. In case of antihydrogen first hyperfine transitions have been recently observed using appearance mode annihilation spectroscopy in a magnetic gradient trap by the ALPHA collaboration \cite{ALPHANature}. In parallel the ASACUSA collaboration reported on the first production of a beam of antihydrogen atoms \cite{ASACUSANautreComm} using a cusp trap. This is an important milestone towards antihydrogen spin transition spectroscopy using Rabi's molecular beam technique and to measure the GS-HFS transition frequency with sub-ppm precision.

\begin{figure}[htb]
        \centerline{\includegraphics[width=0.85 \textwidth,,keepaspectratio]{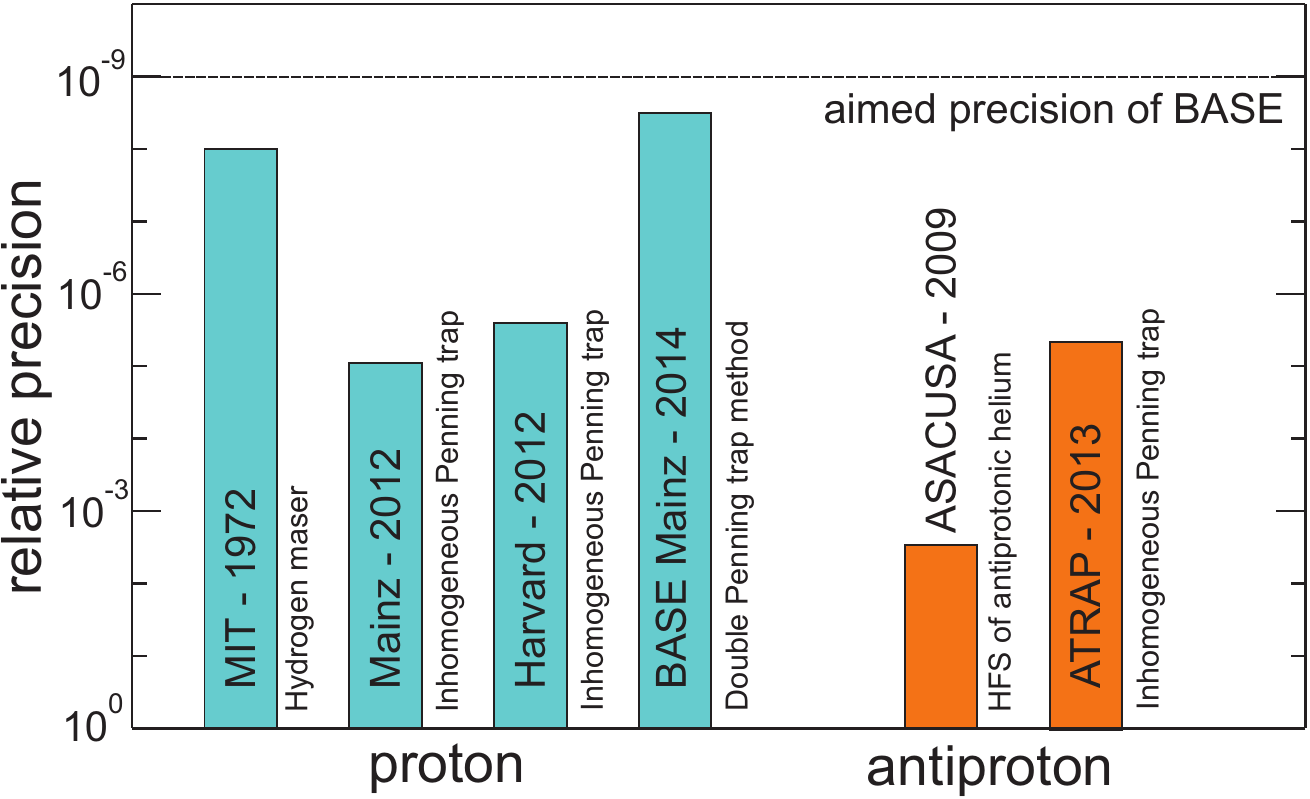}}
            \caption[Precision]{Overview on measurements of the proton (blue) \cite{CCRodegheri2012,Jack2012Proton,MooserNature2014,winkler1972magnetic} and the antiproton magnetic moments (orange) \cite{Jack2013Antiproton,Pask2009Antiproton} and their relative uncertainties. BASE aims to reach a relative uncertainty of $\delta g/g=$10$^{-9}$ for the antiproton by using the double-trap method. For details see text. }
						\label{fig:Precision}
\end{figure}

In our experiment, we aim at performing a sensitive test of CPT invariance by a high-precision comparison of the magnetic moments of the proton $\mu_p$ and the antiproton $\mu_{\overline{p}}$
\begin{eqnarray}	
	\mu_{p/\overline{p}} = \pm \frac{g_{p/\overline{p}}}{2} \mu_N.
\end{eqnarray}
The determined quantity in our measurements is the dimensionless $g$-factor, which expresses the magnetic moment in units of the nuclear magneton $\mu_N = q_p \hbar / 2 m_p$, with $q_p$ and $m_p$ being the proton's charge and mass, respectively.\\
For this purpose, we developed an apparatus capable of a statistical detection of spin-flips of single protons in a Penning trap by utilizing the non-destructive continuous Stern-Gerlach effect \cite{DehmeltCSG}. Using this apparatus, we reported on the first observation of spin flips with a single trapped proton \cite{UlmerPRL2011}, which resulted in a direct measurement of $g_p$ with a fractional precision of 8.9$\,$ppm \cite{CCRodegheri2012}, see Fig.~\ref{fig:Precision}. A measurement of $g_p$ with a relative uncertainty of 2.5$\,$ppm was reported by another group \cite{Jack2012Proton}. These measurements were carried out in a Penning trap with a strong superimposed magnetic inhomogeneity, which ultimately limits the achievable precision. To overcome these limitations we developed methods to observe single transitions of the proton spin \cite{MooserPRL2013} and demonstrated the application of the double Penning-trap technique with a single proton \cite{MooserPLB2013} for the first time. This series of developments culminated in the first direct high-precision measurement of the $g$-factor of the proton with a fractional precision of 3.3$\,$ppb \cite{MooserNature2014}, which is the most precise measurement of $g_p$ to-date. As the measurements are based on spectroscopy of a single particle in a Penning-trap system, the same methods can directly be applied to measure the magnetic moment of the antiproton, which is only known with a fractional precision of 4.4$\,$ppm \cite{Jack2013Antiproton}. Thus, by applying the double Penning-trap technique \cite{haeffner2003double} to the antiproton, a thousand-fold improved test of CPT invariance with baryons will be provided.

Lorentz- and CPT-violating terms are introduced into the SM in the framework of the Standard Model Extension (SME) \cite{KosteleckySME}. There, the sensitivity of different CPT tests is discussed using the measure $r_j=\Delta E/E$, where $\Delta E$ is the upper limit for the energy difference between given conjugate matter/antimatter systems and $E$ the energy-eigenvalue of the full relativistic Hamiltonian describing the system. For the comparison of K-meson masses this figure of merit is $r_K=|m_{K_0}-m_{\overline{K_0}}|/m_{K_0}\approx 0.6\cdot10^{-18}$ \cite{KosteleckyKaon}. In Penning trap experiments $r$ translates to $r = h \Delta\nu/mc^2$, where $\Delta\nu$ is the limit on the difference of the measured frequencies for the particle/antiparticle pair under investigation \cite{KosteleckyPenning}. By applying this figure of merit to the electron/positron $g-2$ comparison, for example, $r_{e,g}=1.2\cdot10^{-21}$ is obtained. Although the fractional precision achieved in the experiment is less precise than in the case of $\Delta m_{K_0}/m_{K_0}$, this lepton $g-2$ comparison is 50 times more sensitive with respect to CPT violation in the SME framework. The figure of merit for a 10$^{-9}$-comparison of the magnetic moments of the proton and the antiproton would lead to $r_{p,g}\approx10^{-25}$, and thus provide one of the most stringent tests of CPT invariance performed with baryons.

To achieve this appealing goal we commissioned a new experiment called BASE (Baryon Antibaryon Symmetry Experiment) at the antiproton decelerator facility of CERN. The BASE apparatus \cite{Ulmer2013ICPEAC,Smorra2013LEAP} has evolved from our proton double-trap experiment installed at the University of Mainz \cite{CCRodegheri2012}, but contains significant modifications and improvements. In addition, the apparatus has been adapted to allow injecting antiprotons from the AD. The new developments feature a reservoir trap \cite{SmorraIJMS2015}, which allows experimental operation even during accelerator shutdown, as well as a cooling trap for faster $g$-factor measuring cycles. Further, single particle detection systems with greatly improved sensitivity were developed, thus allowing faster single particle frequency measurements.

This paper is dedicated to describe the physics principles and the technical realization of BASE. The experimental principle of the magnetic moment measurement and the double-trap method are explained in Sec.~2. The experimental setup is described in Sec.~3. The methods and procedures to prepare single antiprotons are presented in Sec.~4. First results of frequency measurements with single antiprotons and the measurement method used in \cite{UlmerNature2015} are reported in Sec.~5. Finally, the measurement prospects of the experiment are discussed in Sec.~6.


\section{Experimental Principle}
\label{Sect:2}



The fundamental measurement principles of BASE go back to an elegant set of techniques developed by Dehmelt \emph{et al.} for the high-precision comparison of the electron and positron $g$-2 values in a Penning trap \cite{Dehmelt,DehmeltCSG,Wine}. By using the relation
\begin{eqnarray}
\frac{g}{2} = \frac{\nu_{L}}{\nu_{c}},
	\label{eq:g-factor}
\end{eqnarray}
only the frequency ratio of the spin-precession frequency $\nu_L$, also called Larmor frequency, to the cyclotron frequency $\nu_c$ has to be determined. To measure the cyclotron frequency, highly-sensitive image-current detection systems are used to directly measure the motional frequencies of a single trapped particle. The spin-precession frequency is obtained using the continuous Stern-Gerlach effect, which is a non-destructive measurement technique to observe quantum spin transitions of single trapped particles via a change of the axial oscillation frequency $\nu_z$. Exciting spin transitions with an external drive and measuring the spin-transition probability as a function of the excitation frequency allows to determine the spin-precession frequency.

In the following, the aspects relevant for the proton/antiproton magnetic moment measurements are described. In particular, the challenges of the application of this method to the proton and antiproton and the advantage of applying the double Penning-trap technique \cite{haeffner2003double} to reach a precision on the 10$^{-9}$ level are emphasized.

\subsection{Image-current measurement of the free cyclotron frequency}
\label{Sect:2.1}
\subsubsection{The Penning trap}

Using only single particles with low motional amplitudes and long observation times, the Penning trap is a well-suited tool for high-precision measurements of the fundamental properties of charged particles \cite{Blaum2006}. Such a trap consists of a superposition of a magnetic field $B=B_0 \vec{e}_z$ for the radial confinement and an electric quadrupole potential to constrain the particle's motion along the $z$-axis:
\begin{eqnarray}
V(\rho,z)=V_R\ C_2(z^2-\rho^2/2),
	\label{eq:ElectricPotential}
\end{eqnarray}
which is in the ideal case formed by three perfectly-aligned hyperbolic electrodes, one ring electrode and two endcap electrodes of infinite size. The ring voltage $V_R$ denotes the potential difference between the ring and the endcap electrodes, and $\sqrt{1/C_2}$ is a trap specific length. 

\begin{figure*}[htb]
				\centerline{\includegraphics[width=0.95 \textwidth,keepaspectratio]{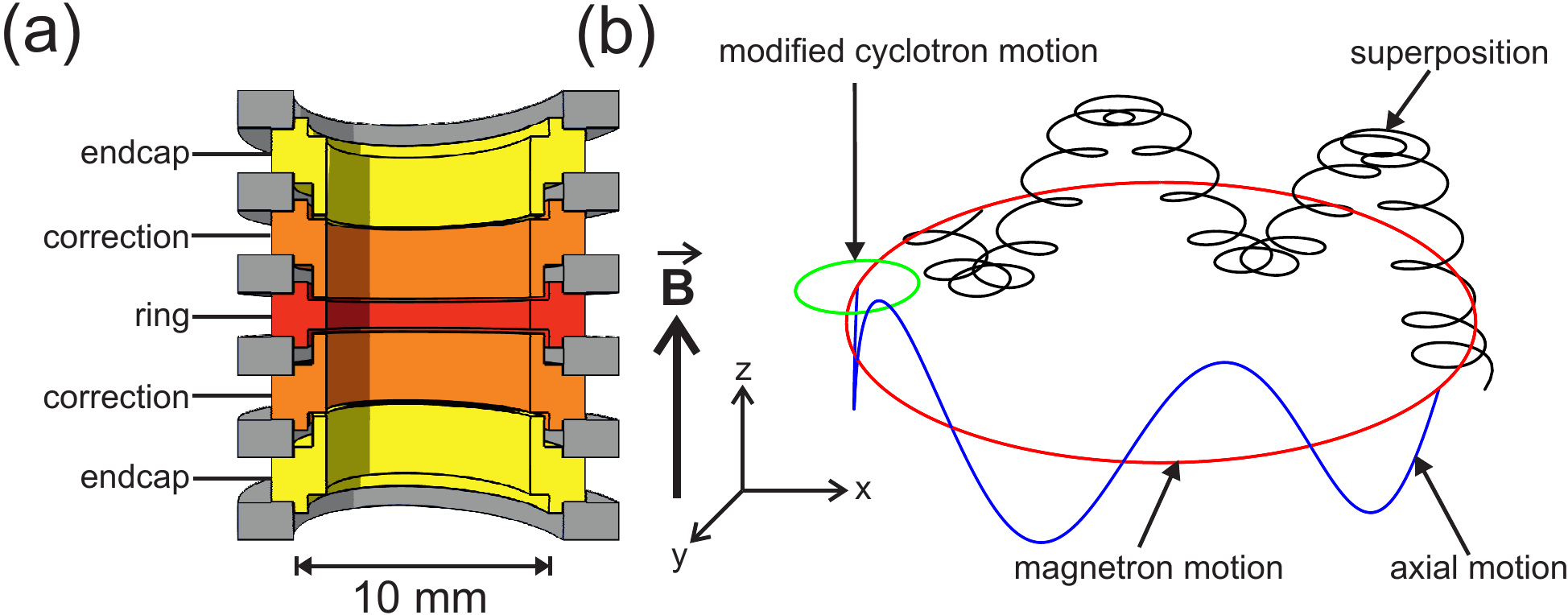}}
            \caption[PenningTrap]{(a) Cut view of a cylindrical Penning trap with five electrodes. The ring electrode (red), the correction electrodes (orange), the endcaps (yellow), and the insulation spacers (grey) are shown. (b) The composition of the motion of a particle inside a Penning trap from its three independent eigenmotions, the axial (blue), the magnetron (red), and the modified cyclotron motion (green), is shown. For details see text. }
						\label{fig:PenningTrap}
\end{figure*}

BASE uses cylindrical five-electrode Penning traps with an orthogonal, compensated design \cite{Jerry5poleTrap} shown in Fig.~\ref{fig:PenningTrap}(a). The two additional electrodes in between the ring and endcap electrodes are called correction electrodes. A fraction of the ring voltage $V_C = \text{TR}\cdot V_R$ is applied to them. By a proper choice of the trap geometry and by adjusting the tuning ratio TR, the next higher-order potential perturbations $C_4$ and $C_6$ in the expansion of the potential $V(0,z)=V_R \sum_j C_{2j} z^{2j}$ are set to zero simultaneously. Orthogonality means that the coefficient $C_2$ is independent from the tuning ratio TR \cite{Jerry5poleTrap}. Using such a compensated trap and cooling the particles to low motional amplitudes ensures that frequency shifts due to higher-order potential terms are negligible and that accurate high-precision measurements can be carried out.

The trajectory of a trapped particle is described by a superposition of three independent eigenmotions \cite{Brown}, see Fig.~\ref{fig:PenningTrap}(b).
The electric potential generates the axial motion, a harmonic oscillation along the magnetic field lines with eigenfrequency:
\begin{eqnarray}
\nu_z=\frac{1}{2 \pi} \sqrt{2 C_2 V_R \frac{q}{m}},
	\label{eq:AxialFrequency}
\end{eqnarray}
where $q/m$ denotes the charge-to-mass ratio of the trapped particle. In the BASE apparatus the axial frequency of protons/antiprotons is in the range of 540 kHz to 680 kHz. The trajectory in the radial plane is composed of two independent motions, the modified cyclotron motion, which is the result of the free cyclotron motion being affected by the radially outwards pulling electric field, and the magnetron motion, a slow drift motion in the crossed static fields. The respective eigenfrequencies $\nu_+$ and $\nu_-$ are given by:
\begin{eqnarray}
\nu_{/pm}= \frac{\nu_c}{2} \pm \sqrt{\frac{\nu_c^2}{4}-\frac{\nu_z^2}{2}}.
	\label{eq:RadialFrequencies}
\end{eqnarray}
$\nu_+$ is mainly defined by the magnetic field $B_0$ = 1.945$\,$T and is 29.65$\,$MHz, whereas $\nu_- \approx C_2 V_r / (2 \pi B_0)$ is typically at 7$\,$kHz. Note that the magnetron motion is meta-stable \cite{Brown}, but the radiative decay rates on the order of 10$^{-15}\,$Hz are insignificant compared to typical measuring times so that the mode can be considered as stable.\\

The free cyclotron frequency of a trapped charged particle is related to the three independent eigenmotions via the invariance theorem \cite{InvarTheo}:
\begin{eqnarray}
	\nu_c^2 = \nu_+^2 + \nu_-^2 + \nu_z^2~.
		\label{eq:InvarTheorem}
\end{eqnarray}
Hence, the free cyclotron frequency can be determined by measuring the three eigenfrequencies of the trapped charged particle. The invariance theorem is robust with respect to first-order perturbations such as a misalignment of the trap axis with respect to the magnetic field, and elliptic distortions of the electric potential.

\subsubsection{Image current detection}
\label{sec:EP:ICD}
\begin{figure*}[ht!]
        \centerline{\includegraphics[width=0.95 \textwidth,keepaspectratio]{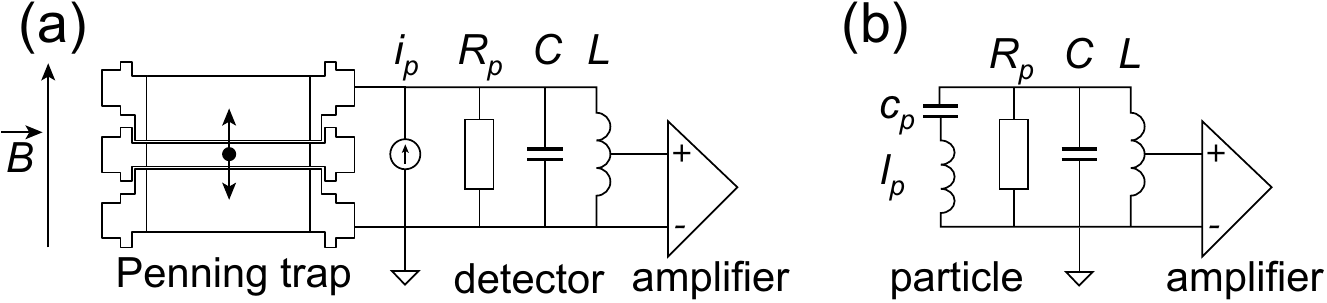}}
            \caption[PenningTrap]{Illustration of the image current detection priciple. (a) The particle induces an image current $i_p$ in the trap electrodes, which is converted to a voltage drop using a resonant tuned circuit. (b) A particle in thermal equilibrium with the detection system can be represented by an equivalent series tuned circuit, which shunts the thermal noise of the detection system at the motional frequency of the particle. For details see text.}
						\label{fig:DetectionPrinciple}
\end{figure*}
To measure the motional frequencies of single trapped protons and antiprotons, detection systems for non-destructive measurements of image currents are used \cite{Wine}. The detection principle is illustrated in Fig.~\ref{fig:DetectionPrinciple}(a) for the axial motion. The oscillating charged particle induces an image current in the trap electrodes, which is on the order of fA \cite{Wine}. By using a large impedance, this small current signal is converted to a detectable voltage. 
To this end, a superconducting inductor $L$ connected in parallel to the trap compensates its parasitic capacitance $C$ to form a tuned circuit with resonance frequency $\nu_{res}$:
\begin{eqnarray}
	\nu_{res} = \frac{1}{2 \pi} \sqrt{\frac{1}{L C}}.
	\label{eq:ResonanceFrequency}
\end{eqnarray}

\begin{figure*}[htb]
        \centerline{\includegraphics[width=0.95 \textwidth,keepaspectratio]{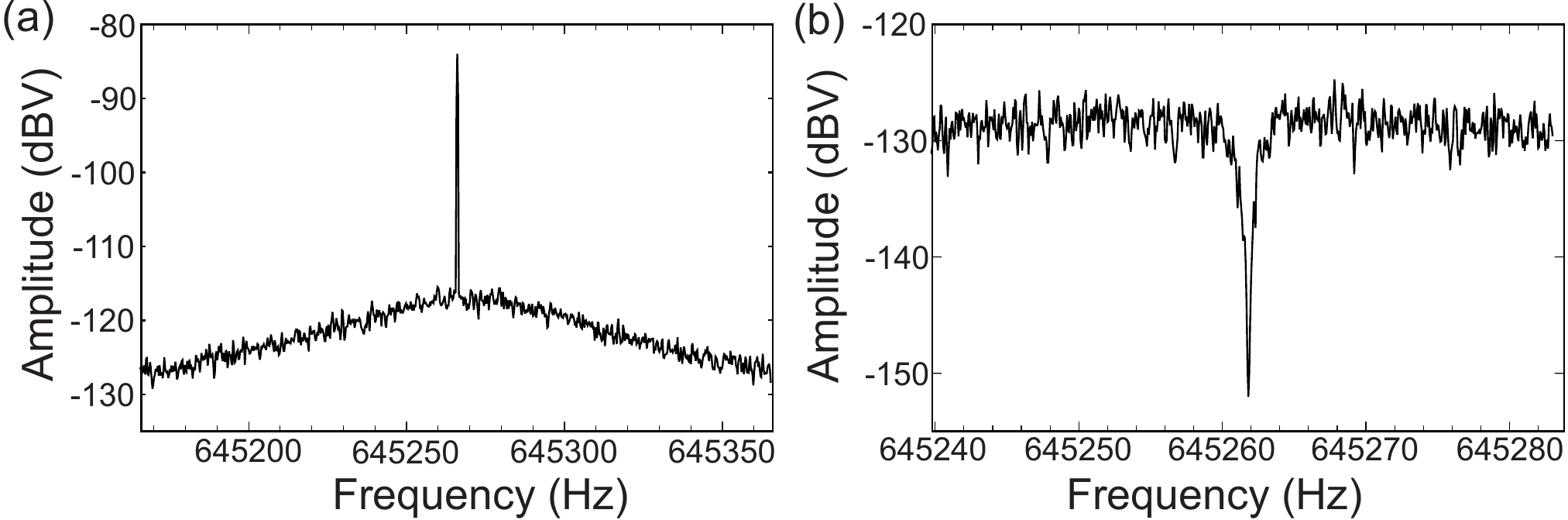}}
            \caption[PenningTrap]{Detection of the image-current signal of single particles in the BASE Penning-trap system. (a) A single antiproton is excited with an rf-drive at $\nu_{rf} $ = 1290538 Hz $\approx 2 \nu_z$ and the power dissipation of the excited particle is observed at the frequency $\nu_{rf} /2 $ in the FFT-spectrum. (b) A single antiproton in thermal equilibrium with the detection system generates a short at its axial frequency. A dip signal with 25 dB signal-to-noise ratio is observed. }
						\label{fig:ParticlePeakDip}
\end{figure*}

On resonance, the circuit acts as a parallel resistance $R_p$:
\begin{eqnarray}
	R_p = 2 \pi \nu_r Q L,
	\label{eq:ParallelResistance}
\end{eqnarray}
where $Q$ is the quality factor characterizing the relative energy loss per oscillation cycle. A particle tuned to the detector's resonance frequency is cooled resistively with the damping constant
\begin{eqnarray}
	\gamma = \frac{q^2 R_p}{D^2 m},
	\label{eq:DampingConstant}
\end{eqnarray}
where $D$, the effective electrode distance, is a specific length depending on the size of the trap and the detailed pick-up geometry. 
The energy dissipation of an excited particle in the detector is used to detect a single particle. For this purpose, the transient signal is amplified with a cryogenic low-noise amplifier and processed by a Fast Fourier Transform (FFT) spectrum-analyzer. Fig.~\ref{fig:ParticlePeakDip}(a) shows the FFT signal of an axially excited single trapped antiproton tuned to resonance with one of our axial detectors.
Once cooled to thermal equilibrium with the detection system, the particle acts as an equivalent $LC$ series circuit with inductance $l_p = m D^2 / q^2$ and capacitance $c_p = 1/l_p(2 \pi \nu)^2$, as shown in Fig.~\ref{fig:DetectionPrinciple}(b) \cite{Wine}. The particle shunts the thermal noise of the detector at its eigenfrequency, thus resulting in a so-called particle dip which is shown in Fig.~\ref{fig:ParticlePeakDip}(b). By performing a best fit of the well-known resonance line to the measured FFT spectrum, the axial frequency $\nu_z$ is extracted. The advantage of this method is that the particle's resonance frequency is measured at thermal amplitudes of about 10$\,\mu$m, which reduces systematic frequency shifts due to trap imperfections \cite{Brown}. For small particle numbers the line-width $\Delta\nu$ of the dip-signal (FWHM) can be calculated from the impedance of the equivalent circuit and is given as \cite{Wine}:
\begin{eqnarray}
	\Delta\nu_z = \frac{N}{2 \pi}\gamma = \frac{N}{2 \pi} \frac{q^2 R_p}{D^2 m},
	\label{eq:DipWidth}
\end{eqnarray}
where $N$ is the number of trapped particles.

\subsubsection{Sideband coupling}
\label{sec:EP:SC}
\begin{figure}[htb]
        \centerline{\includegraphics[width=0.70 \textwidth,keepaspectratio]{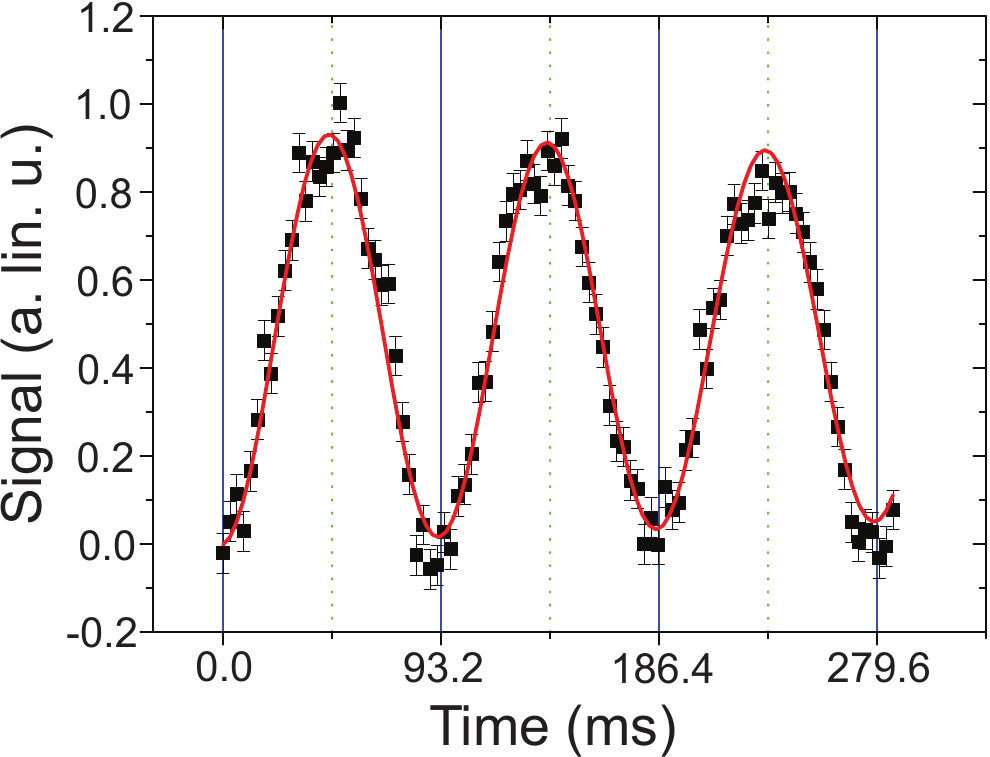}}
            \caption[PenningTrap]{Amplitude in the axial mode as function of the duration of the sideband-coupling drive recorded for a single proton in the trap system at Mainz University. Initially, the axial mode of the particle was cooled and the magnetron mode was excited on a defined radius. The magnetron-to-axial coupling exhibits ``classical Rabi oscillations". For details see text.}
						\label{fig:RabiOscillation}
\end{figure}

To measure the motional frequencies of the radial modes with the axial detection system, sideband coupling is applied \cite{Cornell1990}. A quadrupolar rf-drive with an electric field of the form
\begin{eqnarray}
	\vec{E}_{rf} = E_0\,\mathrm{sin}\left(2 \pi \nu_{rf} t\right) (z \hat{e_\rho} + \rho \hat{e_z})
	\label{eq:QuadrupolExcitation}
\end{eqnarray}
is irradiated into the trap. Here, $E_0$ denotes the electric field amplitude and $\nu_{rf} = \nu_\pm \mp \nu_z$ the drive frequency which couples the modified cyclotron (magnetron) motion to the axial motion. This results in a periodic exchange of energy between the two modes, leading to an amplidute-modulated axial motion:
\begin{eqnarray}
	z(t) = z_0\,\mathrm{sin}\left(\frac{1}{2}\mathrm{\Omega} t\right)\,\mathrm{cos}\left(\omega_z t + \phi_z\right)\,,
	\label{eq:AmplitudeModulatedAxialMotion}
\end{eqnarray}
where $\mathrm{\Omega}$ is the Rabi frequency, which on resonance reads $\mathrm{\Omega_0} = q E_0 /(4 \pi m \sqrt{\nu_\pm \nu_z})$ or in case the coupling drive is detuned by $\delta_{\pm} = \nu_{rf} - \nu_{\pm} \pm \nu_{z}$,  $\mathrm{\Omega} = \sqrt{\mathrm{\Omega_0^2} + 4 \pi^2 \delta_{\pm}^2}$. The periodic exchange of energy between the axial and magnetron mode using a particle with a well-defined initial magnetron radius and thermal energy in the axial mode is shown in Fig.~\ref{fig:RabiOscillation}. Here, the axial amplitude is measured by the signal strength of the peak signal after irradiating the sideband drive at the frequency $\nu_{rf}=\nu_z + \nu_m$ for a certain time. The dynamics depict the amplitude-modulated motion with a full exchange period of 93.2 ms in this case. The quantitative behavior of the coupled motion is comparable to a quantum mechanical two-level system and can be interpreted as ``classical Rabi oscillations". \\

Sideband coupling can be used to cool the radial motions. To this end, a coupling drive is applied until the radial energy has dissipated due to the damping of the axial mode. Thereby, both coupled modes reach equilibrium with the thermal bath of the detection system. This situation is reached when the average quantum numbers $\left\langle n_z\right\rangle$ and $\left\langle n_{\pm}\right\rangle$ are equal \cite{Brown}. By assuming that the axial detection system is at temperature $T_z$, the other modes are cooled to \cite{Cornell1990}:
\begin{eqnarray}
	T_{\pm} = \frac{\nu_{\pm}}{\nu_z} T_z.
	\label{eq:SideBandCoolingLimit}
\end{eqnarray}
Thus, the magnetron mode can be cooled to a small fraction of the temperature of the axial detection system, whereas the sideband cooling limit of the modified cyclotron mode is a factor of 45 larger than $T_z$.\\

The Fourier spectrum of the coupled amplitude-modulated motion is composed of the two sideband frequencies $\nu_l$ and $\nu_r$:
\begin{eqnarray}
	\nu_{l,r} = \nu_z - \frac{\delta}{2} \pm \sqrt{  \frac{\mathrm{\Omega_0^2}}{4 \pi^2} + \delta^2}.
	\label{eq:SideBandFrequencies}
\end{eqnarray}
By combining a measurement of these sideband frequencies and an independent measurement of the axial frequency, the modified cyclotron frequency is obtained as
\begin{eqnarray}
	\nu_{+} = \nu_{rf} + \nu_{l} + \nu_{r} - \nu_{z}.
	\label{eq:DoubleDipModCyclotron}
\end{eqnarray}
The magnetron frequency can either be measured in a separate sideband frequency measurement, or simultaneously by sideband coupling of the `cyclotron-dressed states' to the magnetron mode \cite{UlmerPRL5Dip}. Thereby all three eigenfrequencies of the particle have been determined and the free cyclotron frequency $\nu_c$ is obtained via the invariance theorem in equation (\ref{eq:InvarTheorem}).


\subsection{Measurement of the Larmor Frequency}
\label{Sect:2.2}
\subsubsection{The continuous Stern-Gerlach effect}

\begin{figure}[htb]
        \centerline{\includegraphics[width=0.7 \textwidth,keepaspectratio]{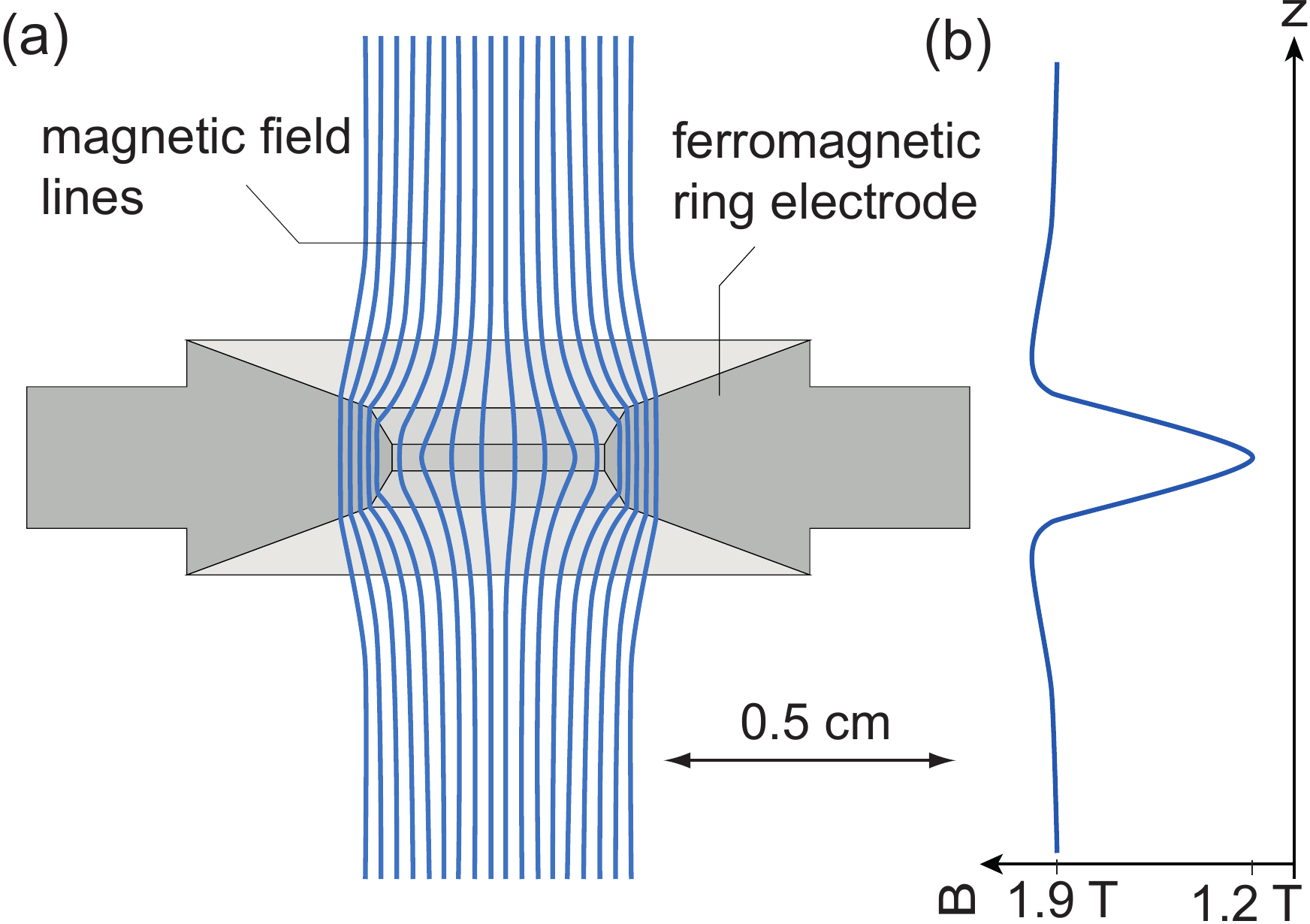}}
            \caption[Experiment]{Schematic of the magnetic bottle: a.) Ferromagnetic ring electrode and magnetic field lines. b.) On axis magnetic field of the magnetic bottle. } \label{fig:Bottle}
\end{figure}

\begin{figure}[htb]
        \centerline{\includegraphics[width=0.7 \textwidth,keepaspectratio]{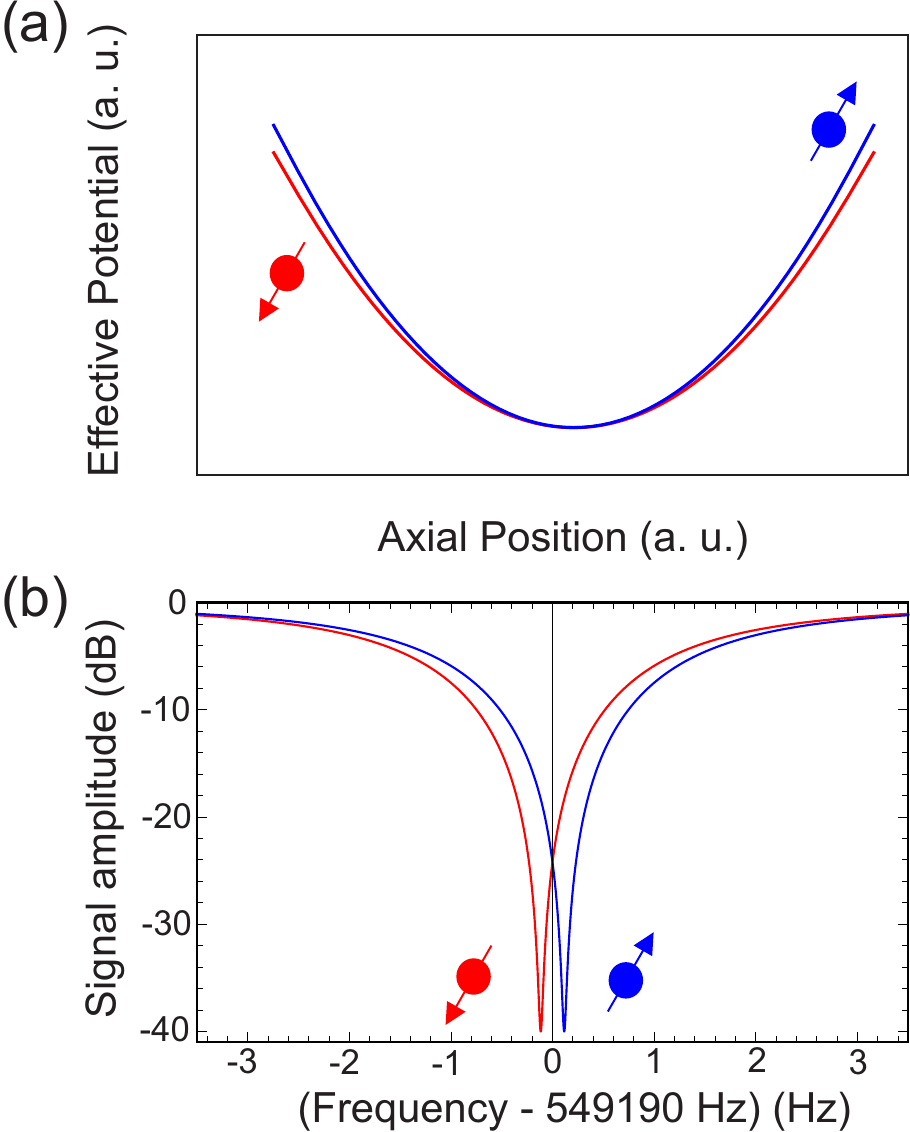}}
            \caption[Experiment]{Illustration of the continuous Stern-Gerlach effect for a single antiproton. (a) The effective potential of the axial motion for the spin-down (red) and spin-up state (blue) of the antiproton in the presence of a magnetic bottle $B(z)=B_0+B_2z^2$ is shown. The magnetic bottle causes the axial oscillation frequency $\nu_z$ to depend on the spin eigenstate. (b) Calculated signal amplitude relative to the noise amplitude of the detection system at its resonance frequency with particle in spin-down (red) and spin-up state (blue). In the BASE apparatus, the axial frequency difference due to a spin flip is only 231 mHz out of 549190 Hz. For details see text.} \label{fig:CSG}
    \end{figure}

The spin precession of trapped particles is not accompanied by a detectable image current, thus the Larmor frequency $\nu_L$ cannot be directly observed with our detection systems. A solution for the non-destructive measurement of $\nu_L$ is provided by the continuous Stern-Gerlach effect, which was introduced by Dehmelt and collaborators \cite{DehmeltCSG}. In this scheme a magnetic field inhomogeneity of the form
\begin{eqnarray}
B_z(\rho,z)=B_0+B_2(z^2-\rho^2 /2),
  \end{eqnarray}
a so-called ``magnetic bottle'' is superimposed on the trap by introducing a ferromagnetic ring electrode, see Fig.~\ref{fig:Bottle}. 
Thereby, a magnetic potential $\Phi = -\vec{\mu}\cdot\vec{B(z)}$ is added to the axial electrostatic potential of the trap. This couples the magnetic moment of the trapped particle to its axial oscillation frequency, and thus, a measurement of $\nu_z$ enables the non-destructive detection of the spin-eigenstate. This is illustrated in Fig.~\ref{fig:CSG}. In the ``spin-down''/``spin-up'' states the antiproton experiences an effective axial potential represented by the red and the blue solid line, respectively. A spin flip causes an axial frequency shift of
    \begin{eqnarray}
    \Delta\nu_{\mathrm{z,SF}}=\frac{1}{2\pi^2}\frac{\mu_{\bar{p}} B_2}{m_{\bar{p}} \nu_{\mathrm{z}}},
    \end{eqnarray}
where $\mu_{\bar{p}}$ and $m_{\bar{p}}$ denote magnetic moment and mass of the antiproton, respectively. 
The axial frequency shift caused by the continuous Stern-Gerlach effect scales linearly with the strength $B_2$ of the magnetic bottle and the ratio $\mu/m$, the latter being extremely small for the proton/antiproton-system. Compared to measurements dealing with magnetic moments on the level of the Bohr magneton, such as in the electron-positron system \cite{Dehmelt} or electrons bound to highly charged ions \cite{Hartmut,Verdu,SvenSi28}, the proton magnetic moment is approximately 660 times smaller. Therefore, we superimpose a strong magnetic bottle of $B_2$ = $30\,$T/cm$^2$ to one of our traps, called \emph{analysis trap}, which is a 2000-fold and 30-fold increase in the magnetic bottle strength compared to the electron/positron g-2 measurements \cite{Dehmelt} and g-factor measurements of electrons bound to highly-charged ions \cite{Hartmut,Verdu,SvenSi28}, respectively. Under these extreme magnetic conditions, an antiproton spin flip changes the axial frequency in the analysis trap by $231\,$mHz out of about $549\,$kHz. \\
Once the non-destructive detection of spin transitions is established, the Larmor frequency is obtained by measuring the spin transition rate as a function of the frequency $\nu_{rf,s}$ of an irradiated spin flip drive. The Larmor frequency $\nu_L$ can be extracted from the well understood line shape \cite{BrownPRL1984,BrownGeoniumLineshape}.\\

\subsubsection{Axial frequency stability in the magnetic bottle}
\label{Sect:2.2.2}
In addition to the spin magnetic moment $\mu_{\bar{p}}$, a trapped antiproton has a magnetic moment due to its radial angular momentum $\mu_{\pm}=q/2m L_{\pm}$, thus the total axial frequency change in presence of a magnetic bottle results in \cite{UlmerPRL2011}:
\begin{eqnarray}
\Delta\nu_{z}&=& \frac{1}{4 \pi^2 m_{\overline{p}}\,\nu_z} \frac{B_2}{B_0}  h \nu_+ \left[\left(n_+ + \frac{1}{2}\right) - \frac{\nu_-}{\nu_+} \left(n_- + \frac{1}{2}\right) + s \frac{g}{2}\right],
\end{eqnarray}
where the classical energies $E_\pm$ have been replaced by the energy terms of the quantum harmonic oscillators $h \nu_\pm \left(n_\pm+1/2\right)$ with the quantum numbers $n_+$ and $n_-$ for the modified cyclotron and magnetron mode, respectively, while $s = \pm$ 1 denotes the spin-quantum number. 

A single quantum transition in the modified cyclotron mode $\Delta n_+=\pm1$ changes the axial frequency by $83\,$mHz, and the axial frequency shift caused by three cyclotron transitions is already larger than the one induced by a spin transition. As cyclotron transitions are electric dipole transitions, fluctuations in $n_+$ caused by an electric-field noise density on the level 300$\,$nV$\,$m$^{-1}\,$Hz$^{-1/2}$ makes the clear identification of single spin-transitions impossible. Reducing the noise amplitude to carry out the spin-state identification out at a constant $n_+$ constitutes a major challenge in proton/antiprotons magnetic moment measurements. \\ 

In the trapped-ion quantum information community, heating rates of single particles in rf-traps and Penning traps have been observed to exceed the ones expected from the density from thermal resistive noise present on the electrodes \cite{WineAnomHeat2000,goodwin_sideband_2014}. The increased noise density is referred to as anomalous heating and originates from surface adsorbates, which was experimentally demonstrated by in-situ cleaning of the electrodes of an rf-trap with Ar$^+$ ion beam bombardment, which decreased the heating rate by a factor of 100 \cite{HitePRL2012}. The details of this mechanism is however not yet understood as the observed heating rates and the experimental parameters such as trap surface properties, temperatures and the particle-surface distance $d$ vary over a large range. Several surface-adsorbate based models have been developed which exhibit different heating rates scaling with $d$ \cite{HiteMRS2013}, and 1/$d^4$ is a frequently quoted distance scaling \cite{oneoverd4scaling}. 


Our analysis trap in the Mainz proton experiment shows also anomalous heating in the cyclotron mode, which can be observed by axial frequency fluctuations in the magnetic bottle \cite{MooserPRL2013}. We extract a normalized electric-field noise density $\omega S_E\left(\omega\right)$ of 3$\times$10$^{-9}\,$V$^2$/m$^2$ at $n_+\approx$ 600 and 1.8 mm trap radius, which follows the 1/d$^4$ scaling trend of the spectral noise density observed by the rf-trap community \cite{WineAnomHeat2000,HiteMRS2013,oneoverd4scaling}. 

For proton/antiproton spin-flip experiments, we determine the standard deviation of the frequency difference of subsequent axial frequency measurements in the magnetic bottle $\delta\nu_{z,i}$=$\nu_{z,i} - \nu_{z,i-1}$, further called axial frequency fluctuation $\Xi_z$:
\begin{eqnarray}
 \Xi_z = \sqrt{\sum_{i=2}^N \frac{1}{N-1} \left(\delta\nu_{z,i}\right)^2}.
\end{eqnarray}
This quantity is compared to the frequency shift induced by a spin-flip $\Delta\nu_{\mathrm{z,SF}}$. In case $\Xi_z$ has about the same magnitude as $\Delta\nu_{\mathrm{z,SF}}$ unambiguous detection of the spin state is not possible. However, in this case the Larmor frequency can be obtained by using a statistical method to measure the spin transition probability, which was developed by members of the BASE collaboration \cite{UlmerPRL2011}. 

\subsubsection{Statistical spin-flip detection}

\begin{figure}[htb]
  \centerline{\includegraphics[width=0.70 \textwidth,keepaspectratio]{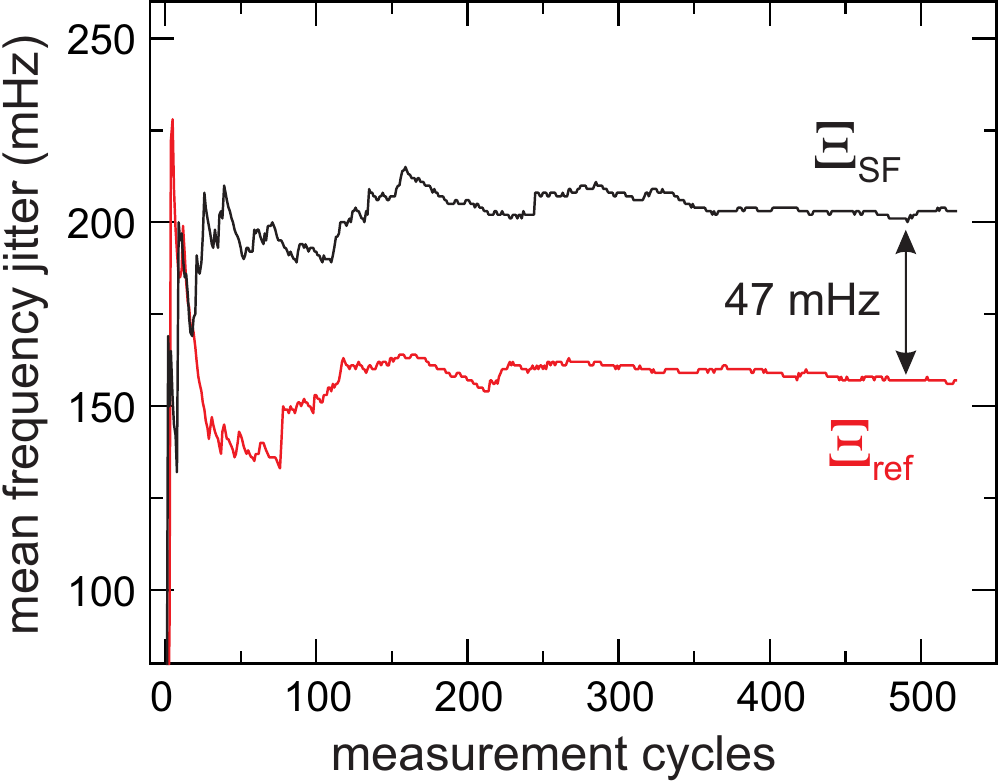}}
  \caption[ATLarmorResonance]{First indirect detection of proton spin-flips from Ref.~\cite{UlmerPRL2011}. The cummulative evolution of the frequency fluctuation $\Xi_{\mathrm{SF}}$ and $\Xi_{\mathrm{ref}}$ with and without spin-flip drive is shown  as function of the number of measurement cycles. The difference of the converged values of $47 \pm 4.5\,$mHz corresponds to a spin flip probability of $P_{\mathrm{SF}}$ = 47$\pm7\%$.} \label{fig:StatisticalSpinflips}
\end{figure}

\begin{figure}[htb]
  \centerline{\includegraphics[width=0.70 \textwidth,keepaspectratio]{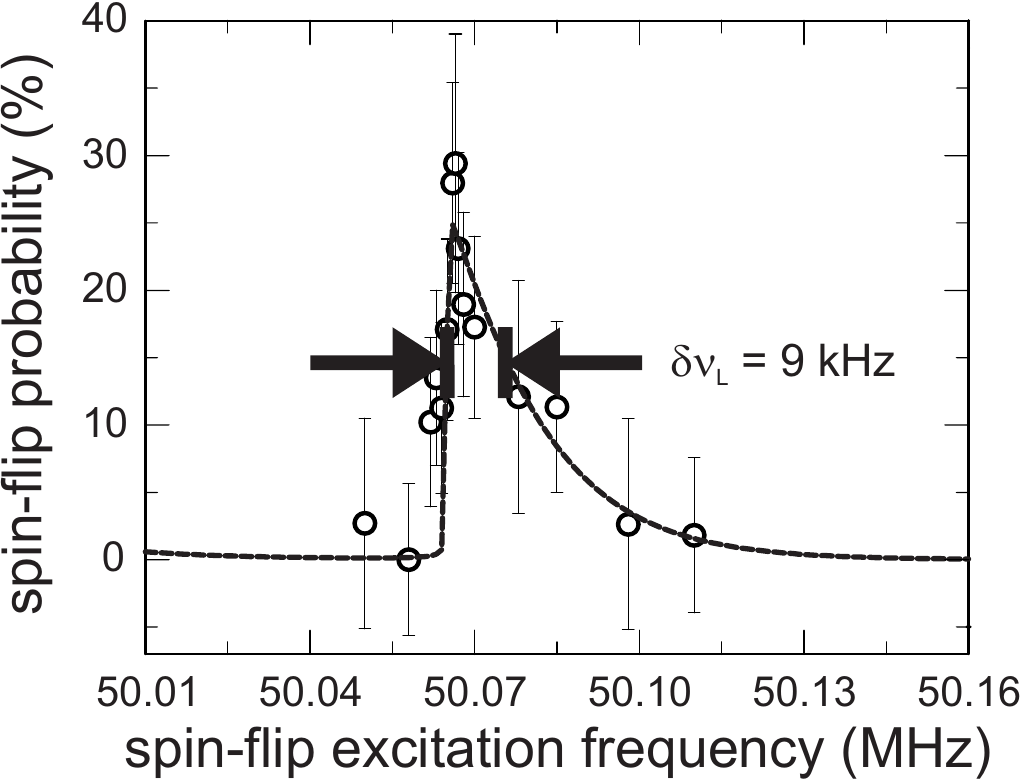}}
  \caption[ATLarmorResonance]{Larmor resonance of a single proton measured with the statistical method in the analysis trap \cite{CCRodegheri2012}. The solid line represents the fit of the resonance line shape to the data. The line width parameter of the resonance $\delta\nu_L$ is 9 kHz and the Larmor frequency $\nu_L$ = 50.064 971(91) MHz is extracted with a relative precision of $1.8\times 10^{-6}$.}
	\label{fig:ATLarmorResonance}
\end{figure}

Driving spin transitions in between axial frequency measurements increases the background axial frequency fluctuation and allows an indirect observation of spin transitions as shown in Fig.~\ref{fig:StatisticalSpinflips}. Here, the increase of the axial frequency fluctuation from $\Xi_{ref} \approx$ 150 mHz to $\Xi_{\mathrm{SF}} \approx 190\,$mHz while irradiating a resonant spin-flip drive allowed to observe spin-transitions of a single proton for the first time \cite{UlmerPRL2011}.
To obtain a Larmor resonance, the axial frequency fluctuation $\Xi$ is determined for different spin-flip drive frequencies $\nu_{rf}$ close to $\nu_L$ in addition to a reference measurement without rf-drive $\Xi_{\mathrm{ref}}$. This allows to extract the spin-transition probability as function of $\nu_{rf}$ by the increased axial frequency fluctuation \cite{UlmerPRL2011,CCRodegheri2012}:
\begin{eqnarray}
 \Xi(\nu_{rf}) = \sqrt{\Xi_{\mathrm{ref}}^2 + P_{\mathrm{SF}}(\nu_\mathrm{rf}) \Delta\nu_{\mathrm{z,SF}}^2}\,,
\end{eqnarray}
where $P_{\mathrm{SF}}$ denotes the spin-transition probability at the drive frequency $\nu_{rf}$:
\begin{eqnarray}
 P_{\mathrm{SF}}(\nu_\mathrm{rf}) = \frac{1}{2}\left(1-\mathrm{exp}\left(-\frac{1}{2} \Omega_R^2 t_0 \chi(\nu_\mathrm{rf})) \right)\right).
\end{eqnarray}
Here $\Omega_R = 2 \pi \nu_L b_\mathrm{rf} / B_0$ denotes the Rabi frequency with the magnetic drive amplitude $b_\mathrm{rf}$, irradiation time $t_0$ and the line shape of the Larmor resonance $\chi(\nu_\mathrm{rf})$, details are discussed in \cite{BrownPRL1984,BrownGeoniumLineshape}. In general, the particle coupled to the axial detection system passes through all thermal states with the rate $\gamma$, which causes a variation of the amplitude and thereby also the average magnetic field experienced by the particle in the magnetic bottle. Thus, the resulting line shape becomes a convolution of the thermally distributed axial energy and the unperturbed Rabi resonance.
The statistical average of the Larmor frequency becomes
\begin{eqnarray}
 \left\langle \nu_L \right\rangle = \nu_{L,0} \left(1 + \frac{B_2}{B_0} \left\langle z_0^2 \right\rangle\right),
 \label{eq:StatisticAverageNuL}
\end{eqnarray}
so that the line width generated by the magnetic bottle can be characterized by the parameter $\delta\nu_L$:
\begin{eqnarray}
 \delta\nu_L = \nu_{L,0} \frac{B_2}{B_0} \left\langle z_0^2 \right\rangle = \nu_{L,0} \frac{B_2}{B_0} \frac{k_B T_z}{4 \pi^2 m \nu_z^2},
 \label{eq:LineWidth}
\end{eqnarray}
where $k_B$ denotes the Boltzmann constant and the equipartition theorem was used to replace $\left\langle z_0^2 \right\rangle$.
Due to the strong magnetic bottle required to observe antiproton spin flips, the line shape for the weak coupling limit $\delta\nu_L \gg \gamma$ derived in \cite{BrownGeoniumLineshape} holds:
\begin{eqnarray}
 \chi(\nu_{rf}) = \frac{\Theta(\nu_{rf} - \nu_{L,0})}{\delta\nu_L} \mathrm{exp}\left(-\frac{\nu_{rf} - \nu_{L,0}}{\delta\nu_L}\right),
 \label{eq:LineshapeAT}
\end{eqnarray}
where $\Theta$ is the Heaviside step function. In this case, 
the resonance shape directly depicts the Boltzmann distibution of the axial energy of the trapped particle, as shown in the Larmor resonance of a single proton in Fig.~\ref{fig:ATLarmorResonance}. The sharp edge of the resonance corresponds to an axial energy $E_z$ of zero. The resonance curve has a line width parameter $\delta\nu_L$ = 9 kHz resulting from $B_2 = 29.7$ T/cm$^2$ and an effective axial temperature of $T_z$ = 2.5$\,$K. Here, negatively phased feedback cooling was applied to reduce the effective temperature of the particle below the thermal limit \cite{UrsoFeedback2003}, which reduces the line width. From a fit of the line shape to the data, $\nu_L$ was determined to 50.064 971(91)MHz with a relative precision of $1.8 \times 10^{-6}$. In this way, $\nu_L$ was derived in the first direct measurements of the proton and antiproton magnetic moments \cite{CCRodegheri2012,Jack2012Proton,Jack2013Antiproton} using only single inhomogeneous Penning trap.

The resolution of this kind of Larmor-frequency measurements in the analysis trap is fundamentally limited by the line width, which is defined by the product of the magnetic inhomogeneity $B_2$ and the axial temperature $T_z$. Clearly, this precision limit, which is at the ppm level, can be overcome by decreasing $B_2$ significantly during the frequency measurements. This is realized in the double Penning-trap method for the measurement of magnetic moments by transporting the particle adiabatically from the magnetic bottle into a trap with a much more homogeneous field to carry out the frequency measurements \cite{haeffner2003double}.



\subsection{Double-Trap Method}
\label{Sect:2.3}

\begin{figure}[htb]
  \centerline{\includegraphics[width=0.95 \textwidth,keepaspectratio]{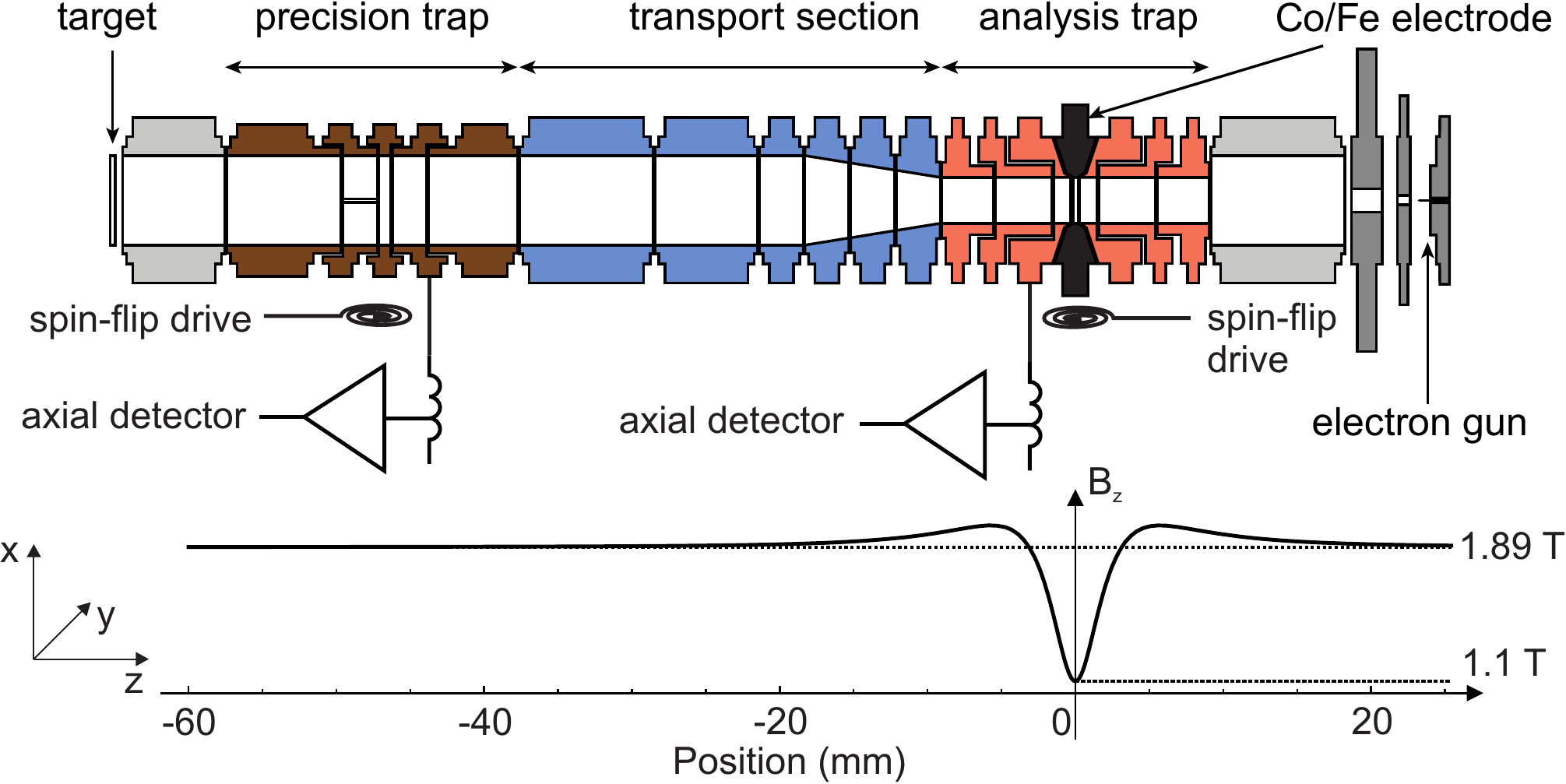}}
  \caption[ATLarmorResonance]{The double trap system used for the proton magnetic moment measurement reported in \cite{MooserNature2014}. An analysis trap with a magnetic bottle of 29.7 T/cm$^2$ is used for the spin-state analysis and the precision trap with a residual magnetic field inhomogeneity of 0.4 mT/cm$^2$ is used for precise frequency measurements. The strength of the magnetic field on the axis of the trap system is shown as well. For details see text.} \label{fig:DoubleTrapMainz}
\end{figure}

In the double Penning-trap measurement scheme \cite{haeffner2003double} one of the two traps used is the analysis trap, as introduced above, with the strong superimposed magnetic bottle for spin-state readout. The magnetic field in the second trap, the precision trap, has an inhomogeneity which is about $100\,000$-fold lower due to the spatial separation from the analysis trap. The double Penning-trap system used in our 3.3$\,$ppb proton magnetic moment measurement \cite{MooserNature2014} is shown in Fig.~\ref{fig:DoubleTrapMainz}. Here, the residual $B_2$ in the precision trap is only 0.4$\,$mT/cm$^2$, a factor of 75000 times smaller than in the analysis trap. Thereby, the line width parameter $\delta\nu_L$ of only 0.55 Hz allows to determine $\nu_L$ with ppb precision.
Due to the small $\delta\nu_L$ in the precision trap, the line shape of the resonance in the strong coupling limit $\delta\nu_L \ll \gamma$ is given in \cite{BrownGeoniumLineshape}, which can be interpreted as the statistical average of the individual Lorentzian line shapes of all thermal states:
\begin{eqnarray}
 \chi(\nu_{rf}) = \frac{\delta\nu_L^2 / (\pi \gamma)}{(\nu_{rf}-\nu_{L,0}-\delta\nu_L)^2 + 4 \pi^2 (\delta\nu_L^2 / \gamma)^2}.
 \label{eq:LineshapePT}
\end{eqnarray}
Note that the resonance frequency is shifted by $\delta\nu_L$. However, the shift cancels out, as the frequency ratio $\nu_L/\nu_c$ is measured and the cyclotron frequency is shifted by the same relative amount as $\nu_L$.

With the lower line width compared to the single-trap method, it is obviously desired to measure the magnetic moment with this measurement scheme. However, the strong advantage of this method comes with a challenging prerequisite for its application: The unambiguous identification of the spin state of a single proton/antiproton in the analysis trap. In the double trap measurement scheme, first, the spin state is analyzed in the analysis trap. Afterwards, the particle is transported to the precision trap, where the cyclotron frequency $\nu_c$ is measured and a magnetic rf-drive is applied to flip the spin. Subsequently, the particle is transported back to the analysis trap and the spin state is analyzed again, so that the spin-flip information for the Larmor resonance is obtained. Thus, it is essential to know the spin state before moving the particle in to the precision trap, otherwise the application of the double-trap method is impossible. 
\begin{figure}[htb]
  \centerline{\includegraphics[width=0.70 \textwidth,keepaspectratio]{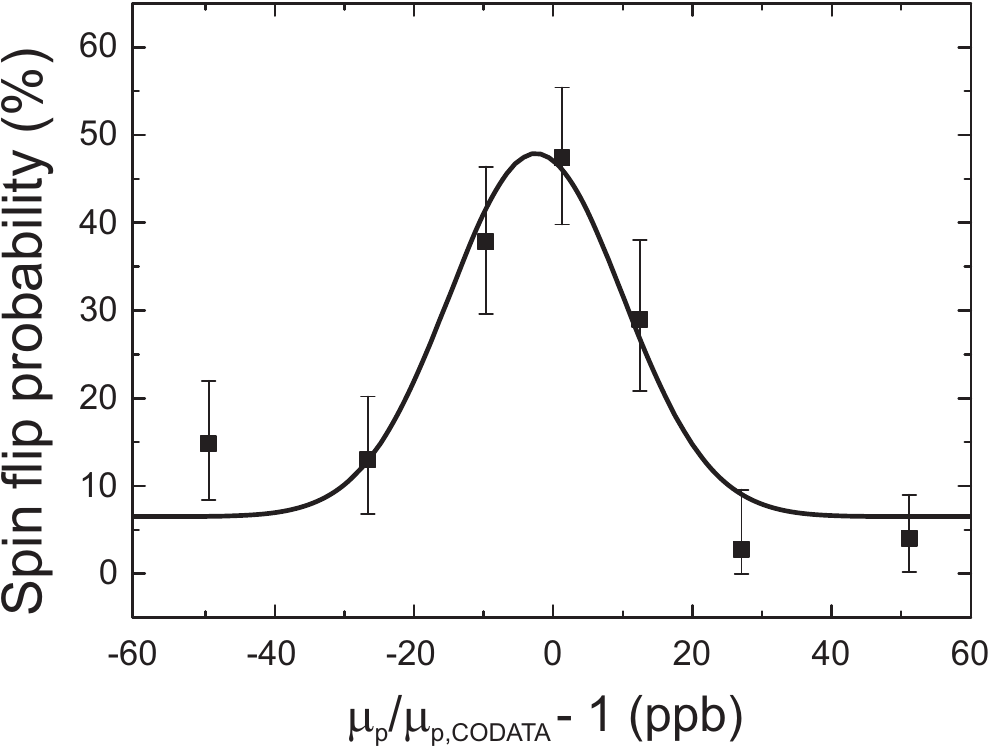}}
  \caption[ATLarmorResonance]{$g$-factor resonance of a single proton measured with the double-trap technique from Ref.~\cite{MooserNature2014}. The black line represents a maximum-likelihood fit to the data. Note that, the data points are shown for visualization and do not explicitly enter the fit. The width of the obtained resonance is 12.5 ppb. The magnetic moment was determined to a precision of 3.3 ppb from this measurement. For details see text.} \label{fig:DoubleTrapResonance}
\end{figure}
By further suppressing spurious noise in the analysis trap combined with resistive cooling of the particle to a sub-thermal cyclotron energy, an axial frequency stability of $\Xi$ = 55$\,$mHz has been achieved \cite{MooserPRL2013}. This allowed the first direct observations of single spin transitions \cite{MooserPRL2013} with a single trapped proton. Shortly after, we reported on the first demonstration of the double-trap method with a single proton \cite{MooserPLB2013}, where spin-flips driven in the homogeneous field of the precision trap were detected in the analysis trap.

The $g$-factor resonance of the first double-trap measurement of the proton magnetic moment is given in Fig.~\ref{fig:DoubleTrapResonance}, showing the spin-flip probability for normalized values of $\nu_{rf}/\nu_c$. The proton $g$-factor is extracted with an uncertainty of only $3.3\,$ppb from this resonance, which is a factor 760 more precise than the measurements carried out in analysis traps \cite{CCRodegheri2012,Jack2012Proton}, and a factor of 3 smaller than the best indirect measurement with the hydrogen maser \cite{winkler1972magnetic}. By applying this scheme to a single trapped antiproton, the current precision in its magnetic moment \cite{Jack2013Antiproton} can be improved by three orders of magnitude.


\section{Experimental Setup}
\label{Sect:3}

\subsection{Antiproton beam production}

The BASE experimental setup is located at the Antiproton Decelerator facility of CERN, which is the worldwide only source for high-intensity pulses of low-energy antiprotons \cite{Maury1999HypInt}. To create antiprotons, protons are accelerated up to a momentum of 26 GeV/c using a linear accelerator, the Proton Synchrontron Booster (PSB), and the Proton Synchrotron (PS) \cite{LHCDesignReport}. After acceleration, an intense pulse of $10^{13}$ protons is focused on an iridium target which creates a highly divergent pulse of antiprotons in pair production processes. The target is followed by a magnetic horn which serves as a collector lens. It allows to transfer about 50 million antiprotons at 3.5 GeV/c momentum from the target into the AD, where antiprotons experience alternating cooling and deceleration steps. To reduce the transverse emittance, stochastic cooling is applied at the initial momentum of 3.5 GeV/c and after the first deceleration step at 2.0 GeV/c, and electron cooling at the lower momenta of 300 MeV/c and 100 MeV/c. Eventually after a cycle length of 120 seconds, a bunch of about 30 million antiprotons with a kinetic energy of 5.3 MeV and a pulse length of about 150 ns is transferred to the experiments. 

\begin{figure}[ht!]
  \centerline{
  	\includegraphics[width=0.95 \textwidth,keepaspectratio]{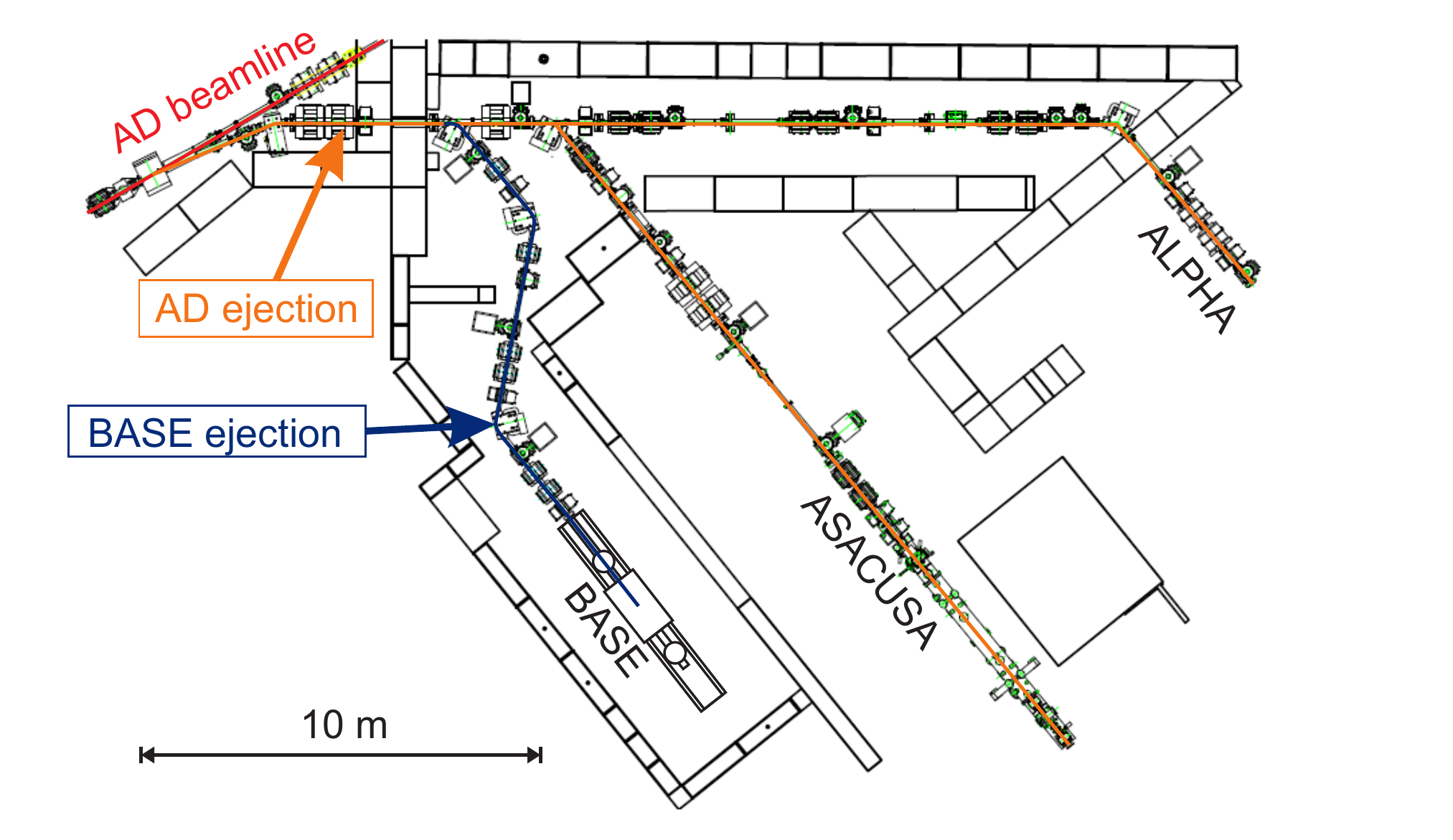}
	}
  \caption[Experiment]{Top view of the integration of the BASE experimental zone in the AD facility is shown.}
	\label{fig:BASE-Zone}
\end{figure}

In order to supply BASE with antiprotons, a new ejection beamline and a new experiment zone were constructed. A top view drawing of the integration of the BASE experimental zone into the AD facility is shown in Fig.~\ref{fig:BASE-Zone}.

\subsection{Overview of the BASE apparatus}

\begin{figure}[ht!]
  \centerline{
  	\includegraphics[width=0.95 \textwidth,keepaspectratio]{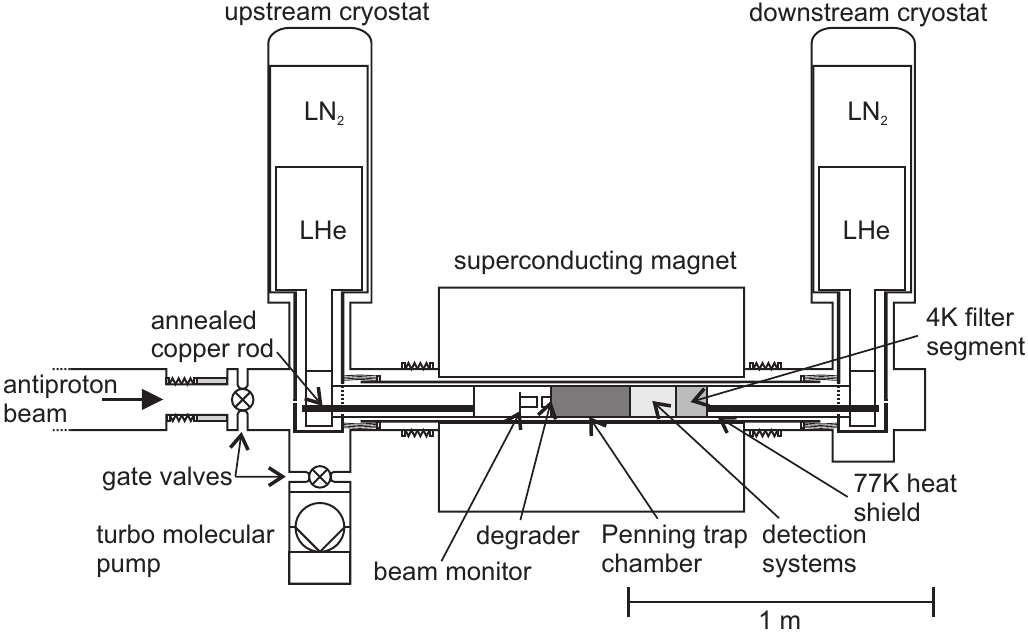}
	}
  \caption[Experiment]{Sketch of the BASE experiment. For details see text. }
	\label{fig:Apparatus}
\end{figure}

The BASE apparatus is an extension of the Mainz proton double-trap experiment which allows the injection and storage of antiprotons. In addition, it has several new features implemented to improve the limitations of the proton double-trap measurement. An overview of the BASE apparatus is shown in Fig.~\ref{fig:Apparatus}. A superconducting magnet is housing the Penning-trap system inside a horizontal bore. The trap system is placed inside a hermetically sealed cryogenic chamber, which is cooled to liquid helium temperature by the two cryostats placed upstream and downstream of the magnet. A horizontal support structure which is anchored on both ends to the liquid helium stages of the two cryostats holds the trap chamber in the magnet bore inside an isolation vacuum. The image-current detection systems for the frequency measurements and a segment with cryogenic electronics and filters for the voltage biasing of the trap electrodes are also located on the liquid helium stage next to the trap chamber. The antiprotons are injected into the Penning-trap system through a vacuum-tight degrader window, which also serves as a separator between the isolation vacuum and the trap vacuum. A cryogenic beam monitor upstream of the traps is used to align the antiproton beam to the trap center.

\subsection{Antiproton transfer beamline}
\label{Sect:3.1}

\begin{figure}[htb]
  \centerline{\includegraphics[width=0.80 \textwidth,keepaspectratio]{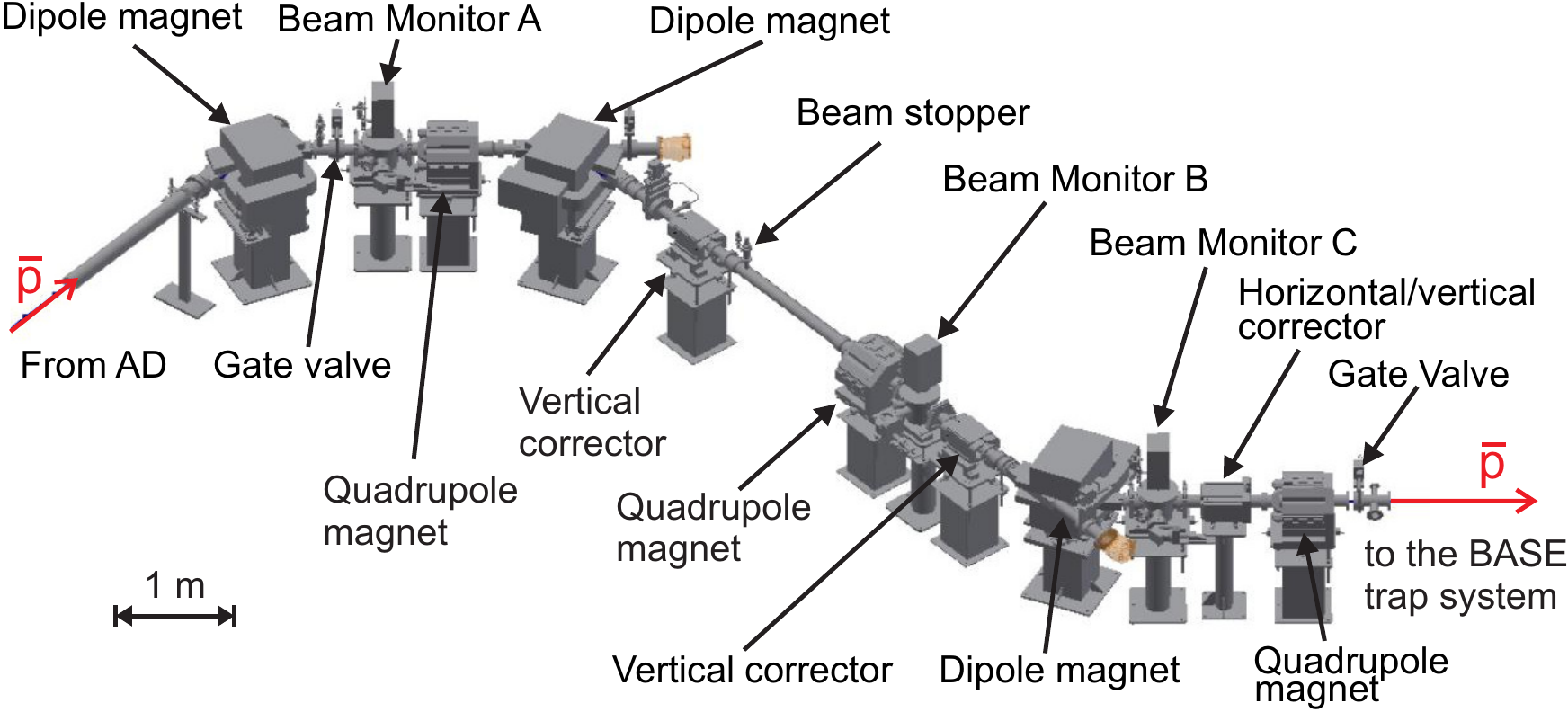}}
        \caption[Zone]{3D model of the AD ejection beamline for BASE showing the magnets, beam monitors, and the vacuum chambers. For details see text.}
\label{fig:BASE-Beamline}
\end{figure}

\begin{figure}[htb]
  \centerline{\includegraphics[width=0.75 \textwidth,keepaspectratio]{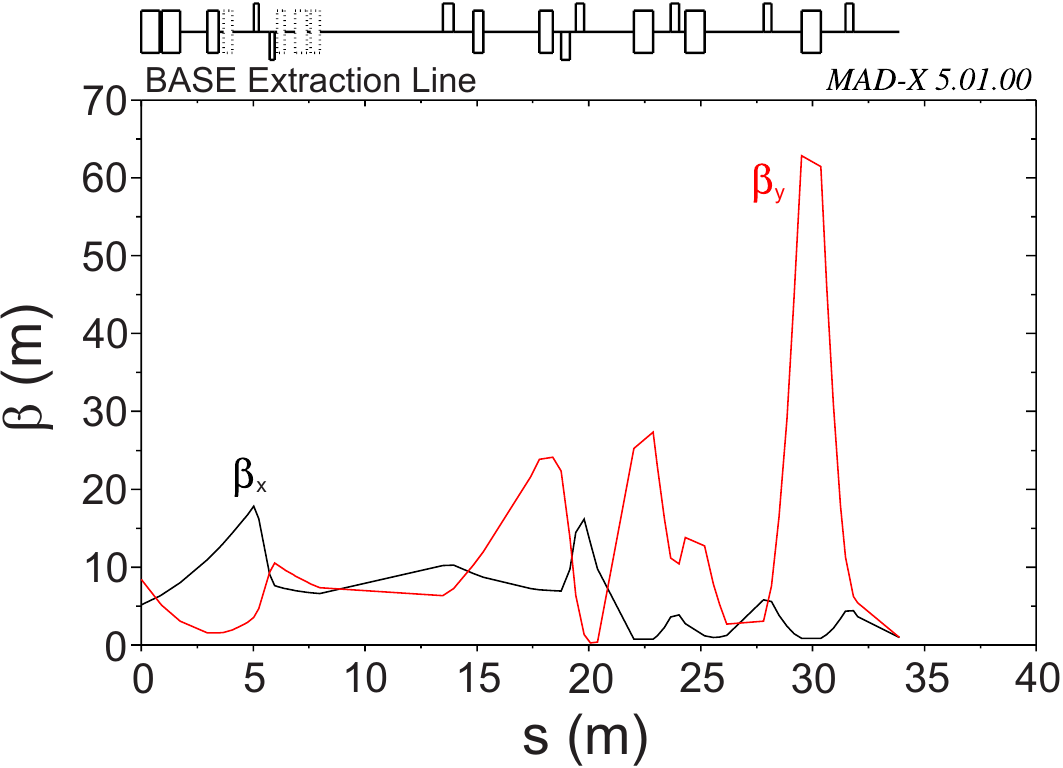}}
        \caption[Zone]{Beta functions $\beta_x(s)$ and $\beta_y(s)$ in the horizonal and vertical plane, respectively, in the AD ejection line for BASE from the AD ejection septum until $s$= 34 m, which is the position of the focal point of the antiprotons in the BASE apparatus. }
\label{fig:BASE-MadX}
\end{figure}

The design of the AD ejection beamline for BASE is shown in Fig.~\ref{fig:BASE-Beamline}. MAD-X was used for its design and to simulate its ion optical properties. The results of these calculations are shown in Fig.~\ref{fig:BASE-MadX}. A dipole magnet is inserted into the AD ejection beamline in order to switch between BASE and the other AD experiments. The AD ejection beamline requires three dipole magnets with 50.4 degrees deflection and three vertically focusing quadrupole magnets to keep the dispersion and the beam diameter at an acceptable level. To optimize the antiproton transport, the beamline has three position-sensitive beam monitors consisting of a phosphor screen and a CCD camera to observe the annihilation signal.

To avoid large magnetron and cyclotron radii of the captured particles, the antiproton pulse has to enter the BASE apparatus parallel onto the axis of the magnetic field with a narrow spatial distribution. Therefore, the focal point generated by the last quadrupole magnet is placed inside the Penning-trap stack at a focal length of 1.5$\,$m. The diameter of the pulse is reduced from a maximum diameter of 50 mm inside the quadrupole to less than 2 mm at the focal point, which is sufficient compared to the inner trap diameter of 9 mm at the injection side. The beam dispersion was matched to zero at the focal point in both planes. 

The last combined horizontal and vertical corrector magnet behind the last dipole magnet is used to steer the beam to the center of the trap. To check the position of the incoming antiprotons inside the BASE apparatus, a cryogenic beam monitor placed in front of the degrader window is used. It consists of four Faraday cups made from a four-fold segmented plate with 50 mm diameter and a 9 mm hole in the center. To ensure high sensitivity of the beam monitor, cryogenic silicon-based low-noise charge amplifiers with a signal strength of 2.5 V/pC are used for the readout. Using this beam monitor, the antiproton beam can be reliably centered to the axis of the Penning-trap system.

\subsection{Superconducting magnet}
\label{Sect:3.2}
A homogeneous magnetic field with high temporal stability is one of the key-components of the experiment, since it defines the measured frequencies $\nu_L$ and $\nu_c$. Further, the suppression of external magnetic field fluctuations by using a solenoid assembly with self-shielding geometry is of great importance for high-precision measurements \cite{JerrySFSHCoil}. The shielding factor $S$ describes the suppression of external magnetic field fluctuations in the center of the superconducting solenoid by $S^{-1}=1+B_i/B_e$. For the measurements reported here, BASE refurbished a spare magnet with a shielding factor $S$ of about 10.  The horizontal room-temperature bore of 150 mm diameter houses the Penning-trap system as shown in Fig.~\ref{fig:Apparatus}. The magnet is operated at a field strength of 1.945 T. By adjusting the shim coils, a spatial homogeneity of 0.25 ppm/cm around the homogeneous center and a homogeneity of 5 ppm/cm in a cylindrical volume of 9 mm diameter and 120 mm length was obtained.

As the experiment is operated in the AD facility, it is exposed to external magnetic field changes caused by the operation of the AD and the neighbouring experiments. To increase the temporal stability for the high-precision measurements, a new self-shielding superconducting magnet from \emph{Oxford Instruments} has been installed after the end of the AD physics run 2014. To further compensate for external magnetic field drifts, a self-shielding coil \cite{JerrySFSHCoil} will be installed to stabilize the magnetic flux in the Penning-trap chamber, and an external stabilization system based on a flux-gate locked pair of Helmholtz-coils as described in Ref.~\cite{FluxGate,Repp} will be used for further suppression. The combined shielding factor is expected to reach 500 to 1000, so that the external magnetic field fluctuations can be reduced by about two orders of magnitude compared to the current system. The new magnet was shimmed to a similar spatial homogeneity with $\Delta B/B <$ 0.3 ppm of 1 cm$^3$ volume at the homogeneous center and to better than $\Delta B/B <$ 50 ppm in the volume covered by the Penning-trap stack. The temporal stability of the magnetic field is better than $(\Delta B/B) (1/\Delta t) <$ 5$\times 10^{-9}\,$h$^{-1}$.

\subsection{Cryo-mechanical setup}
\label{Sect:3.3}
\begin{figure*}[htb]
  \centerline{\includegraphics[width=0.95 \textwidth,keepaspectratio]{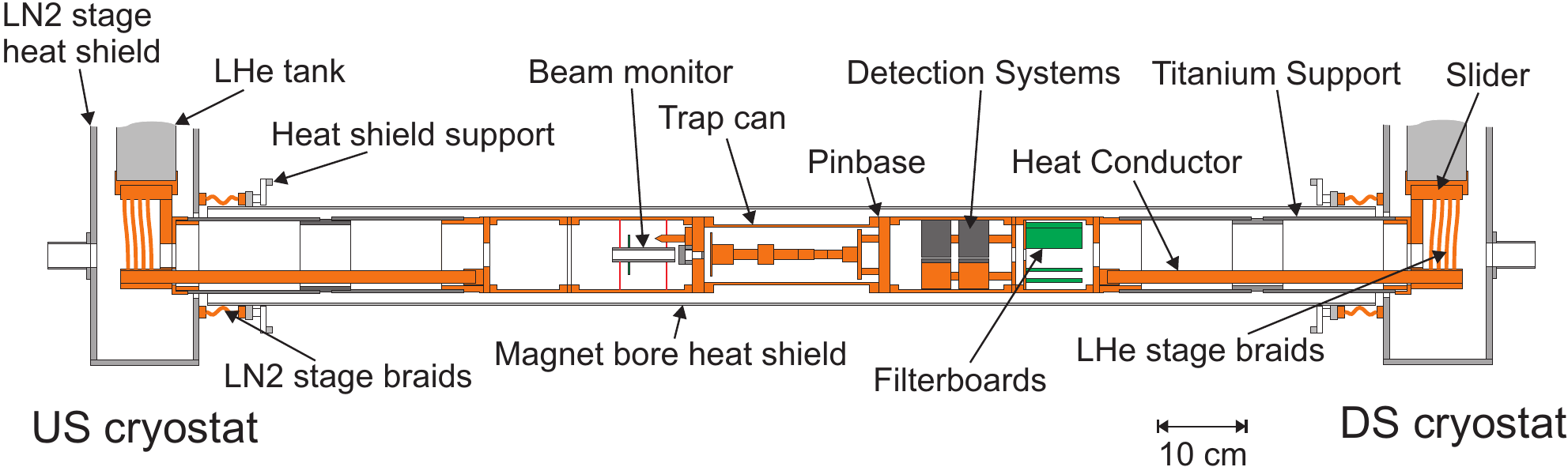}}
        \caption[Zone]{LN$_2$ and LHe stages of the experiment. For details see text.}
\label{fig:CryoInlay}
\end{figure*}

Two cryostats with reservoirs for 35$\,$l liquid nitrogen (LN$_2$) and 35$\,$l liquid helium (LHe) each provide the cryogenic temperatures for the experiment. The assembly of the LN$_2$ and LHe stage of the experiment is shown in Fig.~\ref{fig:CryoInlay}. Using a two cryostat construction for a cryogenic experiment in horizontal geometry has the advantage that the LHe stage can be anchored at both ends to the liquid-helium tanks without the need for an additional support structure. This minimizes the conductive heat load from the LN$_2$ stage on the LHe stage and ensures a low power load on the LHe reservoir.

The radiative load on the LHe stage is suppressed by thermal shields connected to the LN$_2$ tanks of the cryostats. Inside the cryostats, rectangular heat shields made out of 8 mm thick aluminum plates enclose the tail of the LHe tanks and the supports of the 4 K stage. In the magnet bore, an aluminum tube of 127 mm diameter and 3 mm wall thickness is used as a radiation shield. It is mechanically anchored to the vacuum chambers at room temperature using a fibre glass disk as thermal insulation. As thermal link, oxygen-free high conductivity (OFHC) copper braids of 600 mm$^2$ cross section in total form a good connection to the cryostat heat shields. This compensates mechanical stress during cool-down to cryogenic temperatures. The complete LN$_2$ stage is enclosed in 20 layers of multi-layer insulation (MLI) foil. Thereby, a temperature of 80 K at the bottom of the cryostat heat shield and 86 K at the center of the magnet bore heat shield are reached at a total load of 50 W. The standing time of the liquid nitrogen stage is about 70 h and 58 h for the upstream and downstream cryostat, respectively. The downstream cryostat has a higher evaporation rate due to the additional load from the trap biasing lines, in particular by the high-voltage lines.

The inlay of the liquid helium stage consists of a mechanical support, the cryogenic electronics, and the Penning-trap chamber. The latter is a cylindrical indium-sealed cryogenic vacuum chamber (71 mm inner diameter, 234 mm length) located at the center of the 4 K stage enclosing the Penning-trap system. The chamber is made out of high-purity copper. A flange with cyrogenic feedthroughs, the so-called pinbase, closes the chamber at the downstream side. All signals for the single-particle detection systems, trap biasing, particle excitation, spin-flip drive and the catching HV-pulses are connected to the Penning traps via the pinbase. On the upstream side, the Penning-trap chamber is closed by the degrader flange, which has a stainless-steel foil of 25 $\mu$m thickness and 9 mm diameter placed in the center. The foil is vacuum-tight but transparent for the injection of 5.3 MeV antiprotons. In addition, the flange has a connection for a pinch-off tube. To achieve ultra-high vacuum in the Penning-trap chamber, it is pumped out through this connection to a pressure of less than 10$^{-6}$ mbar. Subsequently, the pumping connection is pinched-off with a cold-weld technique and the chamber is installed into the magnet bore. Placed in an isolation vacuum and cooled by the liquid helium cryostats, the Penning-trap chamber forms an independent vacuum system with $\approx$ 5 K wall temperature. The residual gas pressure in the chamber drops below the detection threshold of conventional vacuum gauges. It is below of 10$^{-14}\,$mbar \cite{GoodVacuum} and can be only determined indirectly by the storage time of the trapped antiprotons. We demonstrated that the storage time can exceed one year \cite{SmorraIJMS2015}.

The mechanical support of the Penning-trap chamber has been designed to be symmetric with respect to the magnet's center plane. Thereby, a tilt of the trap axis relative to the magnetic field due to unequal deformation of the support structure can be avoided. Two high-purity copper segments are attached to the Penning-trap chamber on each side. Downstream they contain the single-particle detection systems and cryogenic filters for the trap biasing lines, and upstream the beam monitor and parts of the degrader assembly. As next element, two titanium tubes of 170 mm length and 98 mm diameter with a titanium connection piece are placed on each side around the copper parts in the magnet center. Despite its low heat conductivity at 4 K, titanium was selected for this part of the support structure due to its high stiffness and low weight. At each end of the LHe stage a short copper tube of 30 mm length and 90 mm diameter rests in the cryostat support structure. To prevent mechanical stress due to the contraction while cooling down, the cryostat support structure is attached to a slider on a ball bearing at the bottom of the LHe tank. The movement of the slider compensates the mechanical contraction of the inlay.

To ensure a good thermal link of the trap and the superconducting detection systems to the LHe tanks, the copper segments in the center of the LHe stage are connected to the cryostats with two heat conductors made from annealed OFHC copper rods of 16 mm diameter. On the trap side they are bolted into the last copper segment, and on the cryostat side clamps with OFHC copper braids complete the thermal link to the LHe tank. The braids have a total cross section of 360 mm$^2$ and 125 mm length. The thermal load on the liquid helium reservoirs by the cryogenic inlay is estimated to be 90 mW radiative load, 15 mW conductive load due to wiring, and 20 mW power load from the cryogenic amplifiers. Considering the intrinsic heat load of the cryostats, the LHe stage is designed to have a hold time of 120 h.


\subsection{Degrader system}
\label{Sect:3:D}

To decelerate the 5.3 MeV antiprotons provided by the AD, a system of degrader foils is used. Energetic antiprotons penetrate the degrader material, lose energy in inelastic scattering processes, and are eventually stopped at a certain range. If the degrader foil is chosen thin enough, low-energy antiprotons are transmitted and can be confined partly in the Penning-trap system (see Sect.~\ref{Sect:3.4}) by fast high-voltage pulses. It was shown that the maximum efficiency for low energy antiproton transmission is reached when 50\% of the incoming particles are stopped in the degrader \cite{Holzscheiter1992}. However, the efficiency of the stopping process is quite sensitive to the choice of the degrader material, as well as the thickness and the placement of the degrader components. Moreover, accurate calculations of the stopping power are hampered by the lack of experimental data of the empirical stopping power models at low energies \cite{Holzscheiter1992,Ziegler1999}. 


\begin{figure}[ht!]
  \centerline{
  	\includegraphics[width=0.95 \textwidth,keepaspectratio]{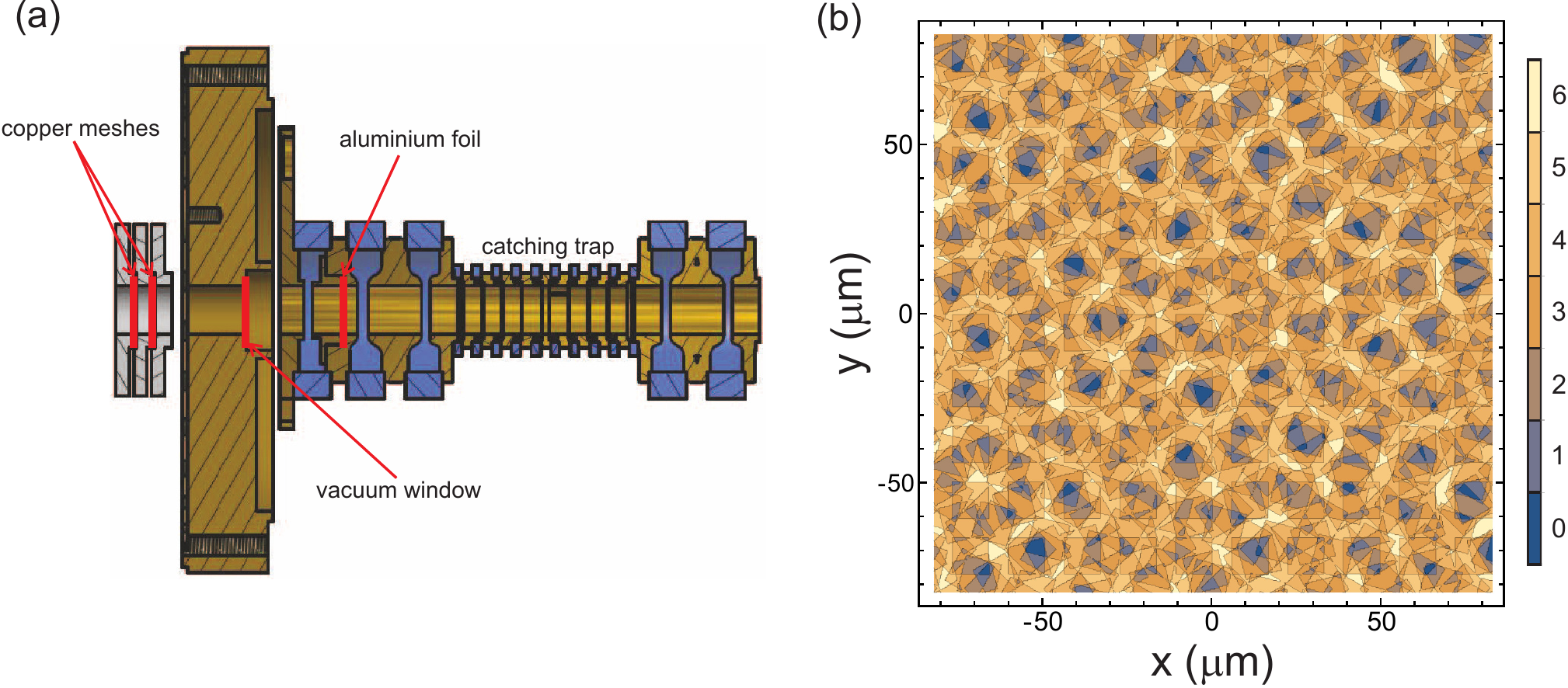}
	}
  \caption[Degrader]{(a) Placement of the degrader foils for the antiproton injection. There are three copper meshes at each of the two indicated locations. (b) Calculation of the structure generated by the copper mesh assembly. The number of meshes passed by antiprotons as function of the location is shown. For details see the text.}
	\label{fig:DegraderDetails}
\end{figure}

\begin{table}
\centering
\begin{tabular}{|c|c|c|c|c|c|c|c|}
  \hline
$N$ &        0 &        1 &         2 &         3 &         4 &         5 &        6 \\ \hline
$p$ & 1.09$\%$ & 5.55$\%$ & 16.51$\%$ & 29.32$\%$ & 30.07$\%$ & 15.03$\%$ & 2.44$\%$ \\
  \hline
\end{tabular}
\caption{Probability $p$ of a particle to hit the massive part of $N$ out of the six copper meshes of the mesh degrader. The meshes have an open area of 44 $\%$, therefore the highest probability is obtained for $N$=4. For details, see text.}
\label{tab:Degrader}
\end{table}
		
\begin{figure}[ht!]
  \centerline{
  	\includegraphics[width=0.6 \textwidth,keepaspectratio]{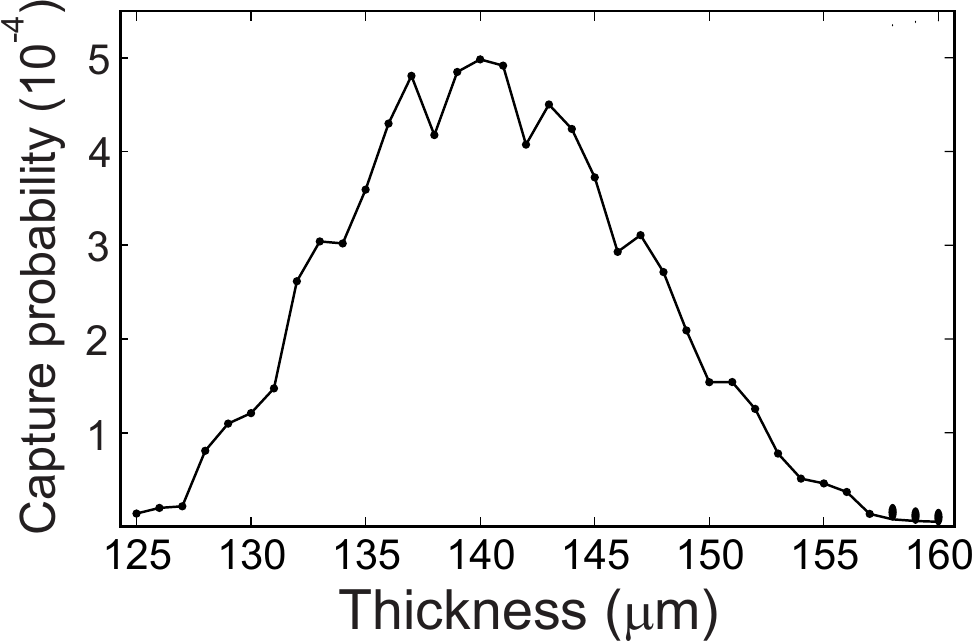}
	}
  \caption[Degrader]{Catching probability of 5.3 MeV antiprotons as function of the thickness of the aluminium degrader calculated with SRIM. The configuration in SRIM consisted of the mesh degrader, the 25 $\mu$m stainless steel window, and the aluminium degrader with variable thickness. Particles with a trajectory not exceeding the trap radius of 4.5 mm and with an axial energy below 1 keV are considered as captured.}
	\label{fig:Degrader}
\end{figure}

To account for this, the degrader system of BASE, shown in Fig.~\ref{fig:DegraderDetails} (a), consists of three elements. The first part provides a variable stopping power to compensate uncertainties in the stopping power calculations and the thicknesses in the production of the degrader foils. It consists out of 6 stacked copper meshes with a thickness of 2.5 $\mu$m rotated by 15 degrees relative to each other. The grid structure of the mesh (15.6 $\mu$m, 44$\%$ open area) is much finer than the antiproton beam diameter, which is typically 2 mm at this position. The pattern generated by the mesh assembly shown in Fig.~\ref{fig:DegraderDetails} (b) adds a large possible variation in stopping power with an equivalent thickness of 0 to 24 $\mu$m aluminium depending on the number of meshes 0 $<N<$ 6 hit by each antiproton. The probability $p$ for an antiproton to hit $N$ of the meshes are given in Tab.~\ref{tab:Degrader}. It is equivalent to the fraction of the area covered by the mesh material of $N$ meshes. As scattering in the degrader foils increases the beam diameter, the mesh assembly is placed directly in front of the Penning-trap chamber. The second degrader is the 25 $\mu$m stainless-steel vacuum window in the degrader flange. The last element is an aluminum foil directly in front of the upstream catching electrode with the purpose of matching the total stopping power of the degrader system for obtaining the maximum number of slow antiprotons. A calculation of the catching efficiency using the simulation code SRIM is shown in Fig.~\ref{fig:Degrader}. Antiprotons transmitted through the degrader system with a kinetic energy below 1 keV and an orbit that does not exceed the trap diameter can be confined by high-voltage pulses on the catching electrodes. The total trapping volume of the catching electrodes is 50 mm in the axial direction and 9 mm in diameter, enclosing the reservoir trap of the Penning-trap stack. In total, this degrader system has a catching efficiency better than 10$^{-4}$ in a broad range around the optimum thickness value. Compared to a single foil with exactly matched stopping power, the maximum efficiency is a factor three less, but the range in thickness with more than 10$^{-4}$ efficiency is a factor of three higher. Compared to tunable gas chambers \cite{JerryBarkas}, this system has slightly lower efficiency but it is robust, simple, reliable and provides enough antiprotons for single particle experiments.

\subsection{Penning-Trap System}
\label{Sect:3.4}
\begin{figure*}[htb!]
       \centerline{\includegraphics[width=0.95 \textwidth,keepaspectratio]{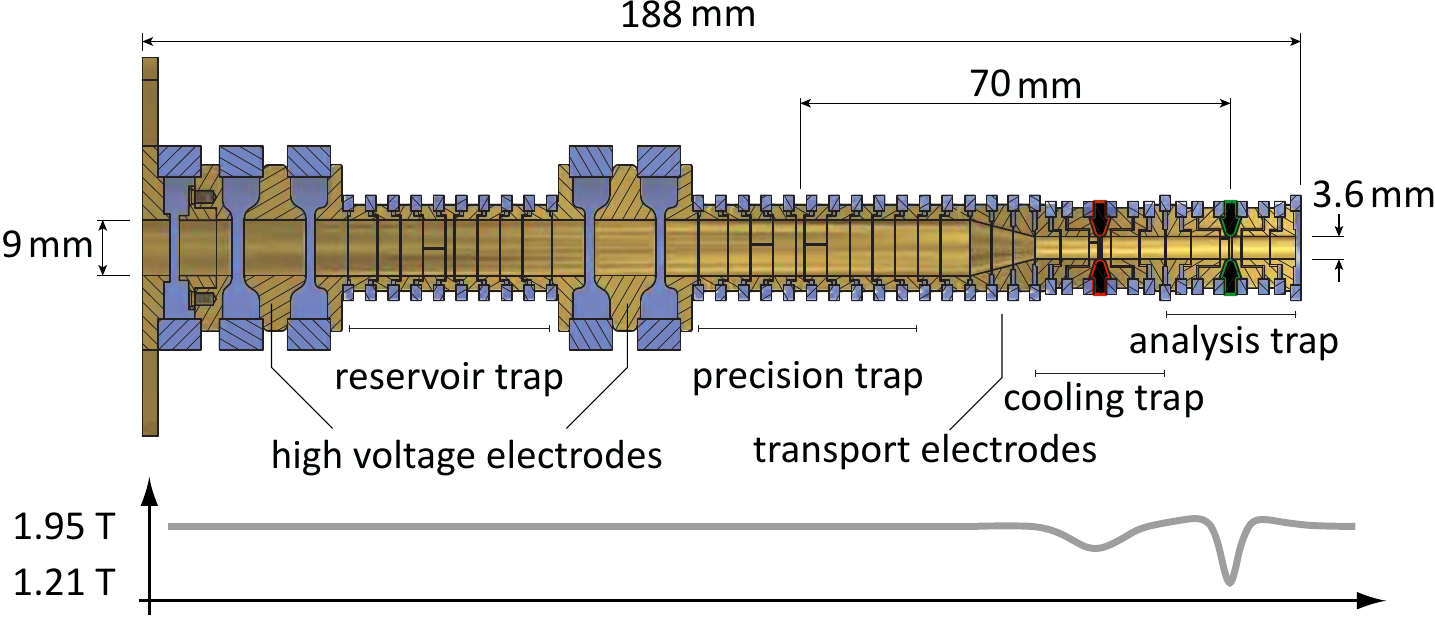}}
           \caption[Penning Trap]{Schematic of the BASE Penning-trap assembly. It consists of four cylindrical traps interconnected by transport electrodes. Two traps, reservoir trap and precision trap, have an inner diameter of 9$\,$mm, while the analysis trap and cooling trap have inner diameters of 3.6$\,$mm. The lower graph shows the on axis magnetic field of the trap assembly. The strong inhomogeneity in the analysis trap is for spin state analysis, the smaller homogeneity in the cooling trap for efficient cooling of the antiproton's cyclotron mode. } \label{fig:Trap}
\end{figure*}

The Penning-trap system, shown in Fig.~\ref{fig:Trap}, is the heart of the experiment. It is installed in the homogeneous center of the superconducting magnet. The trap stack consists of four cylindrical traps in a five-electrode orthogonal and compensated design \cite{gabrielse1989oep}. The individual traps are interconnected by transport electrodes in an optimal length-to-diameter ratio. To prevent oxidation, all electrodes are gold-plated. Compared to a classical double trap which consists of a \emph{precision trap} (PT) and an \emph{analysis trap} (AT), two traps were added: a \emph{reservoir trap} (RT) and a \emph{cooling trap} (CT). PT/RT  and AT/CT have inner diameters of 9.0$\,$mm and 3.6$\,$mm, respectively. All electrodes are machined with an absolute precision better than $5\,\mu$m. The sapphire rings used to separate the individual electrodes and to prevent electrical contacts have a height of 3$\,$mm and a similar machining precision. The crucial parameters of the individual traps including the magnetic properties at each trap center are summarized in Tab.~\ref{tab:TrapParameters}.\\

\emph{Precision (PT) and analysis trap (AT) - }These two traps are used to perform double Penning-trap measurements. The PT is for precision frequency measurements, the AT for analysis of the spin state as explained in Sect.~\ref{Sect:2}. The layout of the AT is an exact copy of the well-working analysis trap used at our experiment at Mainz \cite{CCRodegheri2012}. Compared to this system the inner diameter of the PT was modified from 7$\,$mm to 9$\,$mm. This reduces the systematic shifts in cyclotron frequency measurements caused by anharmonic potential and image charge corrections. In the BASE precision trap, the systematic shift of the cyclotron frequency is only 40$\,$ppt, which is 2.5 times smaller than in our proton Penning-trap system in Mainz. Furthermore, the distance between the centers of the PT and AT is increased from 43.7 mm to 69.7$\,$mm, so that magnetic field inhomogeneities in the center of the PT which are caused by the strong magnetic bottle in the AT are reduced. The magnetic gradient term in the BASE PT is $B_{1,B}=-0.022\,$T/m, the bottle term is $B_{2,B}=0.67\,$T/m$^2$, which is 4 and 6 times smaller than in the trap used in \cite{MooserNature2014}. \\

\begin{table}
\centering
\begin{tabular}{|c|c|c|c|c|}
  \hline
Trap & Inner Diameter & $C_2$ (m$^{-2}$) & $B_1$ (T/m) & $B_2$ (T/m$^2$)   \\ \hline\hline
RT   &      9.0$\,$mm &  18508 & $<$0.010 & $<$1  \\
PT   &      9.0$\,$mm &  18508 & 0.022 & 0.67 \\
CT   &      3.6$\,$mm & 116000 & 1.900  & 16\,000 \\
AT   &      3.6$\,$mm & 116000 & 0.100  & 300\,000 \\
  \hline
\end{tabular}
\caption{Geometry parameters and magnetic field gradients of the four Penning traps in the BASE apparatus.}
\label{tab:TrapParameters}
\end{table}

\emph{Reservoir Trap (RT) - }The RT functions in online operation as catching trap to capture low energy antiprotons from the AD. Therefore, it is placed in between two catching electrodes, which allow the application of DC and pulsed voltages of up to 8$\,$kV needed for capturing of antiprotons. In the period between two injection pulses, the captured particles are cooled by sympathetic cooling with electrons and accumulate in the harmonic potential well of the trap and remain there during the next catching pulse. Thus, several antiproton bunches can be stacked in the RT until an antiproton reservoir of about 1000 particles has been accumulated. Subsequently, the apparatus is disconnected from the ejection beamline and the RT functions as a particle reservoir. Single particles can be non-destructively extracted from the reservoir to supply the magnetic moment measurement cycle with single particles \cite{SmorraIJMS2015}. To maintain the reservoir, the entire trap is operated with uninterruptable power supplies which last for 10$\,$h during power-cuts. Thus, the RT enables long-term storage of antiprotons and allows BASE to operate even during accelerator shut-down periods and perform measurements when the magnetic noise in the AD hall is low.

\emph{Cooling Trap (CT) - } The purpose of the CT is fast and efficient cooling of the cyclotron mode of the trapped antiproton. This is essential for single spin-flip experiments to prepare particles with low cyclotron energies \cite{MooserPRL2013}. In the magnetic bottle, the magnetic moment induced by the motional energy in radial modes $E_{+,-}/B_0$ is coupled to the axial mode. Therefore, spurious noise in the trap at the modified cyclotron frequency, which drives cyclotron quantum transitions, increases the axial frequency fluctuation $\Xi_z$ and reduces the spin-flip detection fidelity \cite{MooserPRL2013}. Three quantum jumps in the cyclotron mode ($\Delta E=$180$\,$neV) contribute axial frequency shifts larger than that induced by a spin quantum jump. The cyclotron heating rates scale with the average quantum number $n_+$ of the cyclotron motion \cite{Ulmer2013ICPEAC}. Thus, for a $g$-factor measurement, efficient cooling of the cyclotron mode to low $n_+$ is crucial. In the proton double-trap system, a cyclotron energy of $E_+$ = 150 $\mu$eV ($n_+\approx$ 1200, $T_+$ = 1.7 K) sets the threshold for single spin-flip experiments. Therefore, it is necessary to cool the particle below the environment temperature \cite{MooserPRL2013}. In our measurements reported in \cite{MooserNature2014}, preparation of cyclotron states with adequately low energy to detect single spin transitions took on average about two hours. This was one of the limiting factors of this experiment.

\begin{figure}[htb!]
       \centerline{\includegraphics[width=0.95 \textwidth,keepaspectratio]{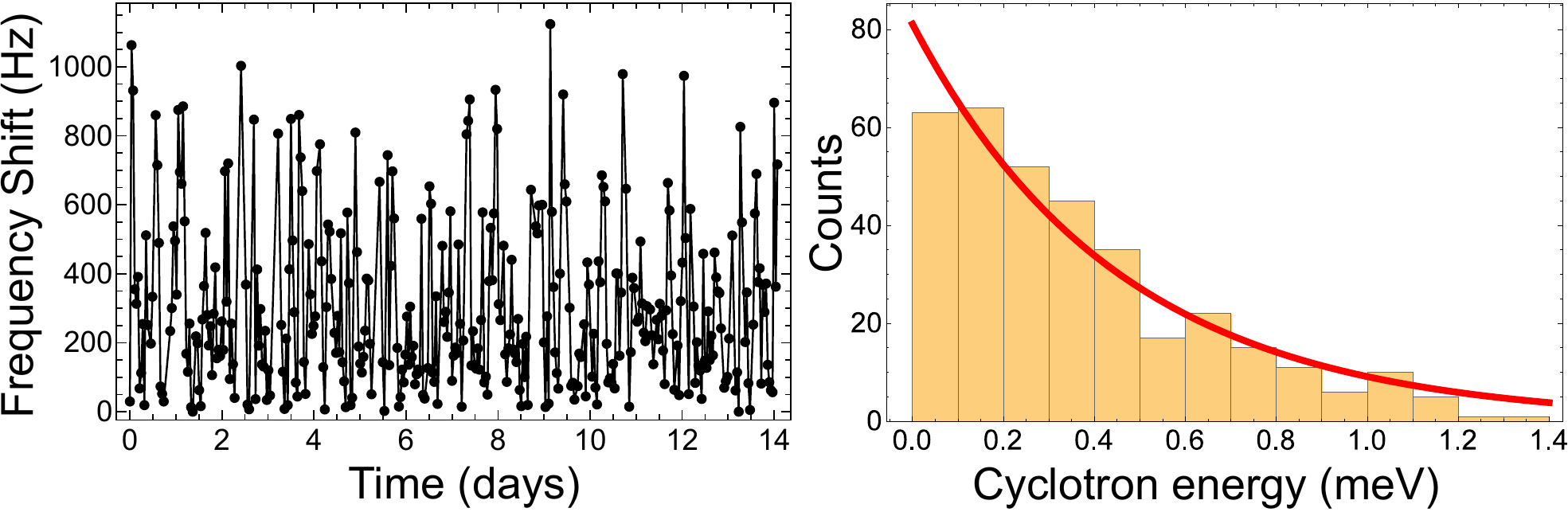}}
           \caption[Penning Trap]{Measurement of the cyclotron energy in the analysis trap of the proton double-trap system. A series of axial frequency measurements in the analysis trap after thermalizing the particle with the cyclotron detector in the precision trap is shown in (a). A histogram of the temperature corresponding to the cyclotron energy of the proton $T=E_+/k_B$ is shown in (b). From this distribution the temperature of the cyclotron detection system was determined to be 5.6(4) K.} \label{fig:ZyklotronTemperatur}
\end{figure}

The CT combines several techniques in one trap to cool the cyclotron motion of the trapped antiproton with high efficiency. It uses a ferromagnetic ring electrode made out of nickel. With the chosen geometry, it provides a magnetic inhomogeneity of $B_{2,CT}=16000\,$T/m$^2$, which allows to measure the cyclotron energy by the frequency shift $\Delta\nu_z$ induced by the cyclotron Energy $E_+$,
\begin{eqnarray}
\Delta\nu_z=\frac{1}{m \nu_z}\frac{B_2}{B_0} E_+.
\end{eqnarray}
Thus, the magnetic bottle of the CT provides temperature resolution of $d\nu_z/dT_+=5\,$Hz/K. The trap is equipped with both an axial and a cyclotron detection system. To provide a cold antiproton a particle is prepared at the center of the trap by cooling the magnetron and axial modes and measuring its axial frequency is measured. Afterwards, the cyclotron detector is tuned to the particle's resonance frequency $\nu_+$ and thermalizes the cyclotron mode. Using a temperature calibration of the magnetic bottle \cite{Djekic}, the cyclotron energy can be determined in a subsequent axial frequency measurement. In case the cyclotron energy is above the threshold for single spin-flip resolution, the cycle is repeated. This principle is also used in the proton double-trap system and the result of a cyclotron temperature measurement in the Mainz apparatus is shown in Fig.~\ref{fig:ZyklotronTemperatur}. There, the cyclotron cooling and the temperature measurement requires the use of both traps, the precision and analysis trap, respectively, and also involves shuttling of the particle between the traps. The large diameter of the precision trap required to reduce systematic shifts in the frequency measurements limits the effective electrode distance of the cyclotron detection system and thereby its cooling time constant. The possibility to perform both procedures in the CT eliminates the delay due to the transport time, and the small inner diameter of the CT provides a strong coupling of the cyclotron detector to the particle with a 4-fold increased coupling constant compared to the proton system. The high quality of the axial detection system used in this trap enables frequency measurements with 100$\,$mHz resolution within averaging times of 10$\,$s. Therefore, a cycle to thermalize the particle and analyze its cyclotron temperature takes about 60$\,$s and preparation of a particle with a cyclotron temperature below 1$\,$K, which is stable enough to observe single spin flips, will take a few minutes only. Thus, compared to a preparation time of two hours in the proton $g$-factor measurements, the CT reduces the particle preparation time by more than one order of magnitude.

\emph{Electron gun - }Another important component which is implemented into the trap stack is the field-emission electron gun. It consists of a sharp tungsten tip with a high aspect ratio, which is placed close to an acceleration electrode. A biasing voltage applied to the tip defines the energy of the extracted electrons. By applying voltages between 500$\,$V and 1.2$\,$kV to the acceleration electrode, electron currents in the range of 10$\,$nA to 350$\,$nA are extracted. On one hand, the electron gun provides particles for electron cooling of antiprotons \cite{JerryElectronCooling}. On the other hand the electron current is used to load the trap with protons. Electron impact on the degrader sputters hydrogen atoms out of the surface. These particles are then electron-impact ionized in the center of the RT, thus enabling the commissioning of the Penning-traps with protons.\\

\begin{figure}[htb!]
       \centerline{\includegraphics[width=0.95 \textwidth,keepaspectratio]{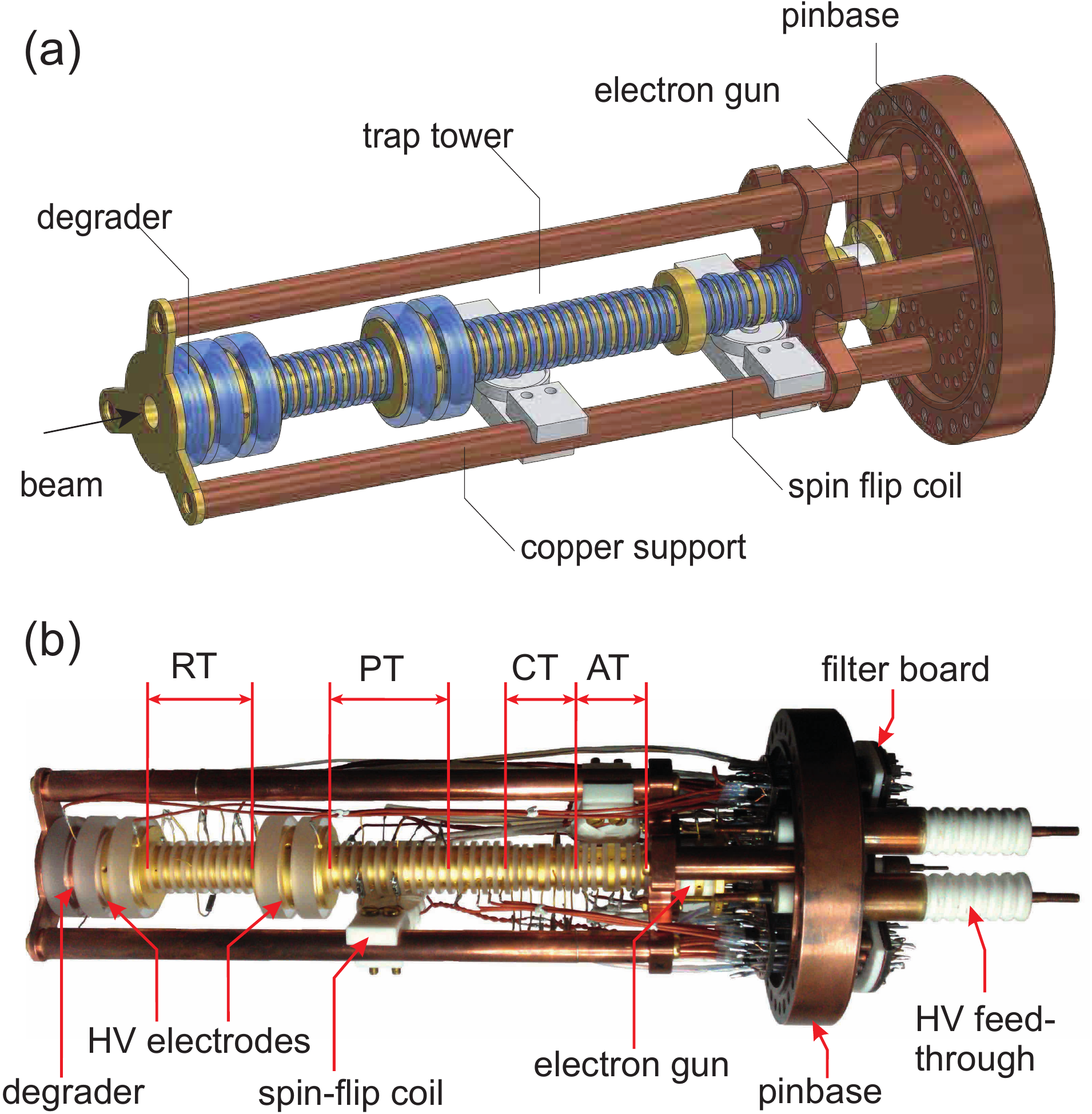}}
           \caption[Penning Trap]{(a) Entire assembly of the core of the experiment including the Penning-trap stack, the electron gun, spin-flip coils and the pinbase flange as 3D model. (b) A picture of the assembled Penning-trap system. } \label{fig:Assembly}
\end{figure}

\emph{Trap assembly - }The mounting of the trap stack is shown in Fig.~\ref{fig:Assembly}. The trap electrodes are pressed together by two plates which are fixed on the upper and lower end of a tripod made out of oxygen-free electrolytic (OFE) copper. The electron gun is connected to the lower plate and the entire assembly is attached to the pinbase flange by three OFE copper spacers. Coils to drive spin transitions are placed on PTFE supports mounted to the tripod. This assembly is placed into the Penning-trap chamber.

\subsection{Single-Particle Detectors}

\begin{figure}[htb!]
       \centerline{\includegraphics[width=0.55 \textwidth,keepaspectratio]{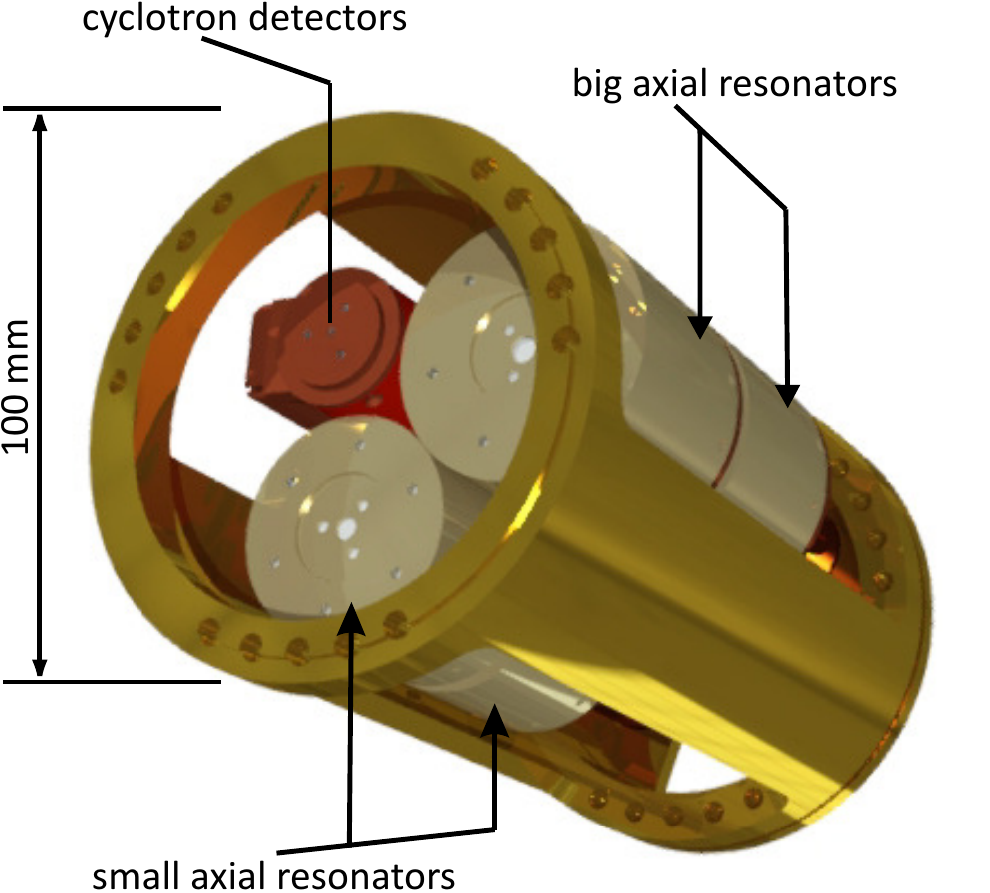}}
           \caption[Penning Trap]{Electronics segment which houses the six cryogenic single particle detectors, four axial detection systems and two detectors for the modified cyclotron frequency.} \label{fig:ESegment}
\end{figure}

All information about the trapped particles is provided by non-destructive detection of image currents induced in the trap electrodes (see Sect.~\ref{Sect:2}). For the BASE apparatus, six highly sensitive superconducting detection systems \cite{Ulm,UlmerPRL2} have been developed, two for the measurement of the cyclotron frequencies at $\approx29.65\,$MHz in the PT and the CT, as well as four axial detectors, one for each trap, operated in frequency ranges between 540$\,$kHz and 680$\,$kHz.\\
Each detector consists of a superconducting NbTi coil in a metal shielding which is attached to a cryogenic ultra-low-noise amplifier. In order to ensure high detection efficiency and to avoid electrical interference, all six detectors are placed in the detection segment shown in Fig.~\ref{fig:ESegment} next to the trap chamber.

\subsubsection{Axial Detection Systems}
The superconducting coils of the four axial detection systems are in toroidal design. The coils are made out of three layers of 75$\,\mu$m PTFE insulated NbTi wire wound on toroidal PTFE cores. They are mounted in a support which is inserted into cylindrical housings made out of NbTi. Due to geometrical constraints defined by the geometry of the apparatus, two different sets of axial resonators were designed. The first set has a diameter of 41$\,$mm and 34$\,$mm length, and the second one with 47$\,$mm in diameter and 40$\,$mm length. The small resonators are used for the PT and RT, while the bigger ones are connected to the CT and AT. The inductances of the resonators are at 1.6$\,$mH and 2.5$\,$mH, for the small and the big coils, respectively, the self capacitances being at 11$\,$pF. The quality factor of all four detection coils have been highly optimized. When cooled to 4$\,$K, quality factors of $200\,000$ are achieved for the small coils, and $250\,000$ and $500\,000$ for the big coils, respectively. The coil with the highest $Q$ value is used for the AT. The $Q$ values correspond to unloaded effective parallel resistances $R_{p,u}$ of several G$\mathrm{\Omega}$.  \\

\begin{figure}[htb!]
  \centerline{
  	\includegraphics[width=0.65 \textwidth,keepaspectratio]{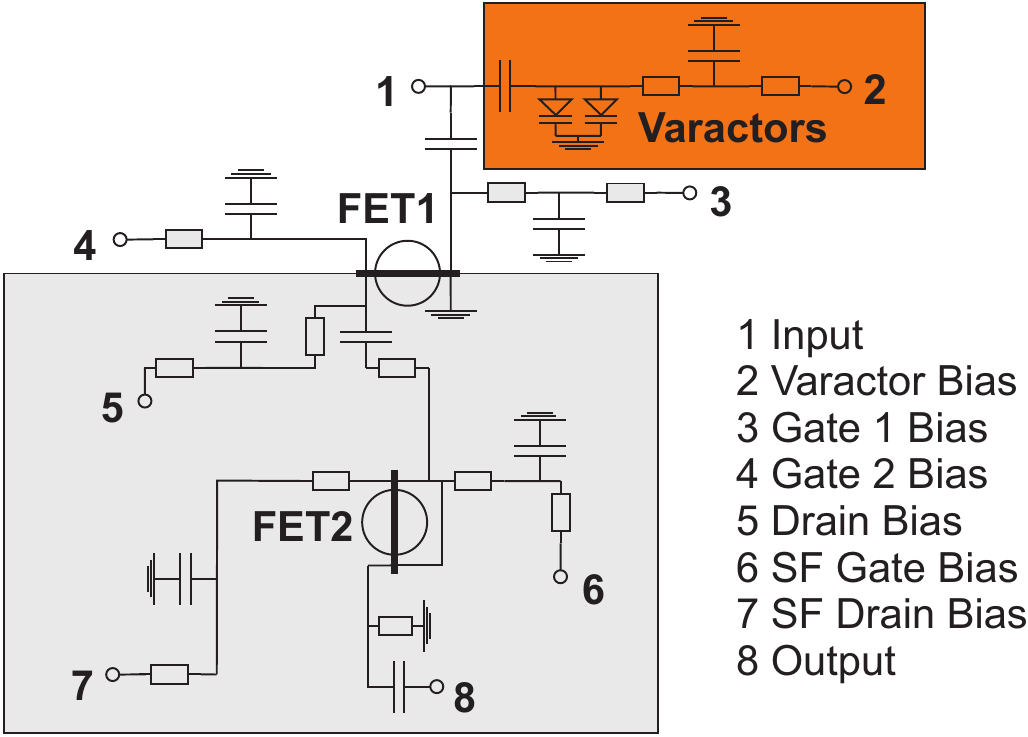}
	}
  \caption[Amplifier Layout]{Layout of the cryogenic amplifiers using two GaAs FETs: FET1 as common-source input stage and FET2 as source follower in common-drain configuration. All connections for biasing are filtered by RC-circuits. The detection systems for the axial and cyclotron mode use the same amplifier layout, except that the cyclotron amplifiers have in addition varactor diodes with biasing circuit (orange box), and a housing which separates the input and output stage (gray box) to prevent feedback effects.}
	\label{fig:AmpLayout}
\end{figure}

\begin{figure}[htb!]
       \centerline{\includegraphics[width=0.5 \textwidth,keepaspectratio]{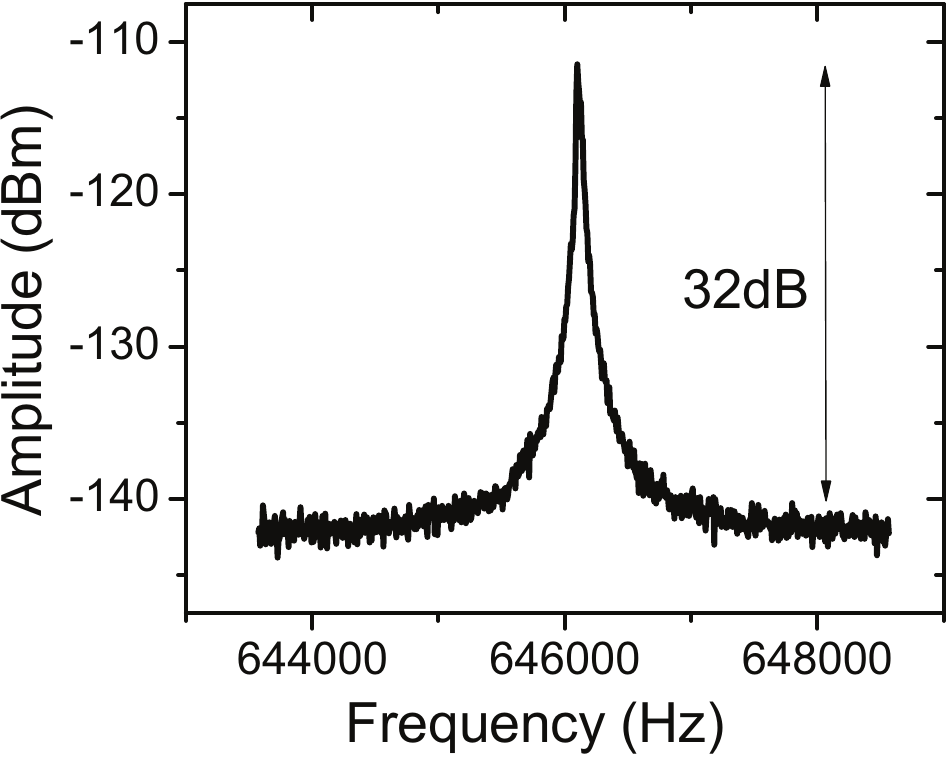}}
           \caption[Penning Trap]{Noise resonance of the axial detection systems used in the RT.   } \label{fig:Axial}
\end{figure}

The layout used for the cryogenic amplifiers of the detection systems is shown in Fig.~\ref{fig:AmpLayout}. The amplifiers are based on GaAs field effect transistors (FET's) with a high impedance common-source input stage and a source follower for impedance matching in common-drain configuration as output stage \cite{UlmerNIMA2012}. The input stage of the amplifiers for the axial detection systems of the cryogenic amplifiers are based on NE25139 FET transistors with an input resistance of $R_{\mathrm{in}}$=8 M$\mathrm{\Omega}$, input capacitance of $C_{\mathrm{in}}$=1.6 pF, and an equivalent input noise of 0.8 nV/Hz$^{1/2}$ at 550 kHz to 0.65 nV/Hz$^{1/2}$ at 1 MHz. For impedance matching of the outputs, CF739 transistors are used. At typical power consumptions of 2$\,$mW to 3$\,$mW, each amplifier provides a gain of about 15$\,$dB. The amplifier boards are made out of low-loss PTFE laminates with a loss tangent $<5\cdot 10^{-5}$ at 4$\,$K to minimize parasitic losses. \\
Inductors and amplifiers are connected and decoupled by a coupling factor $\kappa$, which is adjusted by tapping the coils at a certain winding ratio $N_2/N_1$, where $N_0=N_1+N_2$ is the total number of turns, about 750 for the small and 1100 for the big coils. This allows to adjust the effective parallel resistance of the detector $R_p$ consisting of $R_\text{in}/\kappa^2||R_{p,u}$. To obtain optimal frequency resolution in a short FFT averaging time, we set $\kappa$ in a way that the width of the axial frequency dips of a single antiproton in each trap are in the range of 1$\,$Hz to 3$\,$Hz. \\
When connected to the trap system, the small detection systems have resonance frequencies of 646$\,$kHz (RT) and $689\,$kHz (PT), while the big detectors are at $545\,$kHz (AT) and $585\,$kHz (CT), respectively. A noise resonance of the RT detection system is shown in Fig.~\ref{fig:Axial}. A signal-to-noise ratio (SNR) of 32$\,$dB at a quality factor of $Q=11500$ is achieved. With these parameters, the axial frequency can be determined in a dip measurement with a fit uncertainty of 47$\,$mHz from an FFT spectrum with only 30$\,$s averaging time.
With an independently measured coupling factor $\kappa$ and equivalent input noise $e_n$ of the amplifiers, the noise resonances contain all information required to determine the effective temperatures $T_\text{z}$ of the detection systems. For all detectors the extracted results are $T_\text{z}\approx5.9(1.1)\,$K, which is close to the physical temperature of the apparatus.

\subsubsection{Cyclotron Detection Systems}

The detection systems for the modified cyclotron frequency are designed to match the antiproton frequency of 29.65 MHz at a 1.945 T magnetic field. The main purpose of these detectors is efficient cooling of the modified cyclotron motion. This requires the maximization of the parallel resistance $R_p$, see equation (\ref{eq:DampingConstant}). In addition, the use a low equivalent input noise $e_n$ amplifier is required to reach low effective detector temperatures and a sufficiently high signal-to-noise ratio. This allows the application of negative electronic feedback with a high feedback factor to decrease the effective particle temperature below the thermal limit \cite{UrsoFeedback2003,Dehmelt1986,UrsoSEO2005}.\\
\begin{figure}[htb!]
  \centerline{
  	\includegraphics[width=0.5 \textwidth,keepaspectratio]{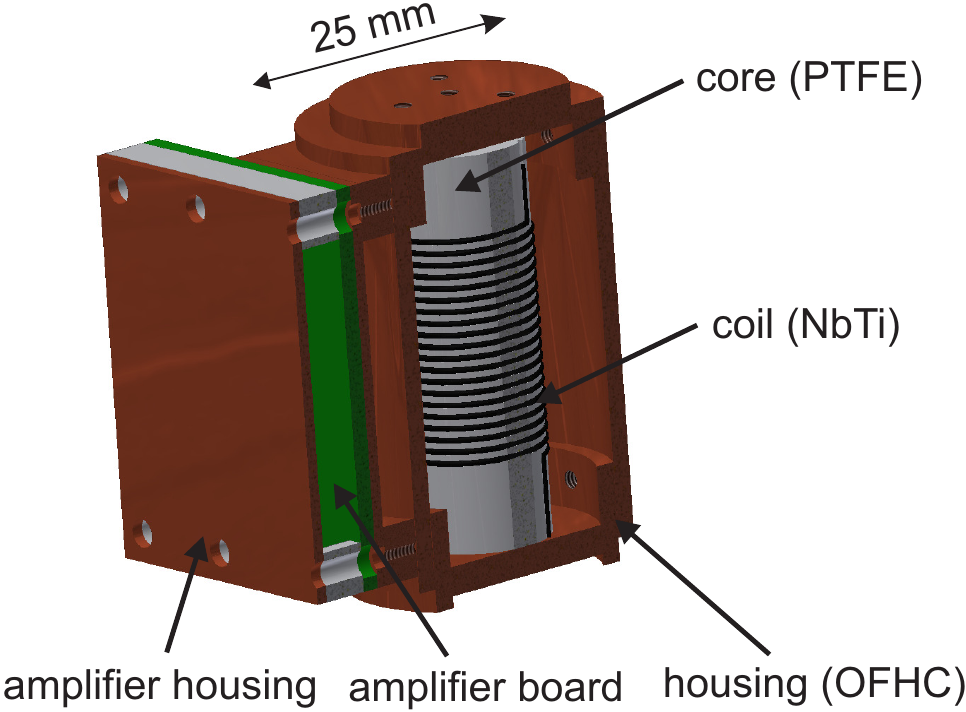}
	}
  \caption[Cyclotron Detector]{3D drawing of the cyclotron detector assembly. For details see text.}
	\label{fig:CycloDet}
\end{figure}
The design of the BASE cyclotron detection systems shown in Fig.~\ref{fig:CycloDet} is based on the general principles reported in \cite{MacAlpine1959} and the work reported in \cite{UlmerNIMA2012}.
Compared to the detector developed in \cite{UlmerNIMA2012}, superconducting NbTi solenoids instead of OFHC copper solenoids are used. The coil is wound on a PTFE core with a diameter of 11.5$\,$mm and a pitch of 1$\,$mm. Inductances are defined by the 14 pF parasitic trap capacitance, and are on the order of 1$\,\mu$H. The coil is pressed into a cylindrical OFHC housing with 23$\,$mm inner diameter and 34$\,$mm length. The unloaded $Q$ values of the solenoids are in the range of 9000 to 11000 at resonance frequencies of about 90$\,$MHz. Tuned to the trap frequency, $Q$ values up to 4500 are achieved. Compared to OFHC copper coils in the same geometry, the series resistances were reduced by almost a factor of 3.

The amplifiers for the cyclotron detection systems are based on dual-gate low-noise GaAs FETs \cite{UlmerNIMA2012}. They use the same concept as the axial detectors, a high-impedance FET in common-source configuration for the input stage and a second FET in common-drain configuration for impedance matching. The input stage FET is a NE25139 transistor with an effective input impedance of $R_{in}$ = 170 k$\mathrm{\Omega}$ at 40 MHz and an equivalent input noise of $e_\mathrm{n,4K}=0.83\,$nV/Hz$^{1/2}$. In addition, the amplifier hosts a MA46H072 varactor diode, which is connected in parallel to the resonator with a $3.6\,$pF capacitor. This allows adjusting the detector's resonance frequency by 650 kHz around 29.65 MHz to precisely match the particle's cyclotron frequency defined by the magnetic field.

Due to space constraints in the experimental setup, the two cyclotron detectors are stacked on top of each other, the CT detector being closer to the trap (see Fig.~\ref{fig:ESegment}). The signal wires to the trap and the amplifier are made from annealed OFHC copper wire, which has a resistance of 300 m$\mathrm{\Omega}$/m for a 30 MHz rf-signal at 4$\,$K. They contribute about 60 m$\mathrm{\Omega}$ series resistance and are a major limitation for the $Q$ value. 
When coupled to the trap and cooled to 4$\,$K, $Q$ values of 1500 are achieved with a 13$\,$dB $S/N$ ratio. A cooling time constant for a single antiproton of $\tau = 1/\gamma = m/R_p D^2/q^2$ of 10$\,$s is obtained in the CT. The cooling time constant is more than a factor of six smaller than in our experiment at Mainz and will thus significantly accelerate preparation of particles with single spin-flip resolution.

\subsection{Electrode voltage biasing}
Another essential component of the experiment are the highly-stable voltage sources and filter stages required for the DC biasing of the trap electrodes. They define the stability of the axial frequency via the noise amplitudes on the trap electrodes. We use commercial power supplies which were specifically developed to match the requirements of BASE (Stahl Electronics - UM1-14/bipolar). Each power supply has ten bipolar channels for biasing of the transport electrodes with 16-bit resolution, and six bipolar high-precision channels with 25-bit resolution. These channels are used to supply ring and correction electrodes, and have a voltage reproducibility of $<$10$^{-6}$. For dip-averaging times of 30$\,$s, the fractional voltage stability is 10$^{-7}$, which causes an additional frequency fluctuation of 30$\,$mHz. This causes much smaller fluctuations than the axial frequency shift induced by a spin-transition.

To significantly suppress noise on the electrodes induced by rf pickup and electromagnetic interference, we use four RC filter stages, one at 300$\,$K, one at 77$\,$K and two at 4$\,$K. The effective corner frequency of the filter assembly is at 15$\,$Hz with an attenuation of 40$\,$dB per decade. The time constant $RC$ of the filters is still sufficiently low to apply voltage ramps for adiabatic particle shuttling within a few 100 ms.

To filter the fast high-voltage lines connected to the two high-voltage electrodes, diode-bridged RC-filters are utilized. When a fast pulse is applied, the diodes open and transmit the pulse signal, in DC-mode the diodes are closed and the RC filter itself is active.

\begin{figure}[htb!]
  \centerline{
  	\includegraphics[width=8.5cm,keepaspectratio]{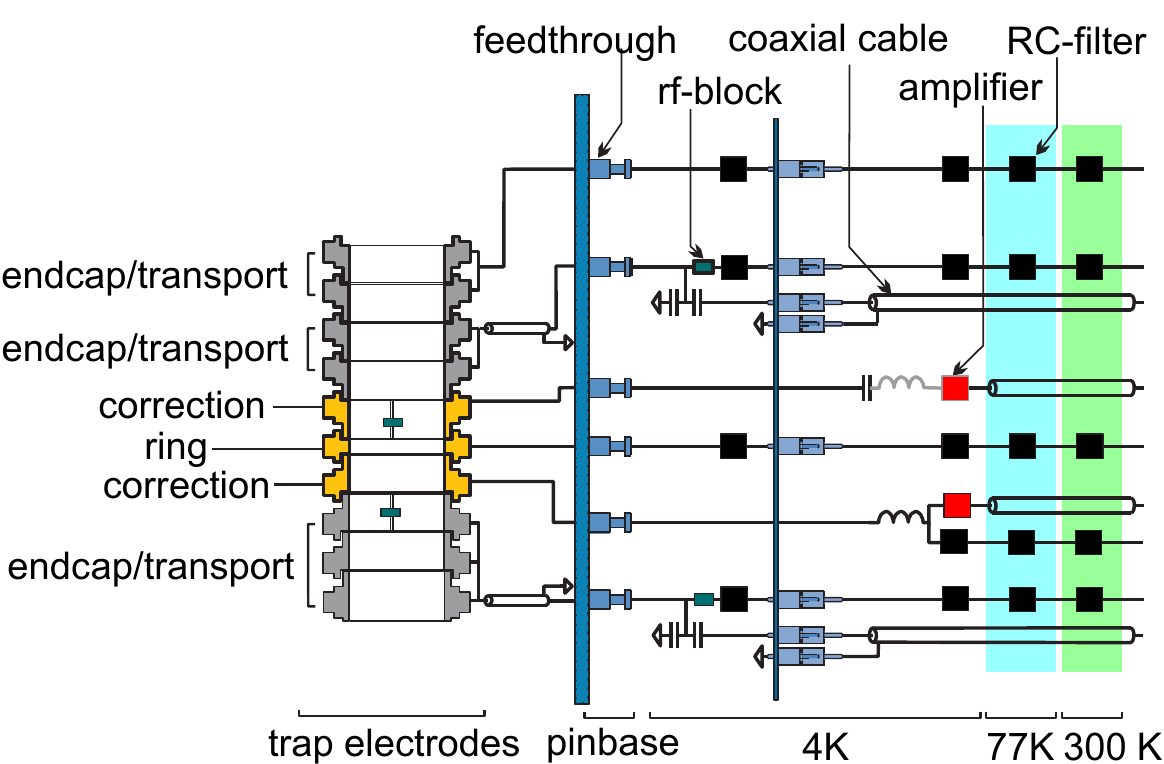}
	}
  \caption[Cyclotron Detector]{Simplified connection diagram of the BASE precision trap. During frequency measurements, adjacent transport and endcap electrodes are set to ground and act as a ``common endcap electrode''. For details see text. }
	\label{fig:Connection}
\end{figure}

A connection diagram of our precision trap is shown in Fig.~\ref{fig:Connection}. All electrodes are DC-biased by the filter stages described above. Coaxial lines for particle excitation are connected to the endcap electrodes by capacitive attenuators. High impedance rf-blocks protect the excitation signal from shorts to ground. The detectors are attached to the correction electrodes, the cyclotron detector to the radially-segmented upper, the axial to the lower one, respectively. The electronics layout of all the other traps is similar to the one shown in the Fig.~\ref{fig:Connection}.

\section{Preparation of single antiprotons}
\label{sec:4}

The BASE trap system has been commissioned in the 2014 antiproton run. Techniques to prepare cold single antiprotons  \cite{JerryElectronCooling,TrappedAntiprotons,JerryAntiprotonCatching} from the 5.3$\,$MeV AD pulses have been established. This includes catching, electron and resistive cooling, cleaning procedures, and single-particle preparation. Details are described in this chapter.

\subsection{Antiproton injection}

\begin{figure}[ht!]
  \centerline{
  	\includegraphics[width=0.75 \textwidth,keepaspectratio]{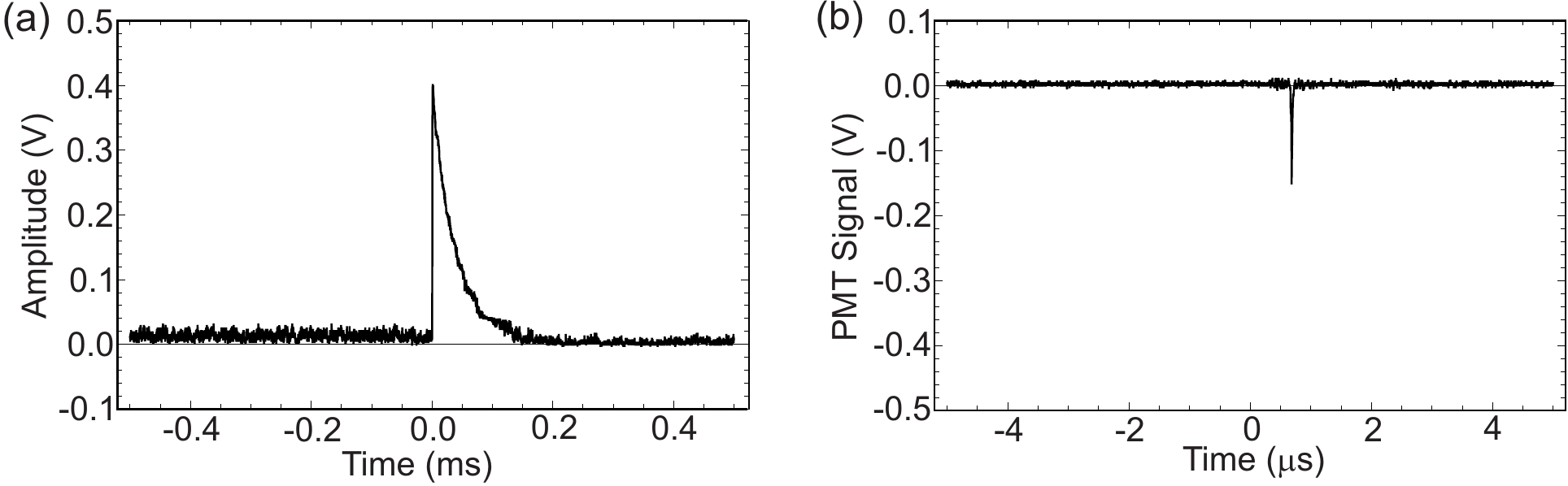}
	}
  \caption[Cyclotron Detector]{(a) Annihilation signal of antiprotons on the beam monitor showing a charge deposition of about $Q$ = 10$^6 e$. (b) After storing about 3000 antiprotons for several seconds in the reservoir trap, the antiprotons are extracted by a second high-voltage pulse and annihilate on the degrader.}
	\label{fig:BeamSteering}
\end{figure}

To inject antiprotons into the trap, the beam is steered to the center axis of the apparatus. The first quadrupole magnet upstream of the apparatus is used to tune the focal point to the degrader. To correct for displacements of the beam with respect to the trap axis, the corrector magnets and the signal strengths on the channels of the four-fold segmented cryogenic beam monitor are utilized. A typical signal from one of the beam-monitor channels is shown in Fig.~\ref{fig:BeamSteering}(a). The peak is due to charge deposition of about 10$^6$ antiprotons. A 9$\,$mm hole in the center of the beam-monitor allows the antiproton pulse to pass.
To confine the incoming antiprotons, the high-voltage (HV) electrodes are used. The static HV electrode is constantly biased with $-1\,$kV. After antiproton injection an adequately timed voltage pulse on the pulsed HV electrode to $-1\,$kV closes the catching trap. The injection timing is obtained precisely from a scintillation detector placed close to the apparatus. To test whether catching was successful, the antiprotons are extracted to the degrader foil. The annihilation signal is observed by triggering the scintillator on the extraction pulse as shown in Fig.~\ref{fig:BeamSteering}(b). To estimate the number of trapped antiprotons we calibrate the scintillation detector by the annihilation signal of a AD pulse with known particle number. Thereby, we estimate the number of confined antiprotons per AD-shot to be about 3000. Comparing this number to the expected efficiency from the degrader simulations indicates that the actual thickness of the degrader is within the desired range close to the optimum value (see Sect.~\ref{Sect:3:D}).

\subsection{Electron cooling}

\begin{figure}[ht!]
  \centerline{
		\includegraphics[width=0.65 \textwidth,keepaspectratio]{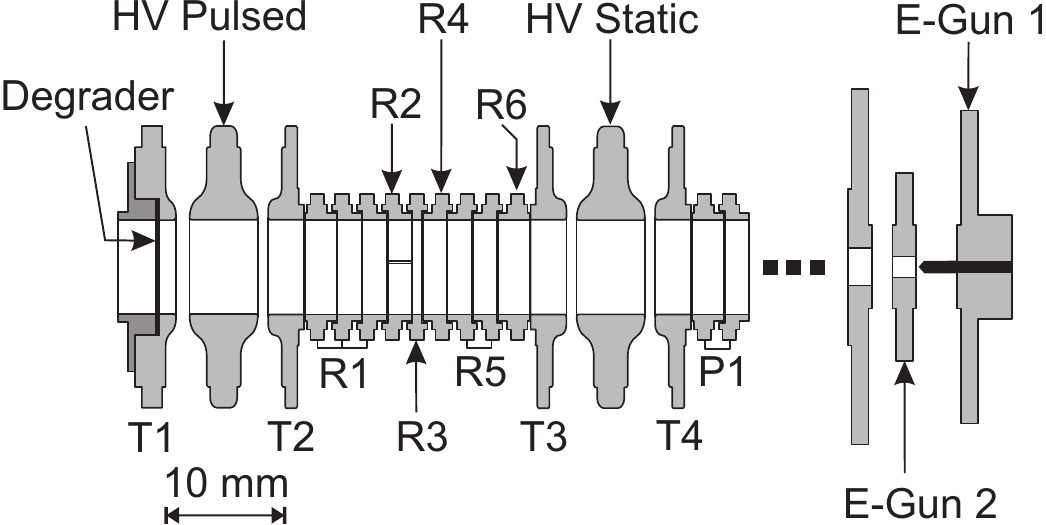}
	}
  \caption[Catching Trap Details]{Detailed sketch of the catching trap and the electron gun. For details see text.}
	\label{fig:CTDetail}
\end{figure}

The trapped antiprotons have kinetic energy up to 1$\,$keV and need to be cooled further. To this end, sympathetic cooling by interaction with cold electrons \cite{JerryElectronCooling} is used. In the strong magnetic field of the Penning trap, electrons are cooled via synchrotron radiation in the cyclotron mode. This process typically takes a few 100 ms and cryogenic particle temperatures are reached. We load electrons into our trap by utilizing our field emission electron gun (see Fig.~\ref{fig:CTDetail}). A 100$\,$nA electron current
is turned on for a few seconds, then the upstream and the downstream high-voltage electrodes are subsequently ramped to -1$\,$kV. After a 3$\,$s waiting time, the electrons thermalize and relax to the center of the trap which is at 14$\,$V. After this procedure, typically $10\,000$ electrons cooled to the environment temperature are prepared. In order to suppress the noise generated by our high-voltage switches, and thus, to avoid spurious heating of the trapped electrons, the high-voltage signals are guided to the trap by fast diode-bridged RC-filters. 

After injection of antiprotons into the cold electron cloud and a thermalization time of 10$\,$s typically several hundred antiprotons per AD pulse accumulate in the harmonic well of the RT. Inside the RT, the antiprotons are further cooled resistively by the detection system at 5.9 K. The number of prepared cold particles is about one order of magnitude less than the initial number of trapped particles. 

\subsection{Cleaning procedures}
\begin{figure}[ht!]
  \centerline{
  	\includegraphics[width=0.9 \textwidth,keepaspectratio]{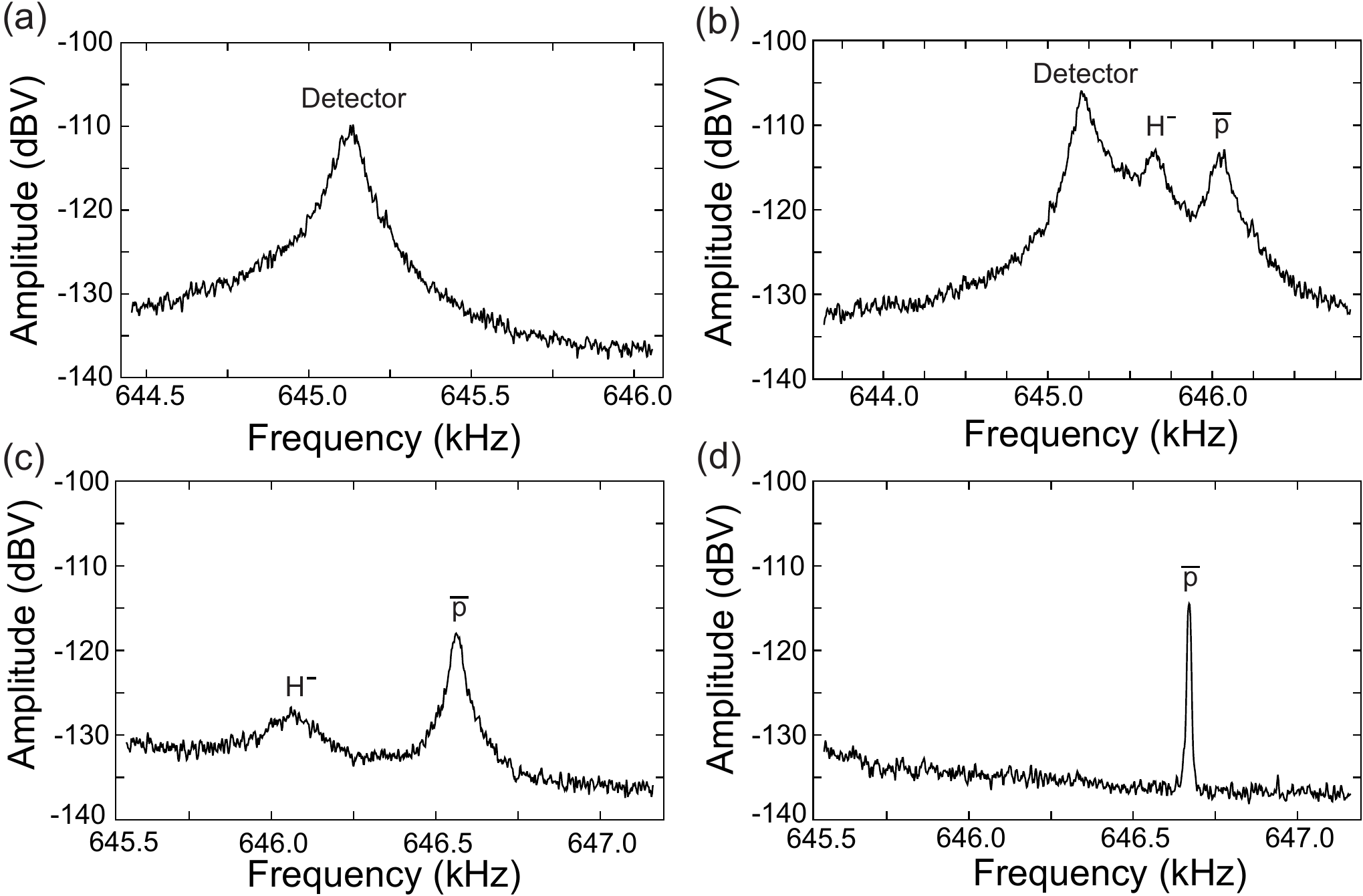}
	}
  \caption[Cyclotron Detector]{FFT spectra taken during the antiproton preparation procedure. (a) After antiproton injection, the electrons shield the low frequency signals of the antiproton motion and only the noise resonance of the detection system is observed.  (b) FFT-spectrum observed after applying the electron axial drive, electron kickout and sideband cooling of the magnetron mode. The ring voltage was detuned by 11.5 mV from the antiproton resonance voltage to observe the H$^-$ and antiproton peak signals 450 Hz and 800 Hz above the resonator frequency, respectively. (c) The FFT-spectrum before the cleaning dipolar drive at the H$^-$ modified cyclotron frequency with 18.5 mV detuned ring voltage exhibits two broad peaks around the axial frequencies of both particle species. (d) After applying the cleaning drive the H$^-$ modified cyclotron frequency, only an antiproton signal is observed which is shifted to a higher frequency and reduced in width.	For details see text.}
	\label{fig:Cleaning}
\end{figure}

After injection, the cloud of trapped particles is composed of electrons, antiprotons and contaminant negative ions. To eventually prepare a single particle, all contaminant particles are removed first. Some FFT spectra taken during the preparation procedure are shown in Fig.~\ref{fig:Cleaning}. The repertoire of cleaning procedures in the sequential order in which they are applied after antiproton injection is listed in the following:\\

{\textit{Electron axial drive}: A strong rf-drive applied to electrode \textit{R1} at the axial frequency of the electrons (28.7$\,$MHz) removes a large fraction of these particles. Subsequently, the magnetron motion of the antiprotons is cooled by a sideband drive at $\nu_z + \nu_-$, which centers them in the trap. }\\

\textit{Electron kickout}: Electrons on large magnetron orbits are not removed by the axial drive and remain in the trap. These particles are efficiently cleaned by opening the trap with a fast voltage pulse in the range of 250$\,$ns to 500$\,$ns duration. Due to their faster acceleration, electrons escape while the 1836-fold heavier antiprotons remain in the trap.
To this end, the cloud of trapped particles is adiabatically transported to electrode \textit{T2} next to the upstream high-voltage electrode.  After the transport, the trap is elevated and the electron ejection pulse is applied to the \textit{pulsed} electrode, as shown in Fig.~\ref{fig:CTDetail}. Electrons escape towards the positively biased degrader. This scheme can be repeated for several times without losing antiprotons during the kick-out pulses. After applying magnetron sideband cooling, signals of the remaining particles are observed. An FFT spectrum recorded after the electron kick-out is shown in Fig.~\ref{fig:Cleaning}(b).\\

\textit{Negative ion cleaning}: To remove contaminant negative ions, such as C$^-$ or O$^-$, a broad-band white noise excitation signal in a frequency band from 20$\,$kHz to 500$\,$kHz is injected to the trap. This covers the axial frequency span of the typically present contaminant ion's axial modes except H$^-$. The signal is applied for 30$\,$s and the trapping potential is lowered subsequently. Thus, the excited ions are released from the trap.\\

\textit{$H-$ cleaning}: After these procedures and subsequent centering of the remaining particles the signals shown in Fig.~\ref{fig:Cleaning}$\,$(c) are detected. At an axial frequency of 646$\,$kHz, the frequency difference between the two species is only 350$\,$Hz. To remove the remaining H$^-$ ions, a resonant dipolar drive at their modified cyclotron frequency $\nu_{+,\text{H}^-}$ is applied and the trap potential is lowered again to release the heated particles. The subsequently recorded FFT spectrum shown in Fig.~\ref{fig:Cleaning}(d) demonstrates that H$^-$ ions are efficiently removed.\\

\begin{figure}[htb]
        \centerline{\includegraphics[width=0.9 \textwidth,keepaspectratio]{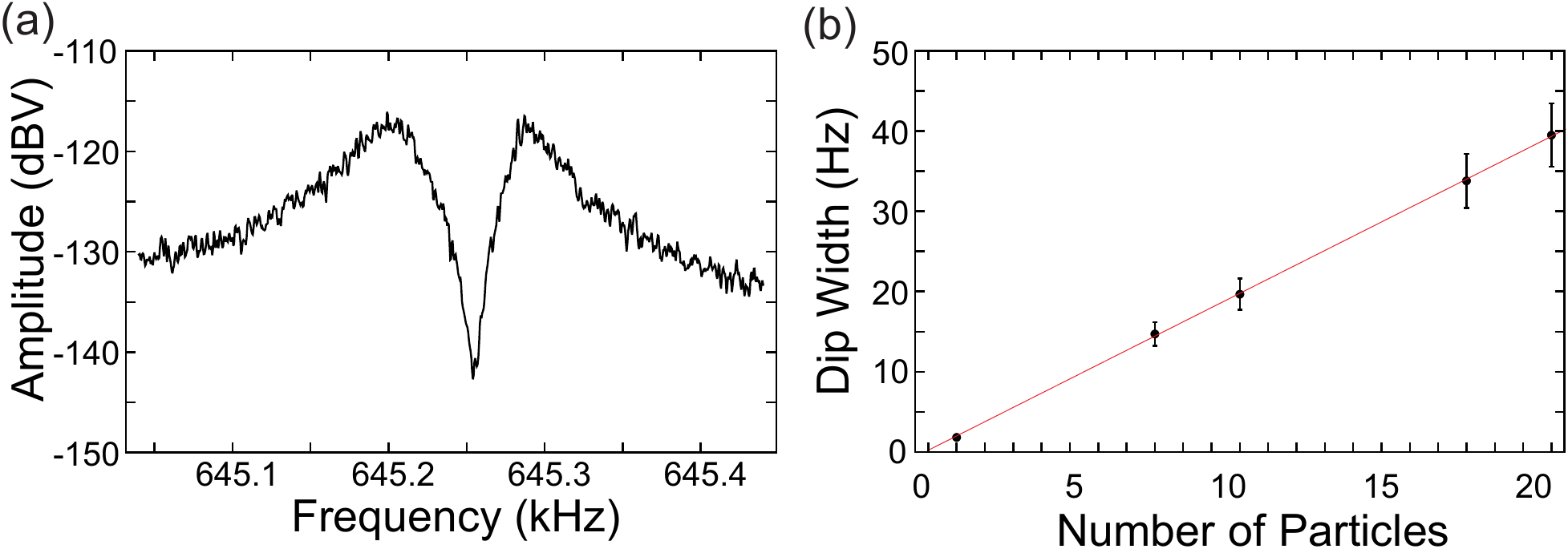}}
            \caption[PenningTrap]{ (a) FFT spectrum of 42$\pm$1 antiprotons in the RT. (b) The dip width scales linearly with the number of particles, which can be used to count the number of particles present in the trap. For details see text. }
						\label{fig:AxialDip}
\end{figure}

\textit{Particle reduction}: After removing all contaminant particles, the dip signal of the remaining antiprotons can be observed by tuning their axial frequency to the center frequency of the detector, see Fig.~\ref{fig:AxialDip}(a). To count the number of particles, the dip width is measured, which is proportional to the number of trapped antiprotons $N$, see equation (\ref{eq:DipWidth}).
To reduce the number of antiprotons to one, the trapping potential is lowered to voltages below 0.5$\,$V to let the hottest antiprotons escape from the trap. The trapping potential is lowered further in each step until eventually a single antiproton remains as shown in Fig.~\ref{fig:AxialDip}(b). After obtaining eventually a single antiproton, frequency measurements with single particles can be carried out.
Note that the particle reduction scheme described above is an established method to prepare single particles in similar precision experiments \cite{CCRodegheri2012,Jack2012Proton,Jack2013Antiproton,Hartmut,Verdu,SvenSi28}. However, it is not suited for our concept of the reservoir trap, as only one particle from the cloud can be extracted. To overcome this, we have recently also developed a non-destructive single-particle extraction scheme to make a more efficient use of the antiproton cloud. The details can be found in reference \cite{SmorraIJMS2015}.\\

\begin{figure}[htb]
        \centerline{\includegraphics[width=0.9 \textwidth,keepaspectratio]{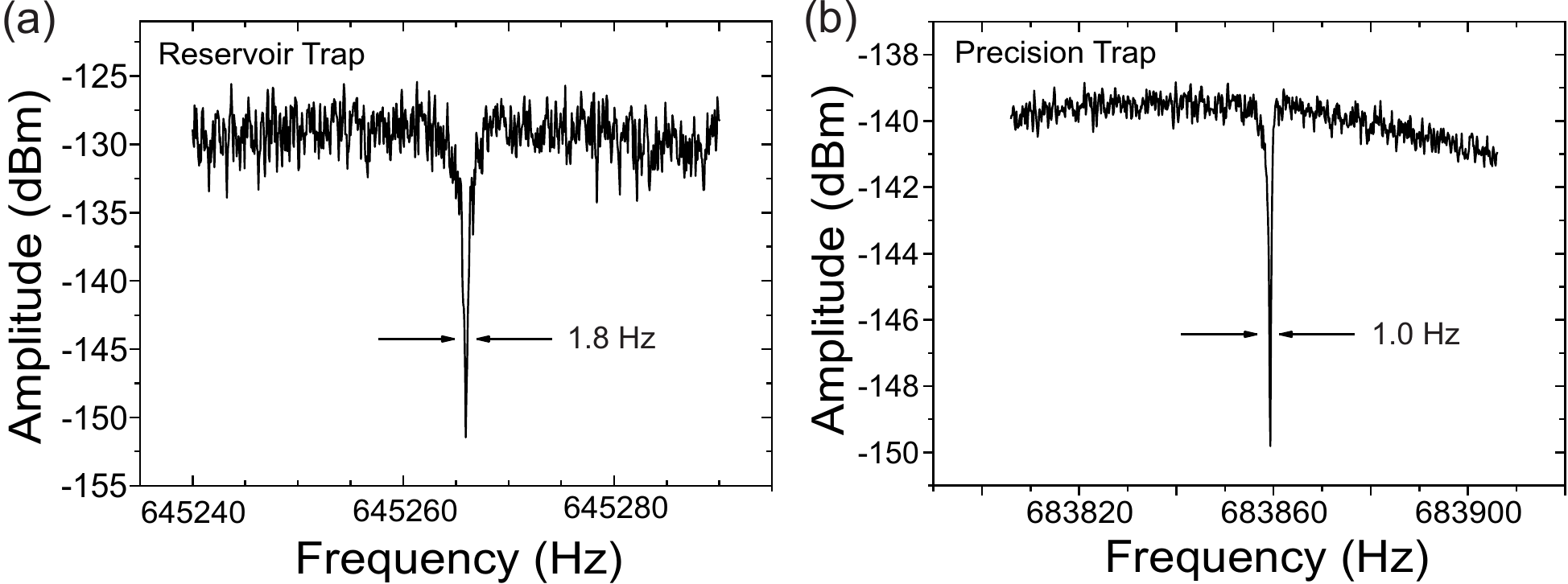}}
            \caption[PenningTrap]{ (a) A single particle in the RT on resonance of the axial detection system at 645.2 kHz with 1.8 Hz dip width is shown after the preparation and cleaning procedures. The FFT spectrum was averaged for 42 s. (b) A particle transported from the RT to the PT is detected with the precision trap detection system at 683.8 kHz with a dip width of 1.0 Hz. A FFT spectrum with 100 s averaging time is shown. }
						\label{fig:DipRTPT}
\end{figure}

\textit{Particle transport}: Having prepared a single antiproton in the RT, we transport it along the trap axis to any other trap by a sequence of slow voltage ramps on the electrodes in between the ring electrodes of the two traps. At the beginning of each voltage ramp, two adjacent electrodes are at 13.5 V and confine the antiproton in axial direction while the other electrodes are on ground potential. Then, the potential of the next electrode in transport direction is ramped to 13.5 V while the the electrode of the potential well in the opposing direction is ramped to ground. Using volatage ramps of 1.5 s duration, the center of the axially-confining potential well can be moved along the trap axis without significant heating of the particle during the transport. Thus, particles can be transported by adiabatic shuttling into the other traps as well, where they can be detected using the detection system of the respective trap. Fig.~\ref{fig:DipRTPT} (a) and Fig.~\ref{fig:DipRTPT} (b) show the comparison of a particle detected with the axial detection system in the reservoir trap at 645.2 kHz and in the precision trap at 683.8 kHz, respectively.

\section{Frequency measurements with single antiprotons}
\label{sec:5}

\subsection{Trap optimization}

\begin{figure}[h!]
\centerline{\includegraphics[width=0.95 \textwidth,keepaspectratio]{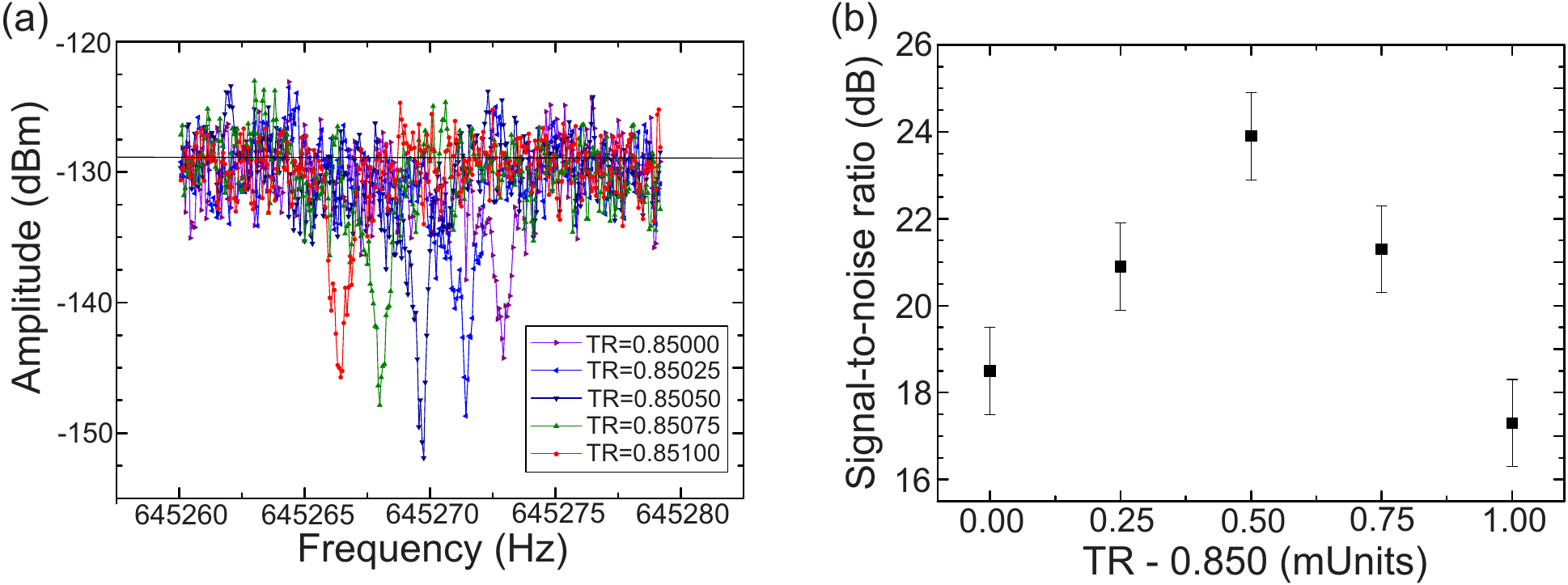}}
\caption{(a) Axial frequency dips measured for different tuning ratios. (b) Dip signal-to-noise ratio for different tuning ratios. } \label{Fig:TrapOptDip}
\end{figure}

The dip detection as discussed in Sect.~\ref{sec:EP:ICD} is used in most measurements of the antiproton's axial frequency $\nu_z$. When tuned to resonance with the detection system, the particle's axial energy $E_z$ performs a random walk within the one-dimensional Boltzmann energy distribution with the temperature $T_z$ of the axial detector as a parameter. The single-particle line-shape is thus a convolution of the particle's unperturbed resonance line and the thermal Boltzmann distribution. In presence of an octupolar trap anharmonicity $C_4$ the axial frequency becomes
\begin{eqnarray}
\nu_z(E_z)=\nu_{z,0}\left(1+\frac{3}{4}\frac{C_4}{C_2^2}\frac{E_z}{qV_R}\right).
\end{eqnarray}
Thus, the axial frequency $\nu_z$ changes with the continuous change of axial energy as well, the thermalization time-scales are given by the axial cooling time constant $\tau_z\approx 50\,$ms. This is about a factor of 1000 smaller than the averaging time typically used for dip detection. Thus, in presence of the $C_4$ anharmonicity the signal-to-noise ratio of the single particle is reduced, which is shown in Fig.~\ref{Fig:TrapOptDip}. The $C_4$ coefficient can be tuned by recording dip spectra for different tuning ratios $\text{TR}=V_\text{CE}/V_\text{R}$, where $V_\text{CE}$ is the voltage applied to the correction electrodes of the trap. Thus, by reducing the $C_4$ coefficient the signal-to-noise ratio of the single particle dip is increased. This allows for an optimization of the tuning ratio to a level of 10$^{-4}$. In $g$-factor measurements performed in the BASE trap the corresponding residual $C_4$ would contribute systematic magnetic moment shifts on the level of 0.1$\,$ppb.

\begin{figure}[h!]
\centerline{\includegraphics[width=0.95 \textwidth,keepaspectratio]{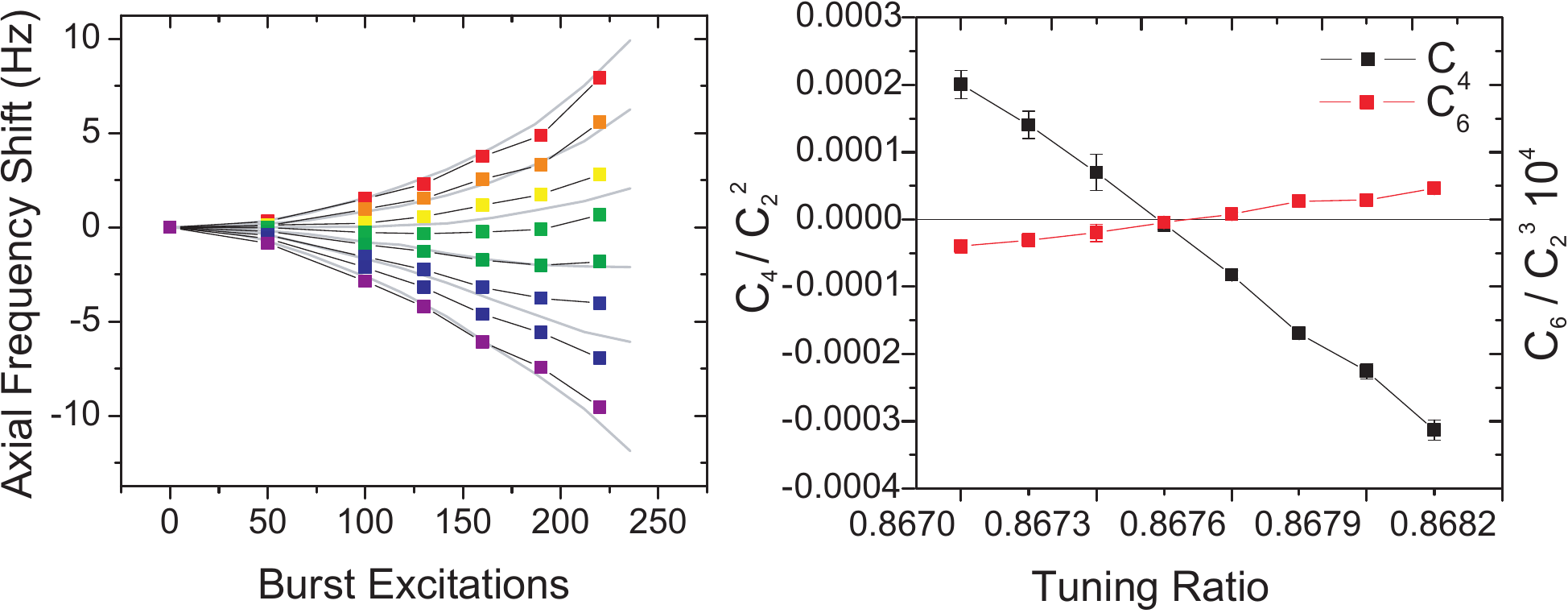}}
\caption{(a) Shift of the axial frequency as a function of resonant burst cycles $N_c$ applied to excite the magnetron mode for different tuning ratios in steps of 1.5$\times$10$^{-4}$ between two adjacent curves. (b) The trap leading-order trap anharmonicity coefficients $C_4$ and $C_6$ obtained from the axial frequency shifts are shown as function of the tuning ratio. For details see text. } \label{Fig:TrapOpt}
\end{figure}

However, the trapping potential can be optimized even further by measuring axial frequency shifts as a function of radial energy and for different tuning ratios. For this purpose, the magnetron motion is excited with a resonant burst drive of $N_c$ cycles in between two axial frequency measurements. Results of such measurements are shown in Fig.~\ref{Fig:TrapOpt}(a). By fitting the data with polynomials in $N_c$, the coefficients $C_4$ and $C_6$ can be extracted for each experimentally applied TR. The extracted coefficients as a function of the tuning ratio are shown in Fig.~\ref{Fig:TrapOpt}(b). In an ideal compensated trap, both anharmonicity coefficients are set to zero $C_4(\text{TR}_\text{opt})=C_6(\text{TR}_\text{opt})=0$ at the same ideal tuning ratio $\text{TR}_\text{opt}$. For the data shown in Fig.~\ref{Fig:TrapOpt}(b), the zero points of $C_4(\text{TR})$ and $C_6(\text{TR})$ are separated by $\Delta\text{TR}=5\cdot10^{-5}$. The deviation is due to the machining precision of the trap electrodes.

To obtain the least systematic shifts in the experiment without exact compensation, we set the tuning ratio to the zero-point of $C_4(\text{TR})$, which can be determined 2$\,$ppm precision. In this case, the uncertainty in $C_4$ will contribute a systematic $g$-factor uncertainty at the level of a few ppt only and the residual hexa-decapolar contribution leads only to systematic shifts on the sub-ppt level if the measurement is carried out at low amplitudes in thermal equilibrium with the detector. If necessary, exact local compensation can be achieved by deliberately superimposing constant electric fields to the trap to further decrease the systematic shifts.

By comparing potential calculations with the measured results, an absolute energy calibration of the magnetron mode can be performed. Using sideband $\pi$-pulses to convert $E_-$ into $E_{+}$ and $E_{z}$, the energy calibration can be extended to the modified cyclotron and axial mode. Thereby, the magnitude of the energy dependent systematic shifts of the measured frequency ratios in all eigenmodes can be determined.

\subsection{Axial frequency measurements}

Precise measurements of the axial frequency $\nu_z$ are required to determine the spin-state of the particle (see Sect.~\ref{Sect:2.2}) and to measure the cyclotron frequency $\nu_c$ by the sideband coupling method (see Sect.~\ref{Sect:2.1}). The measurement precision of $\nu_z$ is determined by two components, the statistical measurement uncertainty $\sigma_{z}$ which is equivalent to a white-noise component and a random-walk noise component. The latter is due to drifts of the trapping voltage or through fluctuations in the radial energy $E_{\pm}$ of the particle that are coupled due to the presence of a magnetic bottle $B_2$ to the axial mode. The dependence of $\sigma_{z}$ on the measurement parameters was investigated through numerical simulations. It is given as:
\begin{eqnarray}
\sigma_z = \alpha_\text{FFT}\sqrt{\frac{1}{2\pi}\frac{\Delta\nu_z}{T\sqrt{\text{SNR}}}}\, ,
\end{eqnarray}
where $\alpha_\text{FFT}$ is a parameter depending on the FFT overlap and windowing functions, $\Delta\nu_z$ the width of the particle dip, $T$ the averaging time and SNR the signal-to-noise ratio of the dip. The white noise component decreases with increasing averaging time as $T^{-1/2}$, whereas the contribution of random-walk noise increases proportional to $T$.

\begin{figure}[htb]
        \centerline{\includegraphics[width=0.75 \textwidth,keepaspectratio]{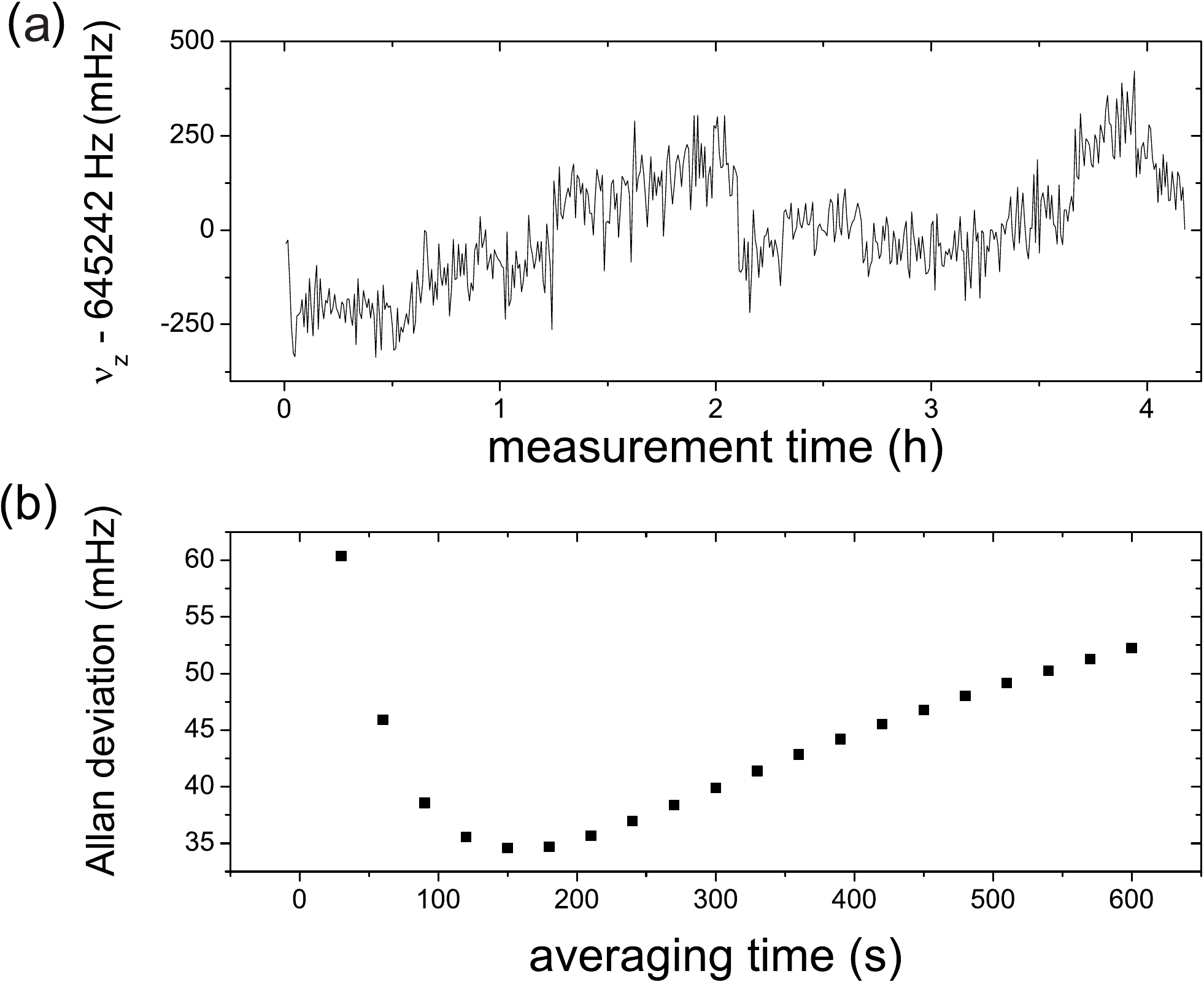}}
            \caption[PenningTrap]{(a) Evolution of the axial frequency of a single antiproton in the reservoir trap during a measurement series of 4 hours. (b) Allan deviation of the data. }
						\label{fig:AllanDev}
\end{figure}

To choose the optimum averaging time $T$ for measurements of $\nu_z$ in the reservoir trap, transient signals of the single particle dip were recorded to determine the Allan deviation as a function of $T$. The absolute values of $\nu_z$ are shown in Fig.~\ref{fig:AllanDev} (a). Within a total measurement time of 4$\,$h the absolute frequency is stable to $\pm250\,$mHz. Fig.~\ref{fig:AllanDev} (b) shows the Allan deviation of the data. For small averaging times, the white-noise component dominates the Allan deviation with the $T^{-1/2}$ scaling. After $T>150\,$s the random-walk component leads to an increase of the Allan deviation. The highest achievable precision for a single $\nu_z$ measurement in the RT is 35 mHz using 150 s averaging time. 

In the RT, a small axial frequency fluctuation compared to the frequency jump of 230 mHz caused by a spin flip can be achieved. Even though, the actual axial frequency fluctuation in the AT which determines the spin-flip detection fidelity has to be determined from a similar measurement in the magnetic bottle, as the $B_2$ coefficient in the AT is more than a factor of 10$^5$ larger. Spurious noise on the trap electrodes heating the modified cyclotron mode defines the magnitude of the random-walk noise component in the AT (see Sect.~\ref{Sect:2.2.2}). We conclude from the measurement in the RT that the stability of the power supplies is sufficiently good to observe single spin transitions, but the heating rate of the cyclotron mode in the AT remains to be determined.

\subsection{Sideband frequency measurements}
\label{sec:FM:SB}
\begin{figure}[htb]
        \centerline{\includegraphics[width=0.95 \textwidth,keepaspectratio]{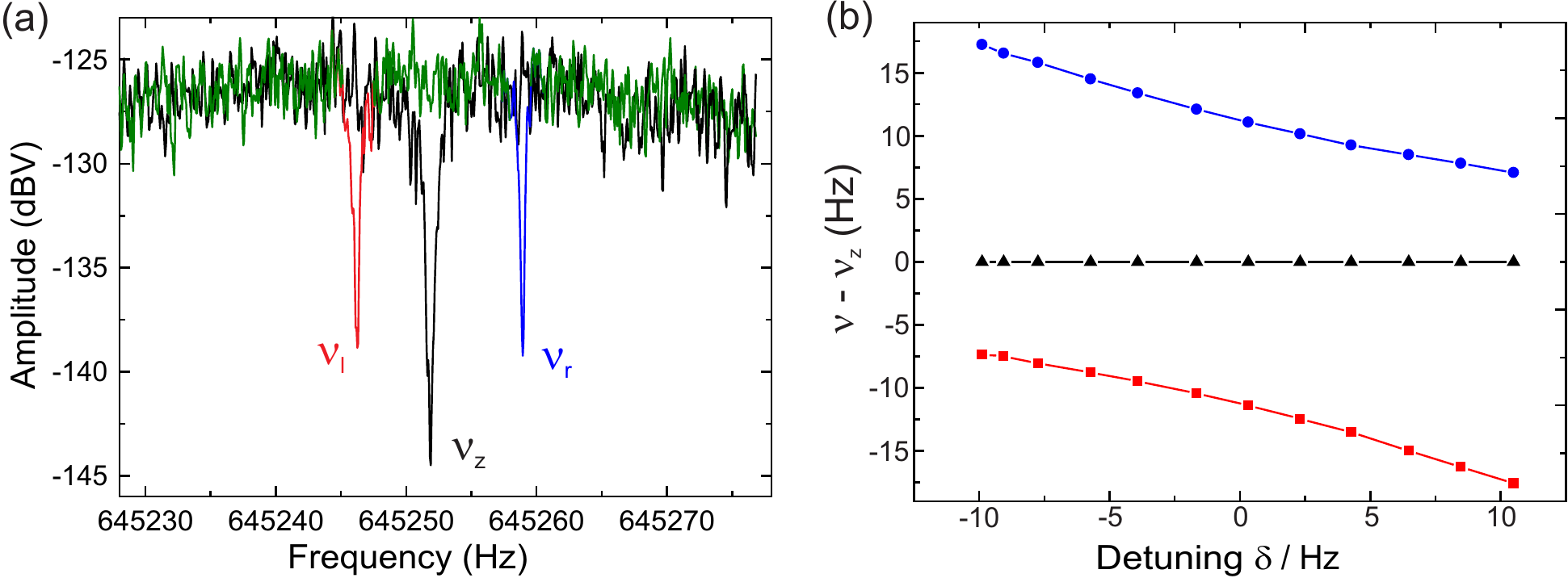}}
            \caption[PenningTrap]{(a) Sideband frequencies from the coupled modified cyclotron and axial motion of a single antiproton in the reservoir trap. The signal of the uncoupled axial motion is shown in black, the signal observed while irradiating a near resonant drive $\delta\approx0$ is shown in green while the sideband signals in this FFT-spectrum of $\nu_l$ and $\nu_r$ are marked in red and blue, respectively. (b) The  sideband frequencies as function of the detuning $\delta$. The upper sideband is shown in blue, the lower sideband in red, and axial frequency in black. Measurement uncertainties are on the order of 10 mHz and are omitted here. }
						\label{fig:RTDoubleDip}
\end{figure}

The sideband coupling method described in Sect.~\ref{Sect:2.1} is used to determine the modified cyclotron or magnetron frequency. In principle, measuring both radial frequencies is necessary to derive the free cyclotron frequency $\nu_c$ via the invariance theorem \cite{Brown}. However, as the uncertainty in the magnetron frequency measurement is suppressed by a factor $\nu_-/\nu_c$, it is sufficient to approximate the magnetron frequency using $\nu_- \approx \nu_z^2 / 2 \nu_+$ to determine $\nu_c$. The error by this approximation is a few 10~ppt for absolute frequency measurements and less than 0.1 ppt for frequency ratios. Thus, we determine $\nu_c$ in alternating measurements of $\nu_z$ and $\nu_{+}$, the resulting Fourier spectra are shown in Fig.~\ref{fig:RTDoubleDip}(a). The axial frequency is directly measured, whereas $\nu_+$ is determined from the sideband frequencies using a near-resonant coupling drive at $\nu_{rf}$=29\,011\,363.7$\,$Hz and the relation $\nu_{+}=\nu_{rf}+\nu_l+\nu_r-\nu_z$, see equation (\ref{eq:DoubleDipModCyclotron}). From the FFT-spectra shown in Fig.~\ref{fig:RTDoubleDip}(a), we obtain the axial frequency as $\nu_z=645\,251.849(14)\,$Hz and the sideband frequencies as $\nu_l=645\,246.199(13)\,$ Hz and $\nu_r=645\,258.198(10)\,$Hz, resulting in $\nu_+ = 29\,656\,616.248(22)\,$Hz and $\nu_-=7\,019.512\,1(8)\,$Hz. By using the invariance theorem the measurement in Fig.~\ref{fig:RTDoubleDip}(a) results in $\nu_c = 29\,663\,635.760(22)\,$Hz, which has a statistical uncertainty of 0.7 ppb. Note that $\nu_c$ can be determined for an arbitrary detuning $\delta$ of the coupling drive but one has to consider that the sideband frequencies change according to equation (\ref{eq:SideBandFrequencies}), which is shown in the 'classical avoided crossing' in Fig.~\ref{fig:RTDoubleDip}(b).

\subsection{Magnetic field stability}
\label{sec:FM:MF}
The magnetic field $B$ defines $\nu_c$ and $\nu_L$, thus a careful characterization of the temporal stability of $B$ is of great importance. To analyze the stability of the magnet the cyclotron frequency of a single trapped antiproton is continuously measured and the Allan deviation is computed.
We fit an empirical model to the measured data which consists of
\begin{itemize}
  \item a linear drift,
  \item an oscillatory term,
  \item a random walk,
  \item and a white noise contribution.
\end{itemize}

\begin{figure}[htb]
  \centerline{\includegraphics[width=0.65 \textwidth,keepaspectratio]{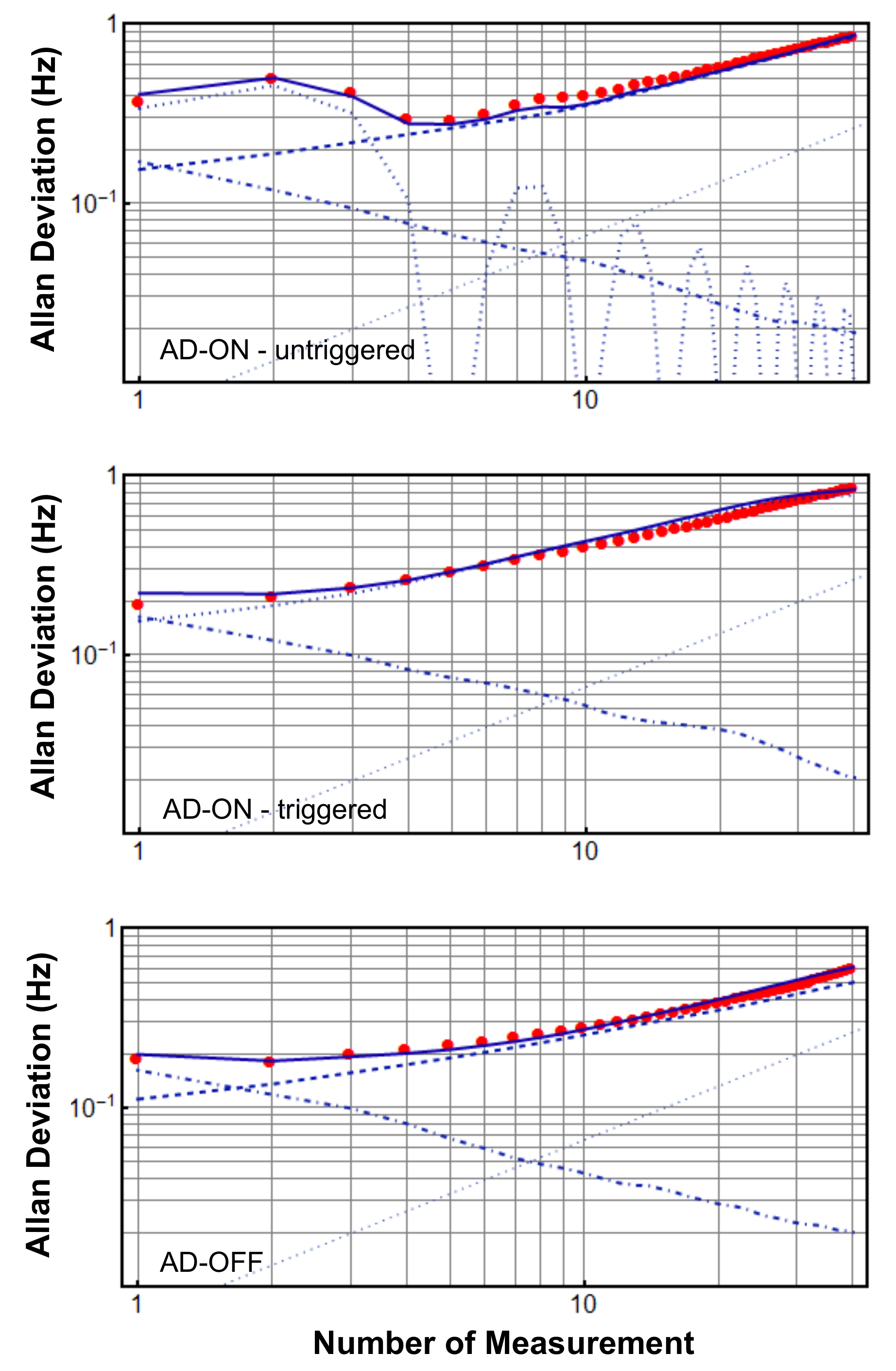}}
        \caption[Zone]{Allan deviation of subsequent $\nu_+$ measurements of a single antiproton for different measurement conditions. The contributions of the white noise, the oscillation term, and the random walk noise contributions are shown as dash-doted, doted and dashed lines, respectively. (a) The Allan deviation of a measurement not synchronized to the AD deceleration cycle is shown. The beating between the AD and measurement cycle gave raise to an oscillation term. (b) Results of a synchronized measurement series during AD operation series where the oscillation term vanished. (c) Results of a measurement series in a AD shut-down period. The random walk noise component is significantly lower compared to the measurement in (b) due to less external magnetic field fluctuations. For more details see text.}
\label{fig:MagnetStability}
\end{figure}

\begin{figure}[htb]
        \centerline{\includegraphics[width=0.65 \textwidth,keepaspectratio]{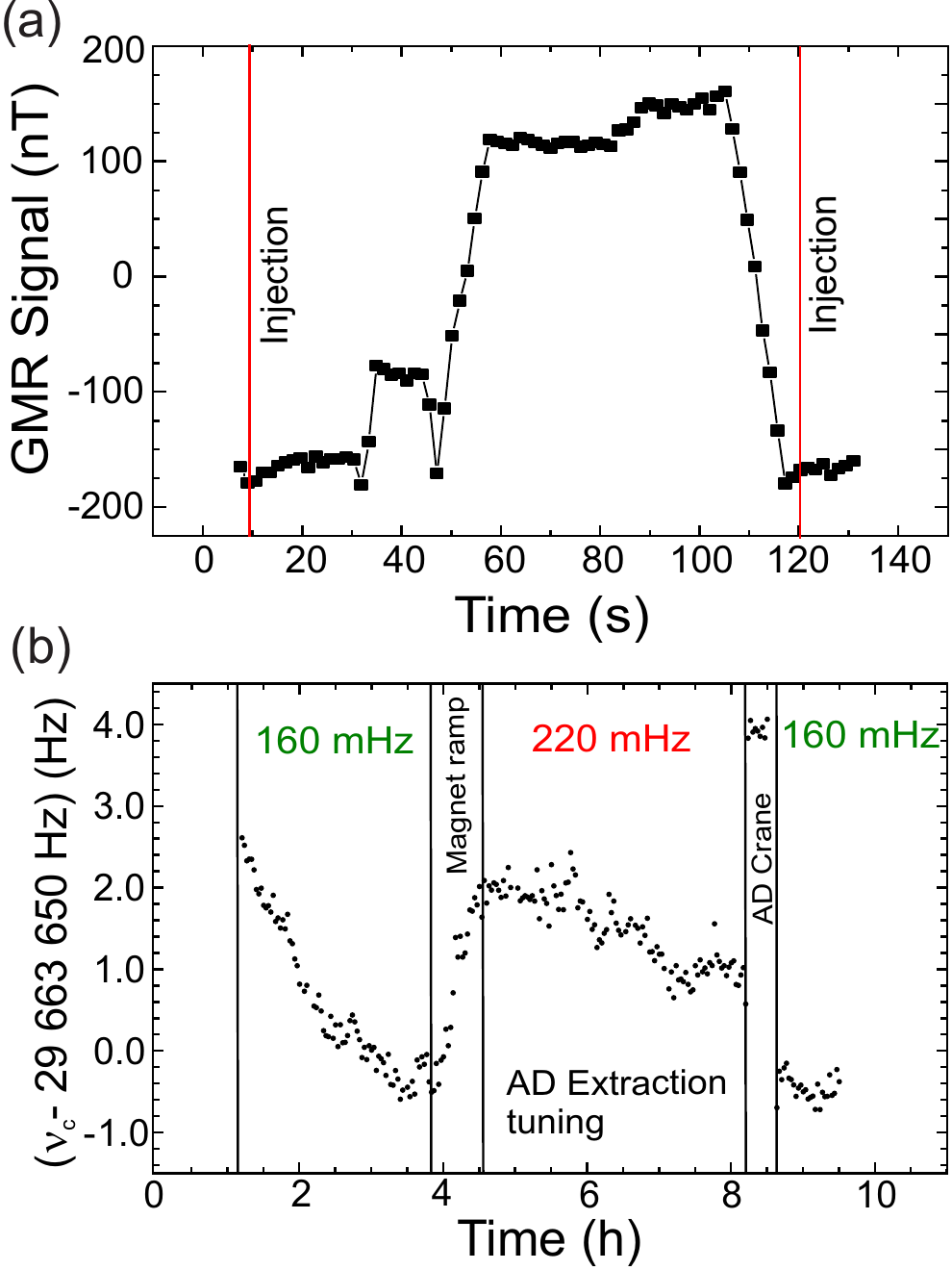}}
            \caption[PenningTrap]{(a) GMR sensor measurements of the background magnetic field fluctuations during the AD cycle. (b) Successive cyclotron frequency measurements of a single antiproton for a 10 hour period showing frequency jumps and drifts due to external perturbations. The average fluctuation of the cyclotron frequency $\Xi_c$ for the different periods are given. }
						\label{fig:ADBeating}
\end{figure}

Results of different series of cyclotron frequency measurements with a duration of 120$\,$s each are shown in Fig$.\,$\ref{fig:MagnetStability}. The red circles represent the measured data points, the blue lines are the individual modelled contributions. In the measurement shown in Fig.~\ref{fig:MagnetStability}(a), our cyclotron frequency measurements are not synchronized to the AD deceleration-cycle. This gives raise to an oscillation term with a 21$\,$min period due to the beating between our measurement sequence and the magnetic field ramps of the AD deceleration cycle. The beat amplitude of 590$\,$mHz corresponds to a magnetic field amplitude of 20$\,$nT, which is consistent with the field amplitude observed outside the magnet using GMR sensors and the magnet's shielding factor of 10, see Fig.~\ref{fig:ADBeating}. The measurement shown in Fig.~\ref{fig:MagnetStability}(b) is triggered to the PS-to-AD antiproton injection trigger, and the duration of one cyclotron frequency determination has been matched to exactly one AD deceleration cycle. Here, the oscillatory contribution vanishes, however a random-walk contribution with standard deviation $\sigma_R=220(20)\,$mHz as well as a Gaussian white-noise component of $\sigma_w=160(15)\,$mHz contribute to the cyclotron frequency fluctuation. For comparison, the same measurement was performed in a shut-down period of the AD, which is shown in Fig.~\ref{fig:MagnetStability}(c). The random-walk component was reduced to $\sigma_R=160(15)\,$mHz, while the white-noise component is not affected within error bars. In all three cases the linear drift component is -5(2)$\,$ppb/h.

Results of a series of cyclotron frequency measurements carried out during AD operation and activities of other experiments in the accelerator hall are shown in Fig.~\ref{fig:ADBeating}(b). Here, magnetic field ramps due to charging and discharging of superconducting magnets and operation of the overhead crane change the cyclotron frequency by several Hz. An array of giant magneto resistance, hall and flux gate magnetic field sensors, which is installed in the experiment zone, is used to verify external magnetic field drifts and jumps. Frequency measurements carried out during such periods are disregarded in any frequency analysis.

\subsection{Cyclotron frequency ratio measurements of two particles}
\label{sec:FM:APE}
Measurements of cyclotron frequency ratio of two charged particles is a general approach used in Penning trap mass spectrometry to compare the charge-to-mass ratios of two particles \cite{Blaum2006}. A variety of excellent cyclotron frequency measurement techniques and schemes for precise mass determination exists \cite{JerryAntiproton,TOFICR,Cornell1989PnPMethod,SturmPhase,PIICR,UWPTMS,Myers2015}. 

In BASE we implemented a loss-less measurement scheme allowing fast comparisons of cyclotron frequency ratios with the aim to compare the proton and antiproton charge-to-mass ratio
. To this end, we first prepare one particle (particle 1) in the HV-pulsed electrode and the second one (particle 2) in the center of the reservoir trap. Subsequently, $\nu_{c,2}$ is measured by the sideband method as described in Sect.~\ref{sec:FM:SB}, which requires typically 100$\,$s. Afterwards, we shuttle the particle 1 to the trap and simultaneously transport particle 2 to the HV-static electrode. This particle exchange takes 15$\,$s. In a next step the cyclotron frequency $\nu_{c,1}$ of particle number 1 is measured. The total duration of the transport and frequency measurements is matched to one AD cycle to avoid the beating effects discussed in Sect.~\ref{sec:FM:MF}. Thus, the entire scheme to measure one single charge-to-mass ratio takes a total measurement time of about 240$\,$s. This corresponds, compared to earlier measurements comparing the antiproton and proton charge-to-mass ratio \cite{JerryAntiproton}, to a 50-fold improved sampling rate. 

\begin{figure}[htb]
        \centerline{\includegraphics[width=0.95 \textwidth,keepaspectratio]{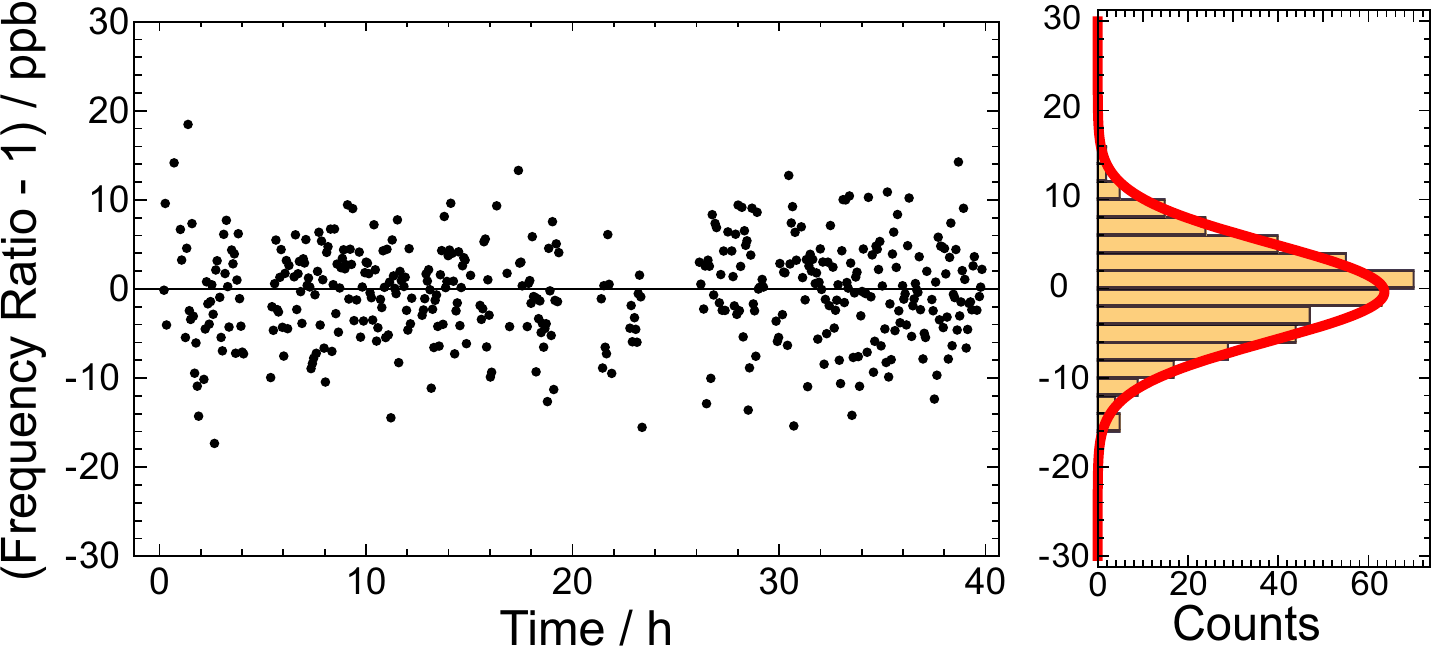}}
            \caption[PenningTrap]{Cyclotron frequency ratio of two particles obtained with the adiabatic shuttling method. For details see text. In total, 433 data points are shown. Data points that were disregarded due to external magnetic field changes are omitted in the plot. The histogram shows a Gaussian distribution with 5.4 ppb standard deviation. The mean frequency ratio obtained is $\nu_{c,1}/\nu_{c,2}-1$ = $55(253)\,$ppt. }
						\label{fig:PbarPbar}
\end{figure}

This scheme was first realized using two antiprotons in order to characterize the measurement performance of charge-to-mass ratio measurements in our apparatus. Afterwards, we applied it using an antiproton and an H$^-$ ion making the so far most precise comparison of the antiproton and proton charge-to-mass ratios with a relative uncertainty of only 69 ppt \cite{UlmerNature2015}. 
One result of a measurement sequence of cyclotron frequency ratios with the adiabatic particle exchange method is shown in Fig.~\ref{fig:PbarPbar}. The scatter of the measured cyclotron frequency ratios has a width of 5.4$\,$ppb caused by white noise and the random-walk noise in the magnetic field. With this magnetic field stability, the average cyclotron frequency ratio of the two particles has been determined in a measurement time of 41 hours to $270\,$ppt. Data that was collected during external magnetic field ramps and crane operation periods has been excluded from the evaluation. The entire sequence was recorded using only two single particles. 

From the above measurement we can conclude on the impact of the magnetic field fluctuations on a $g$-factor measurement. In this measurement sequence, the measurement of $\nu_{c,2}$ is replaced by a measurement of $\nu_L$, i.e. the irradiation of the spin-flip drive \cite{MooserNature2014}, and therefore the magnetic field fluctuations cause a broadening of the $g$-factor resonance. Compared to the line-width contribution expected from the residual $B_2$ inhomogeneity in the precision trap, the contribution from magnetic field fluctuations is dominating, however a factor of 2.5 lower than the total line width in our proton $g$-factor measurement \cite{MooserNature2014}. Thus, this demonstrates that the magnetic field stability of the BASE apparatus is sufficient to perform an antiproton magnetic moment measurement with a fractional precision on the ppb level.

\section{Measurement perspectives}
\label{sec:6}

\subsection{Proton and antiproton magnetic moments}
The highest priority of BASE is the high-precision measurement of the antiproton magnetic moment $\mu_{\overline{p}}$. The best value for the proton magnetic moment $\mu_{p}$ with 3.3 ppb uncertainty, obtained with our proton-double trap system in Mainz, sets the first precision mark for the measurement of $\mu_{\overline{p}}$. Further improvements of this CPT invariance test beyond this level require measurements of both values $\mu_{p}$ and $\mu_{\overline{p}}$ with higher precision.

\subsubsection{Measurement with the state-of-art double trap method}
\label{sec:NP:DT}
Improvements on the major uncertainty limitations of the proton-double trap system \cite{MooserNature2014} have been included in BASE. The increased distance between the precision and analysis trap reduces the residual magnetic bottle in the PT by a factor of six to $B_2 = 0.67$ T/m$^2$ and thus the line-width parameter in equation (\ref{eq:LineWidth}) to 150 ppt/K$\times T_z$. Using axial feedback cooling to reduce the particle amplitude to an effective temperature of 2.5 K, a line width with a fraction of $\sigma_{B2}$ = 390 ppt of the Larmor frequency can be reached. This contribution is not significant compared to the width caused by non-linear magnetic field fluctuations of $\sigma_{B}$ = 5.4 ppb discussed in Sect.~\ref{sec:FM:MF}. Compared to the 12.5 ppb line-width of the $g$-factor resonance in \cite{MooserNature2014}, the line-width is about a factor of 2.5 narrower, so that a measurement of $\mu_{\overline{p}}$ with the established techniques can reach a sub-ppb precision, as the center of the $g$-factor resonance can be determined to a fraction on the 1 $\%$ level of the line width depending on the measurement statistics. 

To further increase the precision, $\sigma_{B}$ needs to be reduced. One way to address this issue is to improve the magnetic field stability by the installation of a self-shielding coil around the Penning-trap chamber and a flux-gate locked pair of Helmholtz coils for further external drift compensation \cite{FluxGate,Repp}. By stabilizing the pressure of the evaporating liquid helium in the magnet, linear drifts of the magnetic field of $(1/B) (\Delta B/\Delta t)$ = 10 ppt/h \cite{UWPTMS} and $\sigma_{B}$ $\approx$ 300 ppt have been achieved in stabilized superconducting magnets \cite{FluxGate}, which is about 500 and 15 times better, respectively, than the magnet currently used. 

\subsubsection{Phase-sensitive detection methods}
It is planned to implement fast phase-sensitive detection methods, which offer several benefits for the magnetic moment measurements \cite{Cornell1989PnPMethod,StahlPhase,SturmPhase}. In the analysis trap, phase detection will be applied to detect the frequency shift $\nu_{z,\text{sf}}\approx230\,$mHz caused by a spin transition.

        \begin{figure}[htb]
        \centerline{\includegraphics[width=10cm,keepaspectratio]{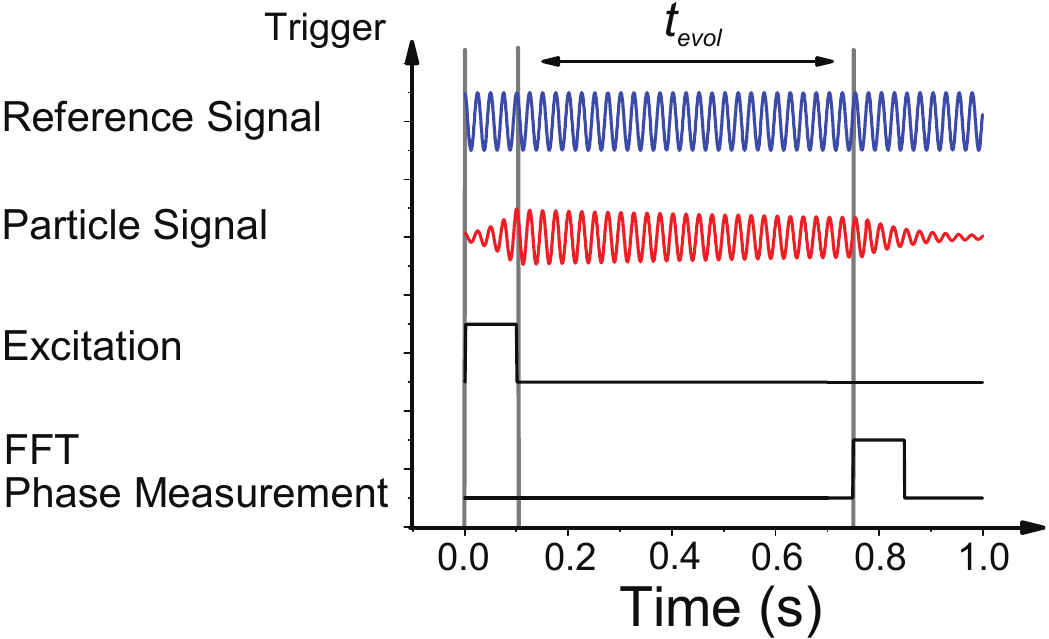}}
            \caption[Idea of Phase-Sensitive Detection Method]{Illustration of the phase-sensitive detection technique. For the details, see text.}
            \label{fig:PhaseSensIdea}
        \end{figure}

The basic principle is illustrated in Fig.~\ref{fig:PhaseSensIdea}. The axial motion of an antiproton with frequency $\nu_z$ is excited with the reference signal. Afterwards, the particle is decoupled from the drive and the axial detection system for a certain phase evolution time $t_{\text{evol}}$. Subsequently, the antiproton is recoupled to the detection system and the relative phase $\Delta\phi$ between the particle signal and the reference signal is measured
\begin{eqnarray}
    \Delta\phi_1(^\circ)=360^\circ(\nu_{\text{drive}}-\nu_z)\cdot t_{\text{evol}}~.
\end{eqnarray}
Once the axial frequency changes by $\Delta\nu_z$ a relative phase
    \begin{eqnarray}
    \Delta\phi_2(^\circ)=360^\circ(\nu_{\text{drive}}-(\nu_z+\Delta\nu_z))\cdot t_{\text{evol}}~
    \end{eqnarray}
will be measured. Finally, the difference of the relative phases of both measurements will be evaluated
    \begin{eqnarray}
    \Delta\phi(^\circ)=\Delta\phi_1(^\circ)-\Delta\phi_2(^\circ)=360^\circ\Delta\nu_z\cdot t_{\text{evol}}~.
    \end{eqnarray}
The comparison of two subsequent phase measurements with respect to the reference oscillator thus enables the resolution of tiny axial frequency differences. By using a phase evolution time of 1$\,$s, the phase difference would amount to $\Delta\phi=360^\circ\cdot\Delta\nu_z/\,$Hz. This would enable to resolve the expected axial frequency change of $\Delta\nu_\text{z,SF}\approx$230$\,$mHz caused by an antiproton spin transition with a phase difference of $\Delta\phi\approx83\,^\circ$ in the BASE analysis trap. 

The proof of principle of this measurement scheme has been carried out in the precision trap of the proton experiment at Mainz. Here, the phase measurement method was used to detect a 200$\,$mHz shift of the reference oscillator and the axial frequency of a proton.
    \begin{figure}[htb]
        \centerline{\includegraphics[width=0.70 \textwidth,keepaspectratio]{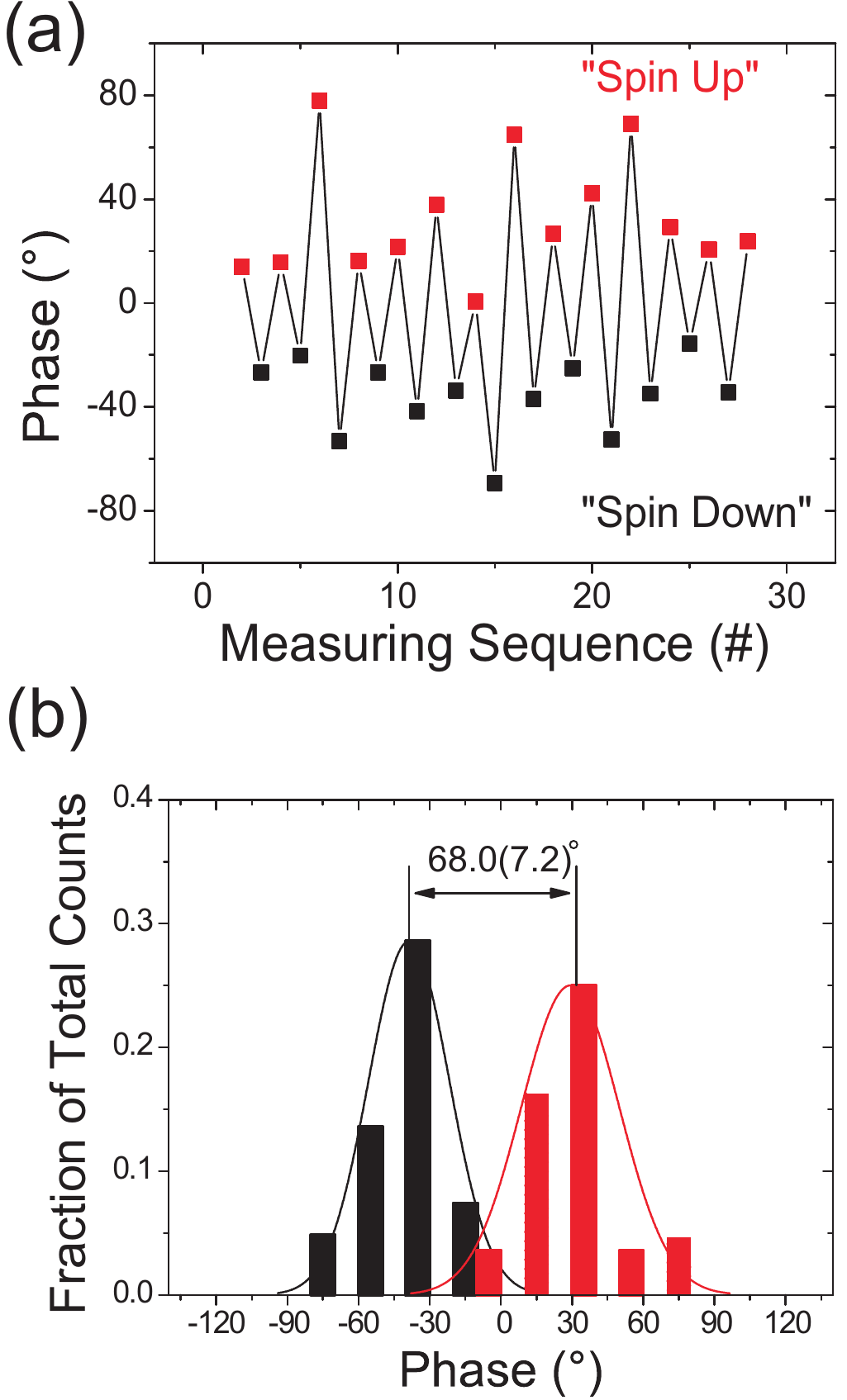}}
            \caption[Spin-Flip Resolution by Phase-Sensitive Detection]{Experimental demonstration of the resolution of a frequency jump of 200$\,$mHz in one second by phase-sensitive detection in the precision trap of the proton experiment in Mainz. (a) Measured relative phase as a function of the measuring sequence for a free phase evolution time of $t_\text{evol}=1\,$s. Two phase levels are observed, corresponding to the $200\,$mHz frequency shift of the reference oscillator. (b) The two data sets with different reference oscillator frequency are shown in a histogram. The overlap of the histograms is only $\approx 3\,\%$. Thus, only in 3 out of hundred measuring sequences it cannot be decided whether the spin has flipped or not.}
            \label{fig:PhaseSpinFlip}
    \end{figure}
In Fig.~\ref{fig:PhaseSpinFlip} (a) the phases measured in a sequence of 28 measurements at a free evolution time of $t_\text{evol}=1\,$s are shown. In the histogram in Fig.~\ref{fig:PhaseSpinFlip} (b) the data is separated into two sets, one for each reference oscillator frequency. The probability to measure the same frequency for the two different reference oscillator frequencies is only about 3$\,\%$, which would correspond to 97$\,\%$ spin-state detection fidelity. In addition, the measurement time to resolve the spin state is only 1 second, and thus 30-fold improved compared to dip detection. This fast detection scheme also allows as well for a much larger cyclotron-temperature acceptance in the analysis trap. Based on this study, we estimate that single spin-flip resolution can be achieved up to $T_+\approx5.4\,$K cyclotron temperature using such short averaging times. Thus, this technique has the potential to accelerate axial frequency measurements and the preparation time for spin flip experiments.

Further, faster phase-sensitive detection methods also exist for measuring $\nu_+$, such as the measurement techniques "pulse-and-phase" (P'n'P) \cite{Cornell1989PnPMethod} or "pulse-and-amplify" (P'n'A) \cite{SturmPhase}, allowing faster measurements with thus reducing the influence of magnetic field fluctuations. As these measurement techniques make direct measurements of $\nu_+$, they reduce the contribution of trapping potential instabilities to the modified cyclotron frequency fluctuation $\Xi_+$ by a factor of $\nu_z/\nu_c$ compared to the double-dip method. It was demonstrated that the modified cyclotron frequency can be resolved to 150 ppt within 10 s \cite{SturmPhase}, which enabled $g$-factor measurements of electrons in highly-charged ions down to 40 ppt uncertainty \cite{SvenSi28}.
Assuming that the magnetic field and voltage fluctuations can be reduced to $\sigma_{B} \leq$ 300 ppt and $\sigma_{V_R} \leq$ 40 ppt in shorter averaging time, respectively, and $\sigma_{r,B2}$ = 390 ppt, a $g$-factor line-width of about 500 ppt can be obtained, which enables measurements of $\mu_p$ and $\mu_{\overline{p}}$ with better than 100 ppt uncertainty. 

\subsubsection{Increased sampling rate and sidereal variation of the magnetic moments}
Another desired improvement for $g$-factor measurements is to increase the data collection rate, which was limited by the time required to cool the modified cyclotron motion to a sub-thermal energy below 150 $\mu$eV (1.7 K) for the spin state read-out \cite{MooserPRL2013}. The implementation of the cooling trap in BASE targets this limitation and removes the most significant time contributions of the cooling procedure due to a lower cooling time constant and by eliminating the transport time between PT and AT for the temperature measurement. Compared to our proton trap system in Mainz, we expect that this improves the data acquisition rate by an order of magnitude, so that a $g$-factor resonance with similar statistics to our measurement in ref.~\cite{MooserNature2014} can be recorded in the BASE setup within 2 weeks. 

A fast sampling of the proton/antiproton magnetic moments would open the possiblity to directly test the Standard Model Extension (SME) \cite{KosteleckyPenning} developed by Kostelecky \textit{et al.}, which introduces Lorentz- and CPT-violating terms into the Dirac equation while maintaining the essential features of local quantum field theories \cite{KosteleckySME}. The implications of the SME on the Larmor and cyclotron frequencies is discussed in detail in the references \cite{KosteleckyPenning,Kosteleckyg-2}. In summary, the cyclotron frequency $\nu_c$ and the anomaly frequency $\nu_a = \nu_L - \nu_c$ for protons and antiprotons are modified in the leading order of the SME as follows:
\begin{eqnarray}
\nu_c^p = \nu_c^{\overline{p}} \approx (1 - c_{00}^p - c_{11}^p -c_{22}^p) \nu_c, \\
\nu_a^{p/\overline{p}} \approx 2 \pi \nu_a \pm 2 b_3^p - 2 d_{30}^p m_p - 2 H_{12}^p,
\end{eqnarray}
where $b$, $c$, $d$ and $H$ denote the Lorentz-violating axial-vector, tensor, axial-tensor, and pseudoscalar fields of the SME, respectively, and the 3-axis points in the direction of the magnetic field. The parameters $c$, $d$, and $H$ are CPT-conserving, but the axial vector $b$ is a CPT-violating parameter. Thus, the cyclotron frequencies of both particles are shifted in the same way, but the difference in the Larmor frequencies of proton and antiproton becomes:
\begin{eqnarray}
\Delta\nu_L = (\nu_L^p - \nu_L^{\overline{p}}) = 4 b_3^p / (2 \pi).
\end{eqnarray}
Thus, a non-vanishing parameter $b^p$ would lead to a difference of the measured frequency ratios $\nu_L/\nu_c$ for protons and antiprotons, whereas the cyclotron frequency ratio $\nu_c^p/\nu_c^{\overline{p}}$ is insensitive to leading-order effects. Thus, by improving the measurement precision of $\mu_p$ and $\mu_{\overline{p}}$ the parameter $b^p$ can be constrained in our experiment in a direct particle/antiparticle comparison. If any difference between $\mu_p$ and $\mu_{\overline{p}}$ would be observed, the SME predicts that $\Delta\nu_L$ would exhibit diurnal variations as the magnet axis moves relative to $b^p$, since it is only sensitive to the component of $b^p$ parallel to the magnetic field. In that case, the diurnal variations would indicate that the $b$-field from the SME framework could describe the CPT violating mechanism. 

\subsubsection{Sympathetic cooling of protons/antiprotons with laser-cooled ions}
Further improvements in measurement precision and sampling rate would be provided by lower motional amplitudes and a deterministic cooling scheme, respectively. This can be achieved using laser-cooled ions for sympathetic cooling. Doppler laser-cooling of the motional states of single $^{40}$Ca$^+$ ions in a Penning trap to less than 10 mK temperatures has been demonstrated \cite{thompson2014}, and cooling to the ground state by optical sideband spectroscopy in a Penning trap has also been recently achieved \cite{goodwin_sideband_2014}. Thus, we consider to implement a sympathetic laser cooling scheme based on readily available laser-cooled ions. For our purpose, $^9$Be$^+$ ions are most suited, since these are the lightest ions relative to the proton/antiproton with good optical access \cite{OspelkausProc}. 

The cooling scheme will rely on a set of techniques proposed by Heinzen and Wineland ~\cite{Wineland1990} which would allow the coupling of the laser-cooled ion to a charged ``spectroscopy particle'', such as protons or antiprotons. This coupling provides cooling, state manipulation and readout of the spectroscopy particle via the laser-cooled ion. Adapting this scheme to single aluminum ions as spectroscopy particles has yielded atomic clocks of the highest accuracy~\cite{wineland_quantum_2001,chou_frequency_2010}. Because many of the steps were inspired by quantum logic experiments with single ions as qubits, this spectroscopy approach is now known as quantum logic spectroscopy~\cite{schmidt_spectroscopy_2005}.

In terms of the remote coupling between the proton/antiproton and the laser-cooled ion, it was suggested to hold the spectroscopy particle and the ion in two spatially separate potential wells with equal axial oscillation frequency, coupled to each other via image currents induced by the motion of the particles in a shared electrode \cite{Wineland1990}. This approach is currently pursued for Paul traps~\cite{daniilidis_wiring_2009} and for Penning traps in the context of mass measurements~\cite{cornejo_quantum_2012-1}.

The general idea was later adapted to a coupling of the proton/antiproton with an atomic ion via the Coulomb force between the two particles held in two spatially separated, but near-by potential wells without the use of a shared electrode~\cite{wineland_experimental_1998}. This direct, double-well approach has been demonstrated using pairs of atomic ions in microfabricated radio-frequency Paul traps in the context of quantum logic experiments \cite{Brown2011,harlander_trapped-ion_2011}. For the double-well coupling, the coupling strength between the two particles of charges $q_i$, masses $m_i$ and trap frequencies $\omega_i$ is given by \cite{Brown2011}
\begin{equation}
\Omega_\textrm{ex}=\frac{q_a q_b}{4\pi\epsilon_0 s_0^3\sqrt{m_a m_b}\sqrt{\omega_a\omega_b}},
\label{eq:coulomb_couling}
\end{equation}
where $s_0$ is the interparticle distance. As the coupling strength scales as $s_0^{-3}$, the distance between the ions needs to be small in order to produce high coupling rates. One of the important possible sources of disturbances, anomalous motional heating~\cite{turchette_heating_2000}, seems to scale as $d^{-4}$~\cite{deslauriers_scaling_2006}, where $d$ is the distance from the ion to the nearest electrode, although a recent study~\cite{goodwin_sideband_2014} has suggested a $d^{-3}$ scaling in macroscopic 3D traps. Depending on absolute heating rates achieved experimentally and on the assumed scaling, this may set a lower technical limit on the desirable trap size.

For demonstration and application to the proton, an electrostatic double-well potential in the axial direction is needed. For a negatively charged particle such as the antiproton, the electrostatic potential would be easier to engineer, as it would consist of a parabolic well for the atomic ion and an inverted parabola for the particle that is to be cooled sympathetically. When designing a potential, it must be kept in mind that the wells need different curvatures to account for the mass ratio between the particles. To minimize asymmetries, the mass ratio between the particles should be as close to unity as possible, making $^9$Be$^+$ the best candidate for our purpose, since it is the lightest ion with cooling transitions accessible by available lasers.

\begin{figure}[htb]
\centering{\resizebox{0.75\columnwidth}{!}{%
\includegraphics{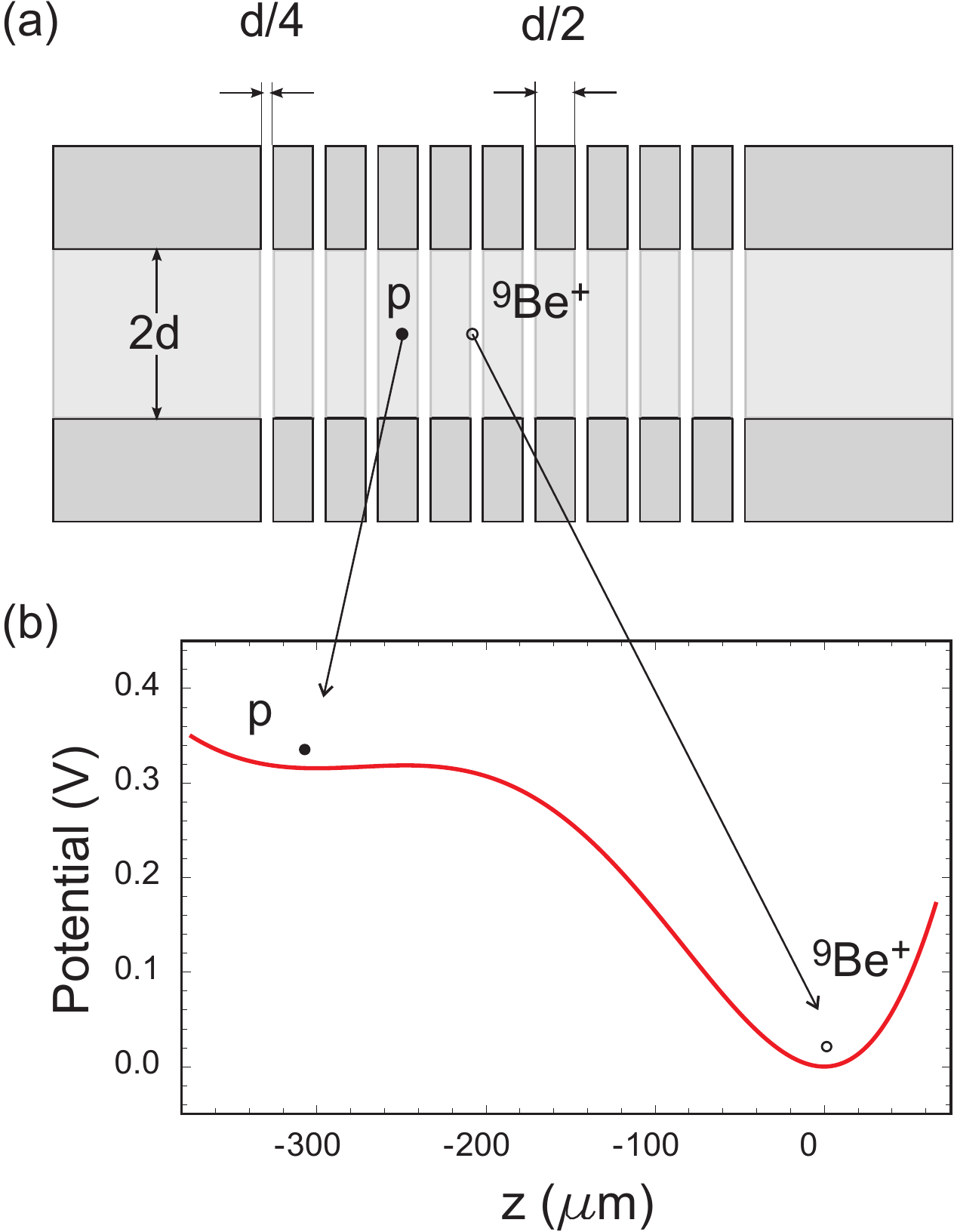} }}
\caption{(a) Cross section of the cylindrical geometry used for simulations. It consists of nine thin stacked electrodes and grounded longer endcaps. Spacers are omitted for clarity but are well receded for a high aspect ratio. The values used for $d$ in the text are 1.8 mm and 400 $\mu$m. The filled circle indicates the trap center for the proton, the hollow circle the atomic ion's location. (b) Double-well potential for a mass ratio of 9:1 in a trap of 800 $\mu$m diameter. For more details see text.}
\label{fig:sim_sketch}
\end{figure}

A possible design for such a double-well potential is a Penning trap consisting of multiple ring-shaped electrodes of a thickness half as large as their inner radii, separated by spacers as shown in Fig.~\ref{fig:sim_sketch}(a). Assuming a trap diameter of 3.6 mm, a well distance of 1.35 mm is achievable at an axial trap frequency of about 890 kHz, yielding $\Omega_\textrm{ex}=2\pi\cdot3.3$ s$^{-1}$. The bump between the potential wells is only 3 meV, which, while still a small value, is an order of magnitude above the thermal energy of a resistively cooled particle in a 4.2 K environment and as large as the value achieved in \cite{Brown2011}.

For higher coupling rates, we examine microfabricated Penning traps where all trap dimensions are decreased linearly, while the voltages are the same to maintain the well depth. In that case, the potential curvature is proportional to $d^{-2}$, and the trap frequencies scale as $d^{-1}$. As a result, reviewing the equation (\ref{eq:coulomb_couling}) for the coupling strength, $\Omega_\textrm{ex}$ scales with $d^{-2}$. Simulations of a trap stack as in Fig.~\ref{fig:sim_sketch}(a) with inner radius $d=400~\mu$m and electrodes of 200 $\mu$m width, separated by 50 $\mu$m spacers each, yield a potential which exhibits two minima with a curvature ratio of 9:1 separated by 300 $\mu$m, see Fig.~\ref{fig:sim_sketch}(b). At an axial frequency of 4 MHz, we obtain a coupling rate of $\Omega_\textrm{ex}=2\pi\cdot68$ s$^{-1}$ and a time of $\tau_\textrm{ex}$ of 3.7 ms for an energy swap between the particles. A trap stack of this design is currently under fabrication at the University of Hannover.

The integration of laser beams for cooling of the axial motion of the trapped ion is challenging in two regards. First, in BASE the laser beams can only be brought in along the axis of the magnet, in the space left available between the antiproton beam and the cryogenic shields. Second, the double-well trap does not provide easy optical access. Adequate geometries to address these challenges are under consideration. While sympathetic Doppler cooling will be one the immediate goals of this sympathetic laser cooling effort, the planned application of quantum logic spectroscopy~\cite{Wineland1990,wineland_experimental_1998} goes much further and also exploits the ion as a detector for both the spectroscopy particle's motional and Larmor frequencies. Beyond sympathetic cooling, this may enable further increases of the sampling rate and additional control over systematic effects. It would also make all quantized degrees of freedom of individual trapped protons/antiprotons amenable to high-fidelity deterministic preparation, control and measurement at the single quantum level.

\subsection{The magnetic moment of $^3$He}
Highly-precise measurements of magnetic fields are desired in various fields of experimental physics. A conventional method for this purpose is using nuclear magnetic resonance (NMR) sensors, where commonly used probes are based on the magnetic moment of protons in water $\mu'_p$. $^3$He can provide an alternative to perform measurements of strong magnetic fields in high-precision magnetometers, which have smaller systematic uncertainties concerning sample shape corrections, impurities and environment dependence. Making use of optical pumping and long relaxation times in low pressure $^3$He gas cells, relative magnetic field measurements up to 10$^{-12}$ precision within 6.6 s have been demonstrated \cite{HeilEPJ2014}. 

The magnetic moment of atomic $^3$He can be expressed in units of the Bohr magneton using the ratios $\mu'_{h}/\mu'_p$ \cite{Flowers} and $\mu'_p/\mu_B$ \cite{Phillips} with a total uncertainty of 12 ppb \cite{CODATA2010}. To provide an independent calibration for magnetic field measurements, the $^3$He magnetic moment $\mu'_{h}$ has to be determined with a better precision. The magnetic moment of the bare $^3$He nucleus $\mu_{h}$ can in principle be determined in the BASE apparatus. The shielding factor $\sigma_h$, which is required to obtain $\mu'_{h}=\mu_{h}(1-\sigma_h)$ contributes to $\mu'_{h}$only at a level of 0.1 ppb uncertainty \cite{CODATA2010}. 

$^3$He based magnetic field measurements are planned in muon storage ring experiments, where the muon anomalous magnetic moment $a_\mu=(g_\mu-2)/2$ is derived from the anomaly frequency $\nu_a = a_\mu q/m_\mu B$. Here, the magnetic field $B$ has been calibrated by measuring the magnetic field experienced by the muon beam using the NMR signal of protons in water \cite{Bennett2006}. Using $^3$He-based sensors, the magnetic field calibration of future experiments may be improved \cite{Chupp}. 

To achieve a high-precision measurement of $\mu_{h}$, an internal source \cite{HeliumSource} or external source \cite{Myers2015} for $^3$He would be required, so that $^3$He$^{2+}$ can be produced inside the Penning-trap system by electron impact ionization. The spin-state detection of the $^3$He nucleus is more challenging compared to the antiproton, since the frequency shift due to a spin-transition is reduced by the ratio $(\mu_{He}/\mu_{\overline{p}})/(m_{He}/m_{\overline{p}})\approx 1/4$ and becomes $\Delta\nu_{z,SF}$= 59 mHz only. However, this is about a factor 2 larger than the frequency fluctuation due to voltage instabilities of the ring voltage, and the frequency shifts due to cyclotron transitions are reduced to 18 mHz as well. Thus, the continuous Stern-Gerlach effect can be applied to measure $\nu_L/\nu_c$ for $^3$He using the double-trap technique to a precision on the ppb level as well.

\subsection{Comparison of the proton and antiproton charge-to-mass ratios}
As discussed in Sect.~\ref{sec:FM:APE}, adiabatic shuttling enables us to perform fast high-precision measurements of the cyclotron frequency ratio of two particles. This technique can be applied using antiprotons and H$^-$ ions to compare the proton and antiproton charge-to-mass ratios by converting the H$^-$ cyclotron frequency with the ratio $R$
\begin{eqnarray}
    R = \frac{\left(q/m\right)_{\overline{p}}}{\left(q/m\right)_{H^-}} = 1.001089218750(2),
\end{eqnarray}
to the one for the proton \cite{JerryAntiproton}, accounting for the two additional electrons and their binding energies. In addition, the polarization shift $\alpha B^2$ of the $H^-$ ion's cyclotron frequency needs to be considered \cite{Pritchard}, which can be regarded as the $H^-$ ion having a lower effective mass. Therefore, $R$ needs to be corrected by 4 ppt in our magnetic field of $B=$ 1.945 T\cite{UlmerNature2015}.

Antiproton/H$^-$ cyclotron frequency ratios were first measured by the TRAP collaboration at LEAR \cite{JerryAntiproton}, where the two particles were simultaneously stored in the same trap. One particle was centered in the trap and its modified cyclotron frequency was measured directly, whereas the other particle was stored on a large cyclotron orbit to avoid a perturbation of the measurement. After each individual frequency measurement the particles were exchanged by resistive cooling and resonant excitation, respectively. In these experiments the particle interchange times were about 2$\,$h. Together with the high-precision cyclotron frequency measurements, which were carried out by direct detection of the modified cyclotron mode, one charge-to-mass ratio comparison took about 3$\,$h. In total, a relative uncertainty of 90 ppt was achieved in this measurement \cite{JerryAntiproton}.

We have recently performed a new measurement of the antiproton and H$^-$ ion cyclotron frequency ratio \cite{UlmerNature2015}. The two particles are stored in two separate potential wells and are alternately placed in the trap center using the adiabatic shuttling method (see Sect.~\ref{sec:FM:APE}). The cyclotron frequency measurement takes place in thermal equilibrium with the axial detection system by using the sideband coupling method (see Sect.~\ref{sec:EP:SC}). In total, about 6500 cyclotron ratios have been measured in 35 days, reaching a statistical uncertainty of 64 ppt. The largest systematic uncertainty in our measurements originates from the so-called trap asymmetry, which causes a difference in the position of the trap center for the two particles. To bring the antiproton and the H$^-$ ion in resonance with the axial detection system, the ring voltage needs to be adjusted by 5 mV. Due to machining tolerances, and patch and contact potentials present on the trap electrodes, this shifts the equilibrium position by as little as about 30$\,$nm. In presence of the residual magnetic field gradient $B_1 = 7.58(42)$\,mT in our measurement trap, the difference in magnetic field experienced by these slightly displaced particles needs to be corrected for. By careful systematic measurements, we were able to determine the uncertainty of the shift to 26 ppt \cite{UlmerNature2015Supplementaries}. In total, we obtained
\begin{eqnarray}
\frac{\left(q/m\right)_p}{\left(q/m\right)_{\overline{p}}}-1=1(69)10^{-12}.
\end{eqnarray}
Thus, according to our measurement CPT invariance holds up to this high level of precision.

This result could be further improved using the phase-sensitive detection methods \cite{Cornell1989PnPMethod,SturmPhase} and the magnetic field stabilization measures as discussed in Sect.~\ref{sec:NP:DT}. The scatter of the cyclotron frequency ratio due to the non-linear fluctuations of $B$ can be as low as 300 ppt, which would tremendously reduce the data collection time required for a high-precision comparison. To reduce the systematic uncertainty due to the trap asymmetry, the magnetic field gradient $B_1$ has to be reduced. To this end, the ferromagnetic ring electrodes of the cooling trap and analysis trap can be replaced by regular copper ring electrodes in a dedicated mass measurement run. A second possibility is to compensate the magnetic field gradient locally in the measurement trap by using the magnet's shim coils or by adding additional shim coils around the trap can. Thereby, $B_1$ can be reduced by at least two orders of magnitude, which reduces systematic corrections to about 1 ppt. Other systematic shifts such as relativistic, image charge or image current shifts contribute less than 0.05 ppt due to the large trap radius and the low kinetic energy. Thus, the cyclotron frequency ratio can be measured in principle down to the limit given by the uncertainty of the frequency ratio $R$ of the H$^-$ to proton $q/m$ ratio of 2 ppt. To further improve the CPT test beyond this value, a more precise measurement of the proton mass is required, which is currently limiting the uncertainty of $R$.



\subsection{High-precision mass measurements}

The principle of comparing cyclotron frequencies of two particles using the adiabatic transport can be also applied to determine absolute atomic masses. To this end, (highly) charged carbon ions can be used as reference particles to calibrate the measurement to the atomic mass standard. Corrections to the reference masses of carbon ions due to the electron mass and binding energies contribute less than 0.1 ppt uncertainty to a measurement. C$^{6+}$ reference ions can be produced by charge breading of carbon in the reservoir trap. In this way, the atomic masses of the proton, the antiproton and other light ions can also be measured in the BASE apparatus. 

The proton mass has been determined in several Penning trap experiments \cite{Pritchard1995,VanDyck1999,Solders2008} with 140 ppt as best single measurement uncertainty. Combining the measurements, the CODATA group has evaluated the proton mass with 89 ppt uncertainty \cite{CODATA2010}. Using cyclotron sideband measurements and adiabatic shuttling, an improvement by a factor of two in a measurement time of two months can be reached with the current performance of the BASE apparatus. For the antiproton, cyclotron frequency ratio measurements with proton and H$^-$ ion \cite{JerryAntiproton,Jerry1995} have been carried out. In addition, the antiproton-to-electron mass ratio has been determined from spectroscopy of antiprotonic helium \cite{Hori}. However, a cyclotron frequency ratio measurement to carbon ions would constitute the first direct atomic mass measurement of the antiproton. Obviously, high-precision mass measurements in the BASE trap system would also benefit from the implementation of the phase-sensitive detection and magnetic field stabilization.

\section{Conclusion and Outlook}

The BASE apparatus, which is dedicated to measure the antiproton magnetic moment with highest precision, has been commissioned at the Antiproton Decelerator at CERN. In the first online run, the essential techniques to catch and cool antiprotons to cryogenic temperatures have been demonstrated, as well as the preparation of single antiprotons for the magnetic moment measurement cycle. In addition, the fast adiabatic exchange of two particles to measure cyclotron frequency ratios with measurement cycle length of four minutes has been demonstrated, which is a 50-fold improvement to the last experiment comparing the proton-antiproton charge-to-mass ratios \cite{JerryAntiproton}. The commissioning of the four-Penning trap system is ongoing with the detection of single antiproton spin-flips being the next milestone. Using the continuous Stern Gerlach effect to detect single antiproton spin transitions for the spectroscopy of the spin-precession frequency, and the sideband coupling technique for the direct measurement of the cyclotron frequency, the antiproton magnetic moment can be determined with a precision below 1 ppb, which will lead to a 1000-fold improved test of the CPT invariance by comparing the proton and antiproton magnetic moments.

\section{Acknowledgements}
We acknowledge financial support of RIKEN Initiative Research Unit Program, RIKEN President Funding, RIKEN Pioneering Project Funding, RIKEN FPR program, RIKEN JRA program, the Max-Planck Society, the CERN fellowship program, the BMBF, the Helmholtz-Gemeinschaft, and the EU (ERC Advanced Grant No. 290870-MEFUCO and ERC Starting Grant No. 337154-QLEDS). N.L. was supported by a Marie Curie International Incoming Fellowship within the 7th European Community Framework Programme. We acknowledge support from CERN, in particular, the AD group, the engineering department, the vacuum-group, the rf-group, the integration service, the HSE department, the magnet group and the survey group of CERN. We would like to express our gratitude towards F. Butin for his strong support regarding integration of the experiment into the AD hall. 

%
%

\end{spacing}

\end{document}